\documentclass[aps,prd,twocolumn,superscriptaddress,groupedaddress,preprintnumbers,nofootinbib]{revtex4}

\usepackage{CJK}
\usepackage[english]{babel}
\usepackage{array}
\usepackage{tabulary}
\usepackage{amsfonts}
\usepackage{amssymb,amsmath}
\usepackage{cancel}
\usepackage{mathcomp}
\usepackage{graphicx}
\usepackage{color}
\usepackage{comment}
\usepackage{enumerate}
\usepackage{natbib}

\def\beq{\begin{equation}}
\def\eeq{\end{equation}}
\def\baq{\begin{align}}
\def\eaq{\end{align}}
\def\d{{\rm d}}

\begin{document}
\begin{CJK*}{}{}
\title{Linear Perturbations in a coupled cosmon-bolon cosmology}

\preprint{}

\author{J.~Beyer \\
{\it Institut f\"ur Theoretische Physik,   
Universit\"at Heidelberg,   
Philosophenweg 16, 69120 Heidelberg, Germany \\
}}

\begin{abstract}
\begin{center}
We investigate linear perturbations in the recently proposed cosmon-bolon model of coupled scalar field dark matter and quintessence. We provide an analytical mechanism to average over the quick oscillations appearing both in the background and at the perturbative level and evolve the effective equations numerically. The resulting matter power spectra are used to predict total halo number counts as well as substructure abundances in a typical galaxy by employing the extended Press-Schechter excursion set approach. We discuss in some detail the ambiguities arising in this formalism, starting from issues with generalizing spherical collapse to our model to filter choices and different barriers. The results are used to put a lower bound on the current bolon mass of roughly $9 \times 10^{-22}$ eV. 
\end{center}
\end{abstract}

\maketitle

\section{Introduction}

For some years now the cosmological standard model (or $\Lambda$CDM-model) has been widely successful in explaining the vast majority of cosmological observations. Still the nature of two of its key components, dark matter and dark energy, remains a mystery. So far both of these substances have eluded any direct detection and can be seen only through their gravitational effects. This leaves plenty of room for speculation, in particular a comparatively strong coupling between the two dark sectors is possible \cite{Amendola:1999er,Pettorino:2012ts,Amendola:2011ie}.

Furthermore the $\Lambda$CDM model faces two severe theoretical problems. In order to explain the current accelerated expansion of the universe, the cosmological constant $\Lambda$ needs to have a tiny value, contradicting expectations from quantum field theory. This is known as the {\it cosmological constant problem}. The other issue concerns the question why the energy density associated with $\Lambda$ is of the same order of magnitude as the energy density of the universe just recently (the {\it why now} or {\it coincidence problem}). 

It is possible to alleviate both of these problems in alternative scenarios where dark energy is dynamical and evolves in time. The first examples of such models, thought up long before the observational discovery of cosmic acceleration, were closely related to the concept of anomalous dilatation symmetry \cite{Wetterich:1987fk,Wetterich:1987fm,Peebles:1987ek,Ratra:1987rm}. Recently, studies of dilatation symmetric theories of gravity in higher dimensions have brought renewed attention to the subject \cite{Wetterich:2008bf,Wetterich:2009az,Wetterich:2010kd}. Amongst such models, those which allow for a dimensional reduction to an effective four-dimensional theory are of particular interest. It was found that for this class, all stable quasistatic solutions of the field equations lead to a vanishing effective four-dimensional cosmological constant. If such a theory is realized as a fixed point of the renormalization group flow of the effective action, it is natural to assume that the cosmological vacuum solution approaches this fixed point for $t \rightarrow \infty$. At the fixed point dilatation symmetry implies the existence of a massless goldstone boson, the dilaton, but for finite times the vacuum solution breaks this symmetry. As a result, the dilaton becomes a pseudo-goldstone boson with a small, time-dependent mass.

In a cosmological setting the dilaton can play the role of a scalar ''cosmon'' field, rolling down its anomalous potential with a slowly decreasing mass and acting as dark energy. The anomalous potentials generated in such models typically give rise to scaling solutions, in which the energy density of the cosmon tracks that of the dominant background fluid, thereby alleviating the coincidence problem \cite{Wetterich:2009az,Copeland:1997et,Liddle:1998xm,Steinhardt:1999nw,Zlatev:1998tr}. 

Besides the cosmon, dimensional reduction of higher-dimensional theories of gravity usually divulges additional scalar degrees of freedom. In a recent paper \cite{Beyer:2010mt} we discussed a class of simple two scalar field models motivated by this scenario and showed that one can obtain a realistic model of coupled dark energy and dark matter already for rather simple anomalous potentials. We named the second field ''bolon'' (from the Greek \textit{bolos} meaning "lump"), since it is responsible for the formation of structure in the universe in this picture. In this work we will refine our previous analysis to include linear perturbations as well as predictions of abundances of cosmic structure using the extended Press-Schechter (ePS) formalism. We will at several points refer to an accompanying paper \cite{EarlyScalings}, which analyzes the evolution of the cosmon and bolon dynamics during the early universe.

This paper is organized as follows: 

In the Sec. \ref{sec:Model} we will introduce the coupled cosmon-bolon model and elaborate on the classes of scalar potentials we are investigating. We continue to study the background evolution of these models in Sec. \ref{sec:Background}. This was already discussed in ref. \cite{Beyer:2010mt}, but here we present a much more detailed analysis, the results of which will be essential in our treatment of the linear perturbations. Those will be dealt with in section \ref{sec:LinearPerturbations}, where we deduce an effective description of linear perturbation growth valid for late enough times and provide numerical simulations,  building on the results of ref. \cite{EarlyScalings}. Finally we calculate some testable predictions of our model in section \ref{sec:PressSchechter}, where we apply the ePS mechanism to predict structure and substructure abundances in the universe and compare our results to warm dark matter models as well as the cosmological standard model.
We present our conclusions in Sec. \ref{sec:Conclusion}.

\section{The model}
\label{sec:Model}

The breaking of dilatation symmetry by the cosmological vacuum solution introduces an anomalous potential $V(\varphi,\chi)$ for the cosmon $\varphi$ and the bolon $\chi$. Dilatation symmetry at the fixed point ensures that the potential vanishes for $\varphi \rightarrow \infty$ (we assume wlog that the cosmon is rolling towards higher field values throughout its evolution). It is this requirement which makes a dark sector coupling an integral part of our model, since asymptotic dilatation symmetry could not be guaranteed otherwise. 
Since the potential arises from anomalies in the quantum effective action, the field equations which can be derived from the principle of stationary action are exact, no additional quantum corrections are present.

We will perform our analysis in the effective four-dimensional theory in the Einstein-frame where the Planck mass is fixed and restrict our attention to a simple class of models for which the scalar potential can be split up as follows:
\beq
V(\varphi,\chi) = V_1(\varphi) + V_2(\varphi,\chi) \, .
\eeq

The quintessence potential we are employing in this work is the often used exponential potential, which arises naturally as an anomalous potential in the context of higher dimensional dilatation symmetric theories \cite{Wetterich:2008bf,Wetterich:2009az,Wetterich:2010kd}, but in other theories of particle physics as well \cite{Wetterich:2008sx}:
\begin{align}
\label{CosmonPotential}
V_1(\varphi) = M^4 {\rm e}^{-\alpha \varphi/M} \, .
\end{align}
The bolon-potential is assumed to have a minimum with a non-vanishing second derivative around which it will stabilize during the later stages of its evolution. Such a behaviour ensures that is will behave like a dark-matter candidate. The specific form of the potential we use is similar to one originally introduced in \cite{Matos:2000ng,Matos:2000ss}, but with an additional coupling to the cosmon field:
\begin{align}
\label{BolonPotential}
V_2(\varphi,\chi) = M^4 c^2 {\rm e}^{-2 \beta \varphi/M} \left( {\rm cosh} \left( \frac{\lambda \chi}{M} \right) -1\right) \, .
\end{align}
The dimensionless constant $c$ in this potential is closely related to the scale of anomalous symmetry breaking of dilatation symmetry, the only mass scale intrinsic to our model other than the Planck scale, as is discussed in more detail in ref. \cite{Beyer:2010mt}.
Asymptotically the bolon-dependence of this potential can be decomposed as follows
\beq
\label{AsymptoticCoshPotential}
{\rm cosh}\left( \lambda \chi / M \right) - 1 \approx \left\{
\begin{array}{l l}
\frac{1}{2} e^{ \left| \lambda \chi / M \right|} \quad \left| \lambda \chi / M \right| \gg 1 \\
\frac{1}{2} \frac{\lambda^2 \chi^2}{M^2} \quad \; \; \, \left| \lambda \chi / M \right| \ll 1 \\
\end{array} 
\right. \, ,
\eeq
which ensures the required quadratic $\chi$-dependence for small field values, with a mass given by
\beq
m_\chi(\varphi)^2 = c^2 M^2 \lambda^2 {\rm e}^{-2 \beta \varphi / M} \, .
\eeq 
For larger field values the potential gets much steeper, which is a necessary feature for a model exhibiting scaling solutions that lead to an insensitivity of the cosmological evolution to initial conditions for a wide range of initial field values. 
Furthermore, at the level of linear perturbations, it ensures the existence of a dominant adiabatic mode in the early-universe evolution, i.e. it suppresses potentially strongly growing isocurvature modes which can be present in the case of a ''frozen'' field or power-law potentials (see accompanying paper ref. \cite{EarlyScalings} for more details on this).

We choose $\alpha > 0$ and $\lambda > 0$ throughout this paper, which can always be achieved by a suitable redefinition of $\varphi$ and $\chi$. Our motivation then requires $\beta > 0$ also. We will however at some stages refrain from this constraint and consider $\beta < 0$ as well.
Interesting generalizations of our model which can also arise from dilatation anomalies include a $\varphi$-dependent exponential $\alpha(\varphi)$ in the quintessence potential, a $\varphi$-dependent coupling $\beta(\varphi)$ or a non-constant minimum of the bolon-potential $\chi_0(\varphi)$. We will discuss some of these possibilities and their connection to an acclerated late-time expansion of the universe below.

The common part of the potential introduces a coupling between the two scalar fields. The form of this coupling can be easily derived from (non-)conservation of the energy-momentum tensor, which reads for a generic component of the cosmic fluid (denoted by subscript $\alpha$)
\beq
\mathcal{D}_\mu T^{\mu \nu}_{\alpha} = Q^\nu_{\alpha} \, .
\eeq 
The total energy-momentum tensor is of course conserved and therefore the sum of all couplings is subject to the constraint
\beq
\sum_{\alpha} Q_{\alpha}^\nu = 0 \, .
\eeq
Furthermore, in an FLRW-background, each coupling is constrained by the usual symmetry assumptions of spatial isotropy and homogeneity, implying
\beq
\label{CouplingDefinition}
Q^{0 \, \mu}_{\alpha} = (-a Q_{\alpha},0,0,0) \, ,
\eeq
where the superscript $0$ denotes a background quantity. We introduce a commonly used dimensionless coupling via
\beq
\label{DimensionlessCouplingDefinition}
q_{\alpha} \equiv \frac{aQ_{\alpha}}{3h(1+\omega_{\alpha})\rho_{\alpha}} \, ,
\eeq
where the background energy-densities and pressure for the scalar fields are defined as follows:
\begin{align}
\label{EnergyDefinition}
\rho_{\varphi} =& \frac{1}{2 a^2} \varphi'^2 + V_1(\varphi) \, , \quad
\rho_\chi =& \frac{1}{2a^2} \chi'^2 + V_2(\varphi,\chi) \, , \\
\label{PressureDefinition}
p_{\varphi} =& \frac{1}{2 a^2} \varphi'^2 - V_1(\varphi) \, , \quad
p_\chi =& \frac{1}{2a^2} \chi'^2 - V_2(\varphi,\chi) \, .
\end{align}
Here and below a prime denotes a derivative with respect to conformal time. We will always assume standard kinetic terms for both scalar fields throughout this work. Non-standard kinetic terms can arise in the process of dimensional reduction \cite{Wetterich:2009az,Wetterich:2010kd} and can have interesting cosmological consequences, for example in k-essence models \cite{Garriga:1999vw,ArmendarizPicon:1999rj,ArmendarizPicon:2000dh,Chiba:1999ka,ArmendarizPicon:2000ah}, but we will not consider them further here.

The equation of energy conservation can now be written as
\beq
\label{GenericEnergyConservation}
\rho_{\alpha}' = -3h(1-q_{\alpha})(1+\omega_{\alpha})\rho_{\alpha} \, ,
\eeq
where (as usual) $\omega_{\alpha} = p_{\alpha}/\rho_{\alpha}$ is the equation of state of the component denoted by $\alpha$ and $h=a'/a$ is the conformal Hubble rate. For our model the dimensionless couplings read
\begin{align}
\label{CosmonCoupling}
q_\varphi & = - \frac{\varphi' V_{2,\varphi}}{3h\rho_\varphi (1+\omega_\varphi)} \, , \\
\label{BolonCoupling}
q_\chi & = - \frac{(1+\omega_\varphi) \rho_\varphi}{(1+\omega_\chi) \rho_\chi} q_\varphi = \frac{\varphi' V_{2,\varphi}}{3h\rho_\chi (1+\omega_\chi)} \, .
\end{align}
We note that a coupling to the trace of the energy momentum tensor, as is often assumed in models of coupled quintessence \cite{Amendola:1999er}, is not inherent to our model. It will however arise as a consequence of the dynamical evolution of both fields during the later stages of the evolution of the universe, as we will see in the next section \ref{sec:Background}. The same holds true for perturbations at the linear level, which we will discuss in section \ref{sec:LinearPerturbations}.

\section{Background evolution}
\label{sec:Background}

\subsection{Basic equations}
\label{sec:BackgroundEquations}

In a standard Friedmann-Lemaitre-Robertson-Walker (FLRW) universe, the field equations for two canonical scalar fields which couple through their potential read
\begin{align}
\label{CosmonFieldEquation}
\varphi'' + 2h\varphi' + V_{,\varphi} &= 0 \, , \\
\label{BolonFieldEquation}
\chi'' + 2h\chi' + V_{,\chi} &=0 \, ,
\end{align}
where $V_{,\varphi}$ and $V_{,\chi}$ are derivatives of the potential with respect to the scalar fields. Einstein's equations give the usual Friedmann equations: 
\begin{align}
\label{FriedmannEquation}
h^2 &= \frac{a^2}{3M^2} \sum_\alpha \rho_\alpha \, , \\
\label{FriedmannEquation2}
h' &= -\frac{a^2}{6 M^2} \sum_\alpha \left( \rho_\alpha+ 3 p_\alpha \right) \, ,
\end{align}
where $M$ is the reduced Planck mass. The index $\alpha$ runs over all particle species present in the universe. In this work we use the common convention that the scale factor today should equal 1.

\subsection{Tracking in the early universe}
\label{sec:TheEarlyUniverse}

The dynamics of the evolution of the cosmon-bolon system in the early universe have been investigated in an accompanying paper \cite{EarlyScalings}, using the approximation of the common potential given by equation (\ref{AsymptoticCoshPotential}) for large field values. A dynamical system analysis revealed that the existing stable scaling solutions split the parameter space for this model into disjunct sections. The scaling solution relevant for our scenario is the one denoted by R4 in ref. \cite{EarlyScalings}, which is the only one allowing for a radiation-like expansion with a continuous range of couplings extending from the negative to the positive realm. It is also the unique stable scaling solution for the range of parameters which are phenomenologically interesting, as we will see in section \ref{sec:LinearPerturbations}. All other stable fixed points do either not provide a realistic early cosmology (i.e. a radiation-like expansion) or require large couplings, which can be observationally excluded, as we discuss below. The constraints on the model-parameters arising from the demands of existence and stability of this fixed point can be read off of Table II in ref. \cite{EarlyScalings} and are given by
\beq
\label{EarlyUniverseParameterConstraints}
\beta/\alpha < \frac{1}{2} \, , \; \frac{\alpha \beta}{\lambda^2 + 4 \beta^2} \leq \frac{1}{2} \, , \;   \alpha>2 \, , \;  \lambda^2 > \frac{4(2\beta  - \alpha)^2}{\alpha^2-4} \, .
\eeq
For the positive range $0 < \beta / \alpha < 1/2$ the additional condition
\beq
\lambda^2 > 2 \beta (\alpha-2 \beta)
\eeq
is required.
We will therefore restrict ourselves to this parameter range from now on.

The evolution of the scalar fields during this phase of the cosmic evolution can be obtained analytically from Table I in ref. \cite{EarlyScalings} and reads
\begin{align}
\label{EarlyPhiEvolution}
\varphi(h) &= - \frac{M}{\alpha} {\rm ln}\left[ \text{f}(\alpha, \beta, \lambda) \frac{h^2}{a^2 M^2} \right] \, , \\
\label{EarlyChiEvolution}
\chi(h) &= \frac{M}{\lambda} {\rm ln} \left[ 8 (1-2\beta/\alpha) \frac{h^2}{a^2 m_\chi(\varphi)^2} \right] \, ,
\end{align}
where
\beq
\text{f}(\alpha,\beta,\lambda) =  \frac{4 (\lambda^2 + 4 \beta^2 -2 \alpha \beta)}{\alpha^2 \lambda^2} \, .
\eeq
The energy densities of the cosmon and the bolon contribute only a fraction of the total early energy density of the universe for this solution and the scalar density parameters are given by
\begin{align}
\label{omegaPhiES}
\Omega_{\varphi,{\rm es}} = \frac{4}{\alpha^2} + \frac{8}{3} \frac{\beta/\alpha (-1+2 \beta/\alpha)}{\lambda^2} \, , \\
\label{omegaChiES}
 \Omega_{\chi,{\rm es}} = \frac{4}{\lambda^2} + \frac{8}{3} \frac{\beta/\alpha (-5+4 \beta/\alpha)}{\lambda^2} \, ,
\end{align}
where the subscript ''es'' stands for ''early scaling''.

The stability of these solutions is of course only guaranteed as long as baryons can be neglected and the exponential approximation for the bolon-potential is valid. Both of these assumptions will eventually be violated and the early scaling solution will be broken by a transition to a matter dominated era, where the bolon quickly oscillates around the minimum of its potential and acts like dark matter. To estimate when this happens, we simply extrapolate the solution of the bolon evolution during the early scaling epoch until it reaches a values of $M/\lambda$. According to equation (\ref{EarlyChiEvolution}) this happens when $h^2/a^2 m_\chi(\varphi)^2 \approx {\rm e}/8(1-2\beta/\alpha)$, which is of order $1$.

\subsection{The late universe}
\label{sec:LateUniverse}

For sufficiently late times the cosmic evolution drives the bolon towards the minimum of its potential, so that it eventually acquires very small field values $\chi/M$. For such small deviations from the minimum we can approximate the bolon potential by
\beq
V_2(\varphi,\chi) =  \frac{m_\chi(\varphi)^2}{2} \chi^2 \, ,
\eeq
with $m_\chi(\varphi) = m_0 \, {\rm exp}(-\beta \varphi / M)$ and $m_0 \equiv Mc\lambda$. The dynamics of the bolon in this regime depend on the ratio $\mu \equiv h/am_\chi(\varphi)$. For $\mu > 1$ the Hubble friction is strong enough to keep the bolon field frozen at some (almost) constant value, whereas for $\mu < 1$ we expect rapid oscillations to occur. As we have seen in section \ref{sec:TheEarlyUniverse}, a scaling scenario in the early universe will deliver the bolon to field values of the order $M/\lambda$ when $\mu \approx 1$. We can track the subsequent behavior of this parameter by employing the following formulae:
\begin{align}
\label{MuHEquation}
\mu =& \frac{h}{a m_0} {\rm e}^{\beta \varphi/M} = \frac{h}{a m_0} \left( 3 \frac{h^2}{a^2 M^2} \Omega_{\varphi,{\rm pot}} \right)^{-\beta/\alpha} \, , \\
\label{dMuEquation}
\mu' =& \mu h \left(-\frac{3}{2} (1+\omega_{\rm eff}) + \beta \sqrt{6 \Omega_{\varphi,{\rm kin}}} \right) \, .
\end{align}
If the quintessence field exhibits a scaling or tracking behavior, i.e. the the density parameter $\Omega_\varphi$ is (almost) constant and not bigger than a few percent, equation (\ref{dMuEquation}) directly tells us that $\mu$ is decreasing (as long as $H=h/a$ is getting smaller, which is a very generic requirement, and $\beta$ is not excessively large). The bound on the coupling $\beta$ we get from equation (\ref{MuHEquation}) in conjunction with the requirement of a decreasing $\mu$ is:
\beq
\beta/\alpha < 1/2 \, .
\eeq
This is a restriction we already found in section \ref{sec:TheEarlyUniverse} by requiring the existence of a radiation-dominated scaling solution in the early universe (see equation (\ref{EarlyUniverseParameterConstraints})) and also not in conflict with the bound coming from the existence of a bolon-dominated scaling solution in the late universe (see below) for our specific model. Furthermore equation (\ref{MuHEquation}) depends crucially on the exponential shape of the quintessence potential, while the conclusion we have drawn from equation (\ref{dMuEquation}) is much more general and is valid for all quintessence fields with a canonical kinetic term and a potential exhibiting scaling or tracking behavior. We have therefore shown that a decreasing $\mu$ is a quite generic feature in realistic quintessence scenarios and we will now consider the dynamics in the regime $\mu \ll 1$.

The evolution of this system has already been described in ref. \cite{Beyer:2010mt}, but we will present a much more detailed analysis here, which we will need later when we treat linear perturbations in section \ref{sec:LinearPerturbations}. The method we apply is derived from one used for the case of a single scalar field in a harmonic potential (see e.g. ref. \cite{Gorbunov:2011zzc}), which we will now generalize to our coupled scenario. 
The basic idea is to first expand all dynamical quantities in a Taylor-series in $\tilde{\mu}=h_0/m_0$. Since $\tilde{\mu}=\mu \, {\rm e}^{\beta \varphi/M} \, a h_0/h$, this quantity is always smaller than $\mu$ for $a<1$ as long as the coupling is not too large, but has the advantage of being time-independent. To give an example, the bolon-field $\chi$ can be expanded as follows:
\beq
\label{MuExpansion}
\chi = \sum_j \tilde{\mu}^j \chi_j \, .
\eeq
We then segment all the Taylor-Coefficients $\chi_j$ into a Fourier-type sum, given by
\beq
\label{FourierTypeExpansion}
\chi_j = \sum_{n \epsilon \mathbb{N}} \chi_{j1,n} {\rm cos}(nx) + \chi_{j2,n} {\rm sin}(nx) \, ,
\eeq
with $n \in \mathbb{N}$.
Here the coefficients $\chi_{i1,n}$ and $\chi_{i2,n}$ are of course time-dependent, but are evolving slowly, i.e. remain almost constant at a time scale $1/m_\chi(\varphi)$. The oscillation frequencies are also time-dependent and given by multiples of the base frequency
\beq
\label{baseFrequency}
x(\eta)=\int_{t_0}^{t(\eta)} m_\chi(\varphi(t')) dt' \, ,
\eeq
with t being the cosmic time and $t_0$ being some suitable early time, chosen such that any phase potentially appearing in the trigonometric functions gets cancelled. The whole expression should of course be read as a function of conformal time. 

Which frequencies appear at which order in $\tilde{\mu}$ can now be seen simply by plugging the most general ansatz into the field equations (\ref{CosmonFieldEquation}) - (\ref{FriedmannEquation}) and comparing coefficients for each trigonometric function at each order in $\tilde{\mu}$. The whole set of resulting differential equations for the Taylor-coefficients can be found in appendix \ref{app:bgExpansion}. Here we simply give the results.

The scale factor $a$, the hubble $h$, the cosmon-field $\varphi$ and all additional energy densities (i.e. photons, neutrinos and baryons) evolve slowly to leading order. We denote the first term in the Taylor-expansion (\ref{MuExpansion}) of such adiabatic quantities with a bar, e.g. $\bar{a}$. The only quantity which is oscillatory at leading order is the bolon field $\chi$, the leading order coefficient is $\chi_{11,1}$ which we denote by $\chi_0$ for simplicity. 
Evaluating the Friedmann equation to leading order gives
\beq
\label{AveragedFriedmannEquation}
\bar{h}^2 = \frac{\bar{a}^2}{3M^2} \left( \bar{\rho}_\chi + \bar{\rho}_\varphi + \bar{\rho}_{\rm ext} \right) \, ,
\eeq
where $\bar{\rho}_\chi$ and $\bar{\rho}_\varphi$ denote the (non-oscillatory) leading order contributions to the bolon- and cosmon energy-densities, respectively. These are given by
\begin{align}
\bar{\rho}_\chi &= \frac{1}{2} \bar{m}_\chi^2(\bar{\varphi}) \chi_0^2 \, , \\
\bar{\rho}_\varphi &= V_1(\bar{\varphi}) + \frac{1}{2 \bar{a}^2} \bar{\varphi}'^2  \, ,
\end{align}
where $\bar{m}_\chi(\bar{\varphi})$ simply denotes the leading order term in the $\tilde{\mu}$-expansion for the mass, which is of course only $\bar{\varphi}$-dependent:
\beq
\bar{m}_\chi(\bar{\varphi}) = m_0 {\rm e}^{-\beta \bar{\varphi}/M} \, .
\eeq
We will drop the explicit $\bar{\varphi}$-dependence of $\bar{m}_\chi$ in all the formulas below. The additional quantity $\bar{\rho}_{\rm ext}$ labels the sum of all energy densities which are present in addition to the scalar fields, in particular photons, neutrinos and baryons.
From the bolon field equation (see appendix \ref{app:bgExpansion}) we can directly see that $\chi_0 \propto \bar{a}^{-3/2} {\rm exp}(\beta \bar{\varphi} / 2M) $ and therefore
\beq
\label{AveragedBolonSolution}
\bar{\rho}_\chi \propto \bar{a}^{-3} {\rm exp} (-\beta \bar{\varphi}/M) \, .
\eeq
The cosmon evolution follows the following equation
\begin{align}
\label{LeadingOrderCosmonEquation}
\bar{\varphi}'' &= -2 \bar{h} \bar{\varphi}' - \bar{a}^2 V_{1,\bar{\varphi}} + \frac{\beta}{M} \bar{a}^2 \bar{\rho}_\chi \, .
\end{align} 

Using our expansion method we have recovered the expected result that to leading order the cosmon and bolon form a system of coupled quintessence governed by the equations (\ref{AveragedFriedmannEquation}), (\ref{LeadingOrderCosmonEquation}) and the equation of energy conservation derived from equation (\ref{AveragedBolonSolution}):
\beq
\bar{\rho}_\chi' + 3 \bar{h} \bar{\rho}_\chi = -\beta \frac{\bar{\varphi}'}{M} \bar{\rho}_\chi \, .
\eeq
If we include additional components into our cosmic fluid, we will of course have to add the corresponding equations of energy conservation for those as well.

This system was already analyzed in ref. \cite{Beyer:2010mt}. One should note that for the range of parameters we are investigating it allows for accelerated expansion only in the case of large negative $\beta$, specifically $\alpha < -2 \beta$. This is not only inconsistent with our original motivation of an asymptotically vanishing bolon mass, but also excludes the possibility of a prolonged matter dominated epoch in our scenario, as we have checked numerically. \footnote{This is of course connected to the fact that the coupled quintessence scenario does not describe our model in the early universe. Within a pure coupled quintessence model realistic accelerated cosmologies are of course possible and were discussed in ref. \cite{Amendola:1999er}.} The universe simply transitions quickly from radiation-domination to accelerated expansion in this case. 
Furthermore current bounds on coupled quintessence models already rule out such strong couplings (see e.g. ref. \cite{Pettorino:2012ts}), as we will discuss further below.
We therefore exclude this possibility as unrealistic and focus on smaller couplings.

For these our cosmology will quickly adjust itself to a matter-dominated scaling solution (see ref. \cite{Beyer:2010mt}), with the bolon energy-density scaling slightly differently than baryons due to the coupling. The parameter bounds resulting from the conditions of existence and stability of this solution can be found in ref. \cite{Amendola:1999er} and read:
\begin{align}
 \left| \alpha - \beta \right| >&\sqrt{\frac{3}{2}} \, ,  \quad \alpha < \left( \beta + 3/\beta \right) \, , \quad \alpha > \sqrt{\frac{2}{3}} 4 \beta \nonumber \\
& {\rm and } \quad \alpha > \frac{1}{2} \left( \beta + \sqrt{12+\beta^2} \right) \, .
\end{align}

None of these constraints are in conflict with the ones we found above from considerations of an attractive radiation dominated era in the early universe or the condition of a decreasing $\mu$. All parameter choices used below will respect all the bounds mentioned in this section.

\subsection{Accelerated expansion}
As was already pointed out, the specific model treated here, with the restrictions on parameters from sections \ref{sec:TheEarlyUniverse} and \ref{sec:LateUniverse},	 does not result in an accelerated expansion. We are not too concerned about this issue, since there are several ways out of it, as we already discussed in ref. \cite{Beyer:2010mt}. One could for example assume a slightly different form of the quintessence potential by considering a $\varphi$-dependent $\alpha$ \cite{Wetterich:1994bg}, alternatively a non-standard kinetic term \cite{Hebecker:2000zb}, a $\varphi$-dependent minimum in the bolon-potential \cite{Beyer:2010mt}, a $\varphi$-dependent coupling $\beta$ \cite{TocchiniValentini:2001ty} or an additional coupling to other components of the cosmic fluids (e.g. neutrinos, see \cite{Amendola:2007yx}) can break the scaling behaviour and effectively stop the evolution of the cosmon, leading to an accelerated expansion. We will choose one of these possibilities in our numerics below in order to get realistic results.

\subsection{Observational parameter bounds}
Current experimental bounds on the model parameters go far beyond the theoretical limits we cited above, originating from the desire to obtain a realistic cosmology independent of model parameters. To our knowledge the most stringent observational bounds on a coupled theory such as ours come from big bang nucleosynthesis (BBN) \cite{Ferreira:1997au,Ferreira:1997hj,Bean:2001wt} and cosmic microwave background (CMB) observations \cite{Calabrese:2011hg,Reichardt:2011fv,Xia:2013dea,Pettorino:2013ia}.

Let us start with the BBN bounds. Adding a tracking quintessence field to the early radiation dominated era in the universe modifies expansion of the universe and thus standard big bang nucleosynthesis. Observations of the abundances of the lightest elements in the cosmos can therefore put an upper bound on the quintessence density parameter, Bean, Hansen and Melchiorri set it at $0.045$, the tightest constraint known to us \cite{Bean:2001wt}. In our model this should be seen as a bound for the combined scalar density parameter $\Omega_{sc,es}=\Omega_{\varphi,es}+\Omega_{\chi,es}$ as given in equations (\ref{omegaPhiES}) and (\ref{omegaChiES}).

Constraints from the CMB on a tracking scalar quintessence component are strong whenever the scalar field makes up a non-negligible fraction of the energy density of the universe at decoupling, which is the case in our model. Recently bounds from this era have been improved to give an upper limit of about $0.02$ \cite{Pettorino:2013ia}, which in our model has to be interpreted as a bound on the quintessence density parameter alone, since the bolon has already started to oscillate at decoupling and does not follow its scaling solution anymore. From ref. \cite{Amendola:1999er} one directly obtains 
\beq
\Omega_{\varphi,cq} = \frac{3-\alpha \beta + \beta^2}{(\alpha-\beta)^2} \lesssim 0.02 \, .
\eeq 
The subscript "cq" stands for coupled quintessence. 
Furthermore CMB analyses of coupled quintessence models also put a bound on the coupling \cite{Pettorino:2012ts}, currently at the order of $\beta^2 \lesssim 0.01$. 

These CMB bounds were derived for standard cold dark matter coupled to quintessence, but we expect similar constraints to hold in our scenario. As we will see below, the evolution of perturbations in our model is different than in coupled cold dark matter models, but we expect these differences to be largely irrelevant for the CMB constraints, as they are only important for scales much smaller than the ones corresponding the multipole moments where the CMB has the most constraining power. 
Future constraints using Planck and Euclid data sets are expected to improve these bounds by about two orders of magnitude \cite{Amendola:2011ie}.

\subsection{Parameter adjustment}

At the background level our model has 4 parameters determining the behavior of the two scalar fields, the exponents $\alpha$ and $\lambda$, the coupling $\beta$ and the mass-parameter $c^2$. To fully determine the background evolution (after some suitable very early time, in particular after neutrino-decoupling and electron-positron annihilation) we also have to fix the current radiation density $\rho_{r,0}$ and the baryon density $\rho_{b,0}$. Adopting a procedure introduced in ref. \cite{Matos:2000ss} we can predict the current density parameters for both the bolon and the cosmon from the fundamental model parameters. 

First, we define a scale factor $a^*$ as the scale-factor at which oscillations start, i.e. when $\chi=M/\lambda$. As an approximation, we then extrapolate the analytically known bolon-evolution for the early scaling solution (eq. (\ref{EarlyChiEvolution})) up to that point, also assumimg that $h \propto 1/a$ still holds. Furthermore we can estimate the cosmon value $\varphi^* = \varphi(a^*)$ by extending the analytic formula for the early scaling solution given by equation (\ref{EarlyPhiEvolution}). This gives the following estimate for $a^*$:
\beq
\label{aStarEq1}
a^* = \left( \frac{\rho_{r,0}}{3 M^4} \right)^\frac{1}{4} \left[ \frac{8 (1-2\beta/\alpha)\, {\rm f}^{-2\beta/\alpha} }{{\rm e} c^2 \lambda^2 (1-\Omega_{sc,es})^{1-2\beta/\alpha}} \right]^\frac{1}{4(1-2 \beta/\alpha)}
\eeq 
where $\Omega_{sc,es}=\Omega_{\varphi,es} + \Omega_{\chi,es}$ and $\text{f}=\text{f}(\alpha,\beta,\lambda)$ as defined in section \ref{sec:TheEarlyUniverse}.

To make contact with current energy densities, we can assume that from $a^*$ onwards, the bolon will follow its evolution determined by eq. (\ref{AveragedBolonSolution}). Since the coupling causes the bolon-density to scale slightly differently than the baryons, we can not simply fix the ratio of the two energy densities, but have to specify $\rho_{\chi}$ at some redshift, say at $z=0$. Once this is set, all that remains to consider is the cosmon energy density, which should dominate the cosmic evolution at late times. In particular we need to stop the evolution of the cosmon at some low redshift $\tilde{z}$ (appoximately $5$). The value of the cosmon $\varphi_0 = \varphi(z=0) \approx \varphi(\tilde{z})$ can the simply be obtained by extending the late time cosmon scaling solution for coupled quintessence to $\tilde{z}$. Extrapolating this evolution back to $a^*$ and estimating the bolon energy-density at $a^*$ by the one given by the scaling solution then gives
\beq
\label{aStarEq2}
a^*=\frac{\rho_{r,0}}{\rho_{\chi,0}} \frac{\Omega_{\chi,es}}{1-\Omega_{sc,es}} {\rm e}^{\beta (\varphi*-\varphi_0)/M} \, ,
\eeq
where we have not yet inserted the cosmon evolution in order to keep the equation simple.
Plugging in this information using the approximations described above and equating the right hand sides of (\ref{aStarEq1}) and (\ref{aStarEq2}) gives us an approximate expression for the current bolon energy density $\rho_{\chi,0}$ given a mass, or vice versa. We can make this and exact expression by including a numerical adjustment factor $N$:
\begin{align}
\label{c2Guess}
\left(\frac{\rho_{\chi,0}}{\rho_{r,0}}\right)^{1-\beta/\alpha}=& N \, \Omega_{\chi,es} (1-\Omega_{sc,es})^{-3/4} \left( \frac{\rho_{r,0}}{M^4} \right)^{-\frac{1}{4}+\beta/\alpha} \nonumber \\
&\left(\frac{8 (1-2\beta/\alpha)}{3{\rm e} \lambda^2 c^2} \right)^{-\frac{1-4\beta/\alpha}{4(1-2\beta/\alpha)}} \left( \frac{{\rm f}}{3} \right)^\frac{-\beta/\alpha}{2(1-2\beta/\alpha)} \nonumber \\
& \left( 6^3 (1+\tilde{z}) \frac{\rho_{\chi,0}}{\rho_{r,0}} \frac{{\rm g}}{2-{\rm g}} \tilde{\rm g} \right)^{\beta/\alpha}
& \end{align}
where
\begin{align}
\text{g}=&\text{g}(\alpha,\beta,\lambda)=  1 + \frac{6 \beta^2 - 6 \alpha \beta + 18}{6 (\alpha - \beta)^2} \quad {\rm and} \\
\tilde{\text{g}} =& \tilde{\text{g}} (\alpha,\beta,\lambda) = \text{g}(\alpha, \beta, \lambda) - \frac{3}{2 (\alpha - \beta)^2} \, .
\end{align}
The adjustment factor $N$ is of order $1$, but always smaller, roughly $0.4$. This is due to the fact that estimating the bolon energy density at $a^*$ by the scaling solution is quite accurate, but using the averaged evolution from that point on leads to an overestimate of $\rho_{\chi,0}$ since the bolon energy dilutes faster than non-relativistic matter during the early oscillatory phase. Furthermore $N$ is a function of the model parameters $\alpha$, $\beta$ and $\lambda$ as well as the energy densities $\rho_{r,0}$, $\rho_{\chi,0}$ and $\tilde{z}$. It of course will also take (slightly) different values depending on which scenario is chosen to achieve late time cosmic acceleration. For practical purposes, we will choose one such scenario in section \ref{sec:Numerics} where we present numerical results and determine $N$ (for fixed $\tilde{z}=5$, $\rho_{r,0}$ and $\rho_{\chi,0}$) by running through a grid on $c^2$ and the remaining model parameters. 

\section{Linear Perturbations}
\label{sec:LinearPerturbations}

In this section we analyze the behaviour of linear perturbations in the late universe, i.e. after the bolon started to perform rapid oscillations around the minimum of its potential. The behaviour of perturbations in the early universe was already analyzed in ref. \cite{EarlyScalings}. As the perturbations also perform rapid oscillations, a numerical evolution of the full field equations is not feasible. We will therefore first deduce a set of effective perturbation equations averaged over one oscillation period and then use the results in our numerical simulations below.

In the interest of brevity we will not discuss our conventions concerning linear perturbation theory here, we merely mention that they precisely coincide with the ones defined in the appendix of the accompanying work ref. \cite{EarlyScalings}. In particular we are only interested in the scalar sector of linear perturbation theory and will employ the following quantities: The gauge-invariant field perturbations $X$ and $Y$, the Bardeen potentials $\Psi$ and $\Phi$, the gauge-invariant energy density and momentum perturbations $\delta \rho_\alpha$ and $\left[ (\rho + p) v \right]_\alpha$, their dimensionless versions, i.e. the density contrasts $\delta_\alpha$ and velocity potentials $\Theta_\alpha$, and the total anisotropic stress $\Pi_{\rm tot}$.

\subsection{Averaged evolution}
We start by considering the exact gauge-invariant linearly perturbed scalar field equations 
\begin{align}
\label{CosmonPEquation}
X'' + 2hX' + k^2 X + a^2 V_{,\varphi \varphi} X + a^2 V_{,\varphi \chi} Y \nonumber \\
+ 2 a^2 V_{,\varphi} \Phi - \varphi' \Phi' - 3 \varphi' \Psi' & = 0 \, , \\
\label{BolonPEquation}
Y'' + 2hY' + k^2 Y + a^2 V_{,\varphi \chi} X + a^2 V_{,\chi \chi} Y \nonumber \\
+ 2 a^2 V_{,\chi} \Phi - \chi' \Phi' - 3\chi' \Psi' & = 0 \, ,
\end{align}
and the Poisson-equation
\begin{align}
\label{PsiEquation}
k^2 \Psi &= -\frac{a^2}{2M^2} \sum_\alpha \left( \delta \rho_\alpha - 3 h \right[(\rho + p)v\left]_\alpha \right) \, .
\end{align}
Furthermore we relate the two Bardeen potentials via
\beq
\label{phiPsiEQ}
\Phi = \Psi - a^2 \Pi_{\rm tot}/M^2 \, .
\eeq
If we add a suitable set of equations describing additional components of the cosmic fluid, these equations form a closed set and uniquely determine the evolution of the scalar linear perturbations. These additional equations would be equations of energy- and momentum-conservation in the case of a fluid description (e.g. for baryons) or the equations for the higher momenta in the multipole-expansion of the phase-space distribution function resulting from the corresponding Boltzmann-equations for more complex descriptions (typically for photons and neutrinos, see e.g. ref. \cite{Ma:1995ey}). In what follows we will ignore such additional equations, but note that the entire averaging procedure presented below can easily be extended to include them, and they emerge unchanged. We explain this in more detail in appendix \ref{app:bgExpansion}.

In order to average the perturbation equations we use the same idea we applied to the background evolution in section \ref{sec:LateUniverse}, i.e. we expand all dynamical perturbative quantities first in a Taylor-series in $\tilde{\mu}$ and then each coefficient in a Fourier-type sum of harmonic functions with time-dependent frequencies, where all occurring frequencies are multiples of a base-frequency (see equations (\ref{MuExpansion}) - (\ref{baseFrequency})). By plugging the results into the linearized field equations (\ref{CosmonPEquation}) - (\ref{PsiEquation}) and comparing coefficients we see which frequencies are present at which order and can then use the results to calculate the evolution equations for the energy density and momentum perturbations for the bolon averaged over one oscillation period. 

The precise details of the calculation can be found in appendix \ref{app:bgExpansion}. 
The procedure reveals that the decompositions of the scalar-field perturbations and the gravitational potential change for different wavenumbers and we have to split up our analysis into two regimes: Large wavelength perturbations for which $\mu k^2/h^2 \ll 1$ and small wavelength perturbations which have $\mu k^2/h^2 \gtrsim 1$.

%Before going into the detailed analysis for the two regimes, a remark concerning the anisotropic stress is in %order. As mentioned above, we ignore it in our analysis below. While this is a simplification, including %anisotropic stress (e.g. from neutrinos) does not change our conclusions, as long as the typical timescale %on which such a contribution evolves is much longer than one oscillation period of the bolon. {\bf Maybe %explain this in more detail below!}

\subsubsection{$\mu k^2/h^2 \ll 1$}

For large wavelengths the equations resulting from the averaging procedure (to subleading order in $\tilde{\mu}$) are
\begin{align}
\label{dChiEQAV}
<\delta_\chi>' =& -\bar{\Theta}_\chi + 3 \bar{\Psi}' - \beta P'/M + \beta \frac{\bar{\varphi}'}{M} \left( \bar{\Psi} - \bar{\Phi} \right) \, , \\
\bar{\Theta}_\chi' =& - \bar{h} \bar{\Theta}_\chi + k^2 \bar{\Phi} + \beta \frac{\bar{\varphi}'}{M} \bar{\Theta}_\chi - \beta k^2 P/M \, , \\
\label{pEQAV}
P'' =& - 2 \bar{h} P'  - \left( k^2 + \bar{a}^2 V_{1,\bar{\varphi} \bar{\varphi}} \right) P \nonumber  - 2 \bar{a}^2 V_{1,\bar{\varphi}} \bar{\Phi} \nonumber \\
& + \bar{\varphi}' \left( \bar{\Phi}' + 3 \bar{\Psi}' \right)
+ \frac{\beta \bar{a}^2}{M} \bar{\rho}_\chi \left( <\delta_\chi> +2 \bar{\Phi} \right) \, .
\end{align}
Here the dynamical quantities describing the bolon are its density contrast defined as $\delta_\chi = \delta \rho_\chi/\rho_\chi$, and its averaged velocity potential, given by 
\beq
\bar{\Theta}_\chi = -k^2 \frac{<\left[ (\rho + p)v \right]_\chi>}{\bar{\rho}_\chi + <p_\chi>} = -k^2 \frac{<\left[ (\rho + p)v \right]_\chi>}{\bar{\rho}_\chi} \, .
\eeq
Triangular brackets denote an oscillatory quantity averaged over one oscillation period, whereas, as before, a bar denotes a quantity which is evolving adiabatically at leading order.
Note that the velocity potential $\bar{\Theta}_\chi$ is a well defined variable, despite that fact that trying to define the same velocity potential with the non-averaged quantities would yield a badly defined (periodically divergent) $\Theta_\chi$. 

The quintessence field is evolving slowly to leading order, as are the gravitational potentials, and we denote the leading order quantities by $P$, $\bar{\Psi}$ and $\bar{\Phi}$ respectively. 

To complete our equations we have to evaluate the Poisson equation, which gives 
\begin{align}
\label{poissonEQAV}
\bar{\Psi} =\frac{-\bar{a}^2}{2 M^2 k^2 - \bar{\varphi}'^2} & \left[\bar{\rho}_{\rm ext}  \left( \bar{\delta }_{\rm ext} + \frac{3 \bar{h}}{k^2} (1+\bar{\omega}_{\rm ext}) \bar{\Theta}_{\rm ext} \right) \right. \nonumber \\
&+ \bar{\rho}_\chi \left( <\delta_\chi> + \frac{3 \bar{h}}{k^2} \bar{\Theta}_\chi \right) + \bar{\varphi}'^2 \frac{\bar{\Pi}_{\rm tot}}{M^2}\nonumber \\
&\left. + V_{1,\bar{\varphi}} P + 3 \bar{h} \frac{\bar{\varphi}' P}{\bar{a}^2
} + \frac{\bar{\varphi}' P'}{\bar{a}^2}  \right] \, .
\end{align}

Here the label "ext" labels all additional components of the cosmic fluid, i.e. neutrinos, photons and baryons. As mentioned above, the corresponding energy densities evolve adiabatically to leading order, the velocity potentials and higher momenta of the phase space distribution function even to subleading order.

Equations (\ref{dChiEQAV})-(\ref{pEQAV}) together with eq. (\ref{poissonEQAV}) and eq. (\ref{phiPsiEQ}) are equivalent to the set of equations (25) - (31) in ref. \cite{Amendola:2003wa} for a constant coupling and $\omega=0$. The apparent differences arise from the fact that the author of ref. \cite{Amendola:2003wa} employed slightly different definitions of the veolocity potential $\Theta$ and the coupling $\beta$, ignored anisotropic stress contributions and used derivatives taken with respect to $\rm{ln}(a)$, not with respect to conformal time. The averaged cosmon and bolon perturbations therefore behave exactly as standard cold dark matter coupled to quintessence with a constant coupling in this wavelength-regime.

\subsubsection{$\mu k^2/h^2 \gtrsim 1$}

For smaller wavelengths the results from the averaging procedure are almost the same, except that pressure perturbations can not be neglected at subleading order. The equations of energy- and momentum-conservation for the bolon now give:
\begin{align}
<\delta_\chi>' =&- \bar{\Theta}_\chi + 3 \bar{\Psi}' - \beta \frac{P'}{M} + \beta \frac{\bar{\varphi}'}{M} \left( \bar{\Psi} - \bar{\Phi} \right) \nonumber \\
& - \left( 3 \bar{h} - \beta \frac{\bar{\varphi}'}{M} \right) c_{s,\chi}^2 <\delta_\chi>  \, , \\
\bar{\Theta}_\chi' =& - \bar{h} \bar{\Theta}_\chi + k^2 \bar{\Phi} + \beta \frac{\bar{\varphi}'}{M} \bar{\Theta}_\chi - \beta k^2 \frac{P}{M} \nonumber \\
& + k^2 c_{s,\chi}^2 <\delta_\chi> \, ,
\end{align}
where the bolon sound speed is given by
\beq
\label{soundSpeed}
c_{s,\chi}^2 = \frac{k^2}{4 \bar{m}_\chi^2 \bar{a}^2} \, .
\eeq
Such a sound speed modifies the growth of perturbations on the dark matter sector considerably, as we will show below. This generalizes the previous result found in refs. \cite{Hu:2000ke,Matos:2000ss}, where an oscillating scalar field but no coupling to the cosmon was considered. Furthermore we have provided a rigorous way to handle the quick oscillations present in both the background and the full set of gauge-invariant perturbation equations, something not provided in references \cite{Hu:2000ke,Matos:2000ss}. The field equation for the cosmon and the Poisson equation are exactly the same as in the large wavelength regime.

A comparison of these equations with those found in equations (25) to (31) in ref. \cite{Amendola:2003wa} yields obvious differences. These go beyond the simple fact that the definitions of the velocity potential, the coupling and the time-variable are different, but originate from the fact that the fluids considered in ref. \cite{Amendola:2003wa} are assumed to be barotropic, i.e. the equation of state is assumed to be a function of the energy density alone, no non-adiabatic pressure perturbations are present. Our results however do correspond to a non-barotropic fluid and the pressure-perturbation is indeed a non-adiabatic one. 

\subsection{Damping of the power spectrum}
The basic features of the evolution of the bolon density contrast can be understood fairly easily. Perturbations with wavenumbers in the regime $\mu k^2/h^2 \ll 1$ follow the evolution of a coupled cold dark matter component, whereas perturbations in the regime $\mu k^2/h^2 \gtrsim 1$ are expected to be strongly surpressed, due to the small sound speed. Since this is a time-dependent condition, let us see how the quantity $\mu k^2/h^2$ evolves during the main stages of the cosmic evolution. We can generally assume that $\bar{h} \propto \bar{a}^\eta$, where $\eta = -1, -1/2$ and $1$ during radiation-domination, matter domination and de-Sitter expansion respectively. The coupling between the bolon and the cosmon will of course change this behavior slightly, but that is not crucial for our argument. We then have
\beq
\mu k^2/\bar{h}^2 \propto \bar{a}^{-1-\eta} {\rm e}^{\beta \bar{\varphi}/M} \, ,
\eeq
which shows that during the eras of bolon- or cosmon-domination (and for not too large couplings), this quantity is decreasing. Therefore modes which have been suppressed in their evolution during some early era will eventually enter the regime where they behave just like coupled cold dark matter. Due to this delayed onset of growth for these modes, the power spectrum exhibits a sharp cutoff at a scale which can be approximated by
\beq
\label{jeansWavenumber}
k_{\rm J}^2 = m_0 {\rm e}^{-\beta \bar{\varphi}^*/M} \sqrt{\frac{\rho_{r,0}}{3 M^2 (1-\Omega_{sc,es})}} \, ,
\eeq
where the value $\varphi^*$ of the cosmon at $a^*$ can once again be approximated by extrapolating the early scaling solution given in eq. (\ref{EarlyPhiEvolution}). This formula generalizes eq. (63) in ref. \cite{Matos:2000ss} and is of course only an estimate, but it is the best definition of a Jeans length available in our model as it describes the smallest wavenumber for which the pressure balances out the gravitational attraction during some stage of the cosmic evolution. We can assign a corresponding Jeans mass in the usual way:
\beq
\label{jeansMass}
M_{\rm J} = \frac{4 \pi}{3}  k_{\rm J}^{-3} \bar{\rho}_\chi \, .
\eeq

\subsection{Numerical evolution}
\label{sec:Numerics}
Now we move on to numerically evolve the linear perturbations in our model. We draw the initial conditions for our simulations from the results of an accompanying paper \cite{EarlyScalings}. There it was shown that for the coupled cosmon-bolon system there exists an adiabatic mode which will evolve to be dominant if a sufficiently long era of tracking is present, which is a quite generic feature in our model. To avoid such an era initial conditions would have to be very skewed (see ref. \cite{Beyer:2010mt}).
We therefore work with purely adiabatic initial conditions (see equation (47) in ref. \cite{EarlyScalings})  and ignore possible isocurvature contributions. Our numerical simulation then simply evolves the evolution equations for all the components of the cosmic fluid, i.e. photons, neutrinos, baryons and the two scalar fields. As is common in Boltzmann-codes, we use several approximations in order to speed up the calculation, which we now quickly describe.

For the neutrino- and the baryon-photon-sector we use a set of approximations that are also implemented in the recent CLASS-code (\cite{Lesgourgues:2011re,Blas:2011rf}) and discussed in detail in ref. \cite{Blas:2011rf}. For the early universe we use the exact first order version of the tight coupling approximation, corresponding to the setting \textit{first\_order\_CLASS} in the CLASS-code. In order to deal with the quick oscillations in the relativistic species (neutrinos and photons) in the late universe we also employ the ultra-relativistic fluid approximation corresponding to the setting \textit{ufa\_class} and the relativistic streaming approximation in its simplest version, corresponding to the setting \textit{rsa\_MD}. 
The thermodynamic history of the universe was calculated using a modified version of the latest fortran-release of recfast \cite{Seager:1999bc,Seager:1999km,Wong:2007ym,Scott:2009sz}.

Since we do not use synchronous gauge but a manifestly gauge-invariant approach (see appendix of ref. \cite{EarlyScalings}) instead, we had to rederive some of the approximations mentioned above. To check the correctness of our equations as well as their implementation and the validity of our choices for the triggers determining the switches between the approximation schemes, we have compared the results our code gives for the standard $\Lambda$CDM-model with the CLASS results. As it turns out, the matter power spectra (which is the quantity we are after) agree to excellent accuracy (see appendix \ref{app:ClassComp}).

In our treatment of the scalar fields, we start by evolving the exact scalar field equations and switch to the effective averaged fluid description for the bolon at some suitably late time, i.e. after the oscillations have started. The initial values for the effective fluid description are obtained by an explicit numerical integration over one oscillation period.
For large wavenumbers and late enough times (i.e. in the subhorizon regime), the cosmon field will exhibit quick oscillations, similar to those present in the photons or neutrinos. We therefore extended the RSA-approximation to include the cosmon by employing equation (43) in ref. \cite{Amendola:2003wa}, which reads for our conventions
\beq
P = 3 \beta M \frac{\bar{h}^2}{k^2} \bar{\Omega}_\chi <\delta_\chi> \, .
\eeq
We have checked that this is a good approximation by varying the RSA-onset trigger.

To address the issue of accelerated expansion we simply change the exponent of the quintessence potential for large field values by making a smooth transition from $\alpha$ to $0.1$ at a suitable point:
\beq
\alpha (\varphi) = \left\{
\begin{array}{l l}
\alpha \quad & \varphi<\varphi_0 \\
\alpha - t(\varphi) \quad & \varphi_0 < \varphi < \varphi_1 \\
1  \quad & \varphi_1 < \varphi \\
\end{array} 
\right. \, ,
\eeq
with
\beq
t(\varphi) = \left( 3 \frac{(\varphi-\varphi_0)^2}{(\varphi_1-\varphi_0)^2} + 2 \frac{(\varphi-\varphi_0)^3}{(\varphi_1-\varphi_0)^3}\right) (\alpha-0.1) \, .
\eeq This is of course slightly ad-hoc and should be taken as only one of many ways to slow down the evolution of the cosmon, several other possibilities have been discussed in section \ref{sec:Background}. We do this merely to get a realistic cosmology for the numerics.

In this section we will always choose the value 20 for $\alpha$. The values for $\lambda$ and $\beta$ will vary, and for each choice the values for $c^2$, $\varphi_0$ and $\varphi_1$ will be adjusted to give the correct current Hubble rate and density parameters. For each set of parameters used will quote the values for $\varphi_0$ and $\varphi_1$ as well as the normalization-factor $N$ (see eq. (\ref{c2Guess})) from which $c^2$ can be deduced.
We adjust the cosmological parameters to the WMAP7-values \cite{Jarosik:2010iu}, i.e. 

\begin{figure}[t]
	\centering
  \includegraphics[width=1.0\linewidth]{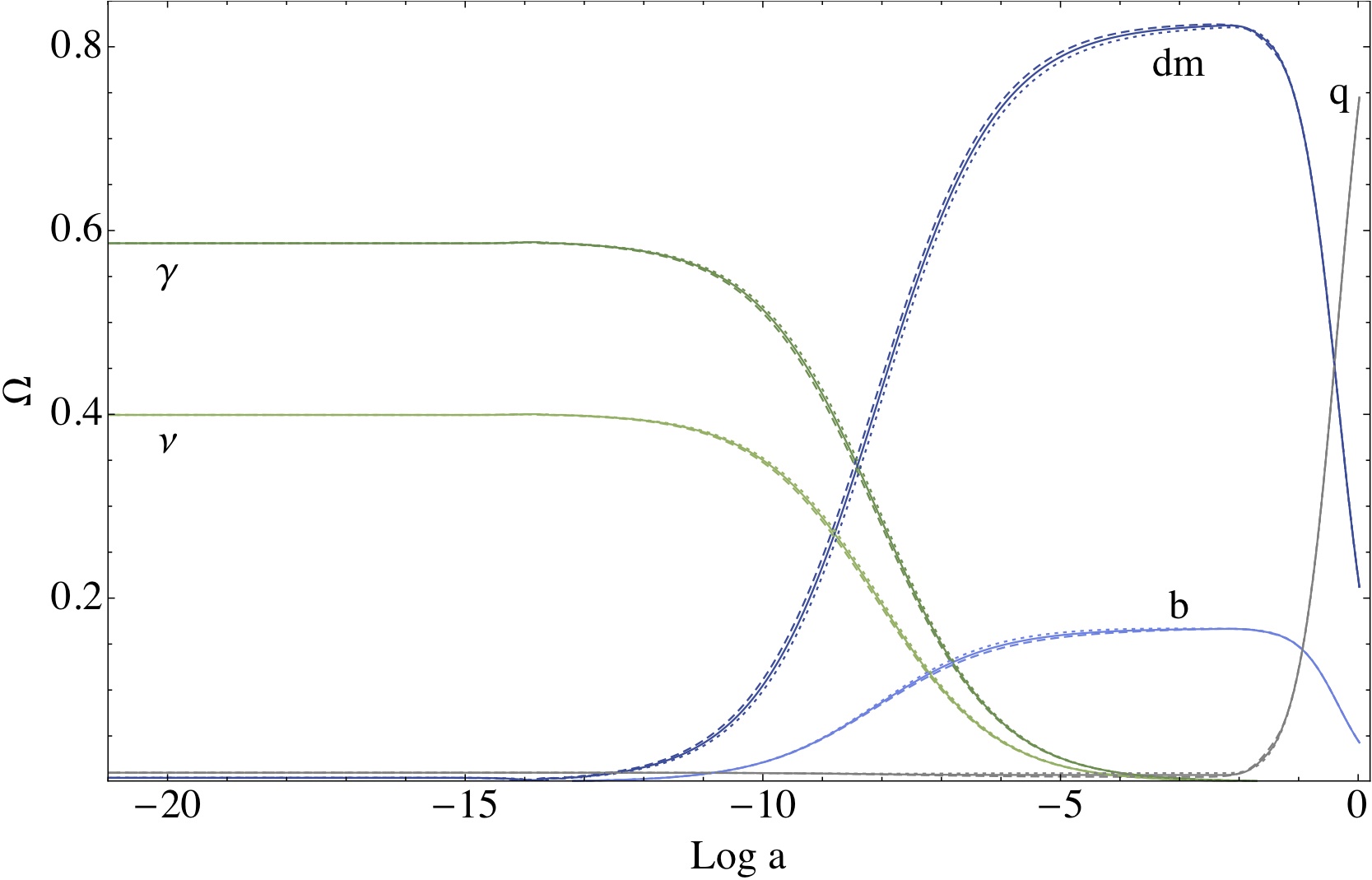}
	\caption{Density parameters for photons (dark green), neutrinos (light green), bolon (dark blue), baryons (light blue) and cosmon (grey). The solid lines represent the uncoupled model (with $N=0.38$, $\varphi_0=13.806$, $\varphi_1=15.493$), the dashed lines $\beta=0.05$ (with $N=0.3727$, $\varphi_0=13.786$, $\varphi_1=16.248$) and the dotted lines $\beta=-0.05$ (with $N=0.378$, $\varphi_0=13.789$, $\varphi_1=16.168$ ).}
	\label{fig:omegas}
\end{figure}

\begin{table}[h]
\label{densityParameters}
\begin{tabular}{ccc}
$h=0.71 \, ,$ & $\Omega_{\chi,0}=0.222  \, ,$ & $\Omega_{b,0}=0.0449 \, ,$ \\
$T_{CMB}=2.728 \, {\rm K} \, ,$  & $n_s=0.961 \, ,$ & $\sigma_8=0.801 \, .$
\end{tabular}
\end{table}
The evolution of the density parameters can be seen in Figure \ref{fig:omegas} for different couplings. We choose initial conditions corresonding to the scaling solution given in equations (\ref{EarlyPhiEvolution}) and (\ref{EarlyChiEvolution}). Different initial conditions do not change the late time cosmology unless they are very skewed (see ref. \cite{Beyer:2010mt}).

\begin{figure}[t]
	\centering
  \includegraphics[trim=0cm 8.7cm 0cm 5.4cm, clip=true, width=1.0\linewidth]{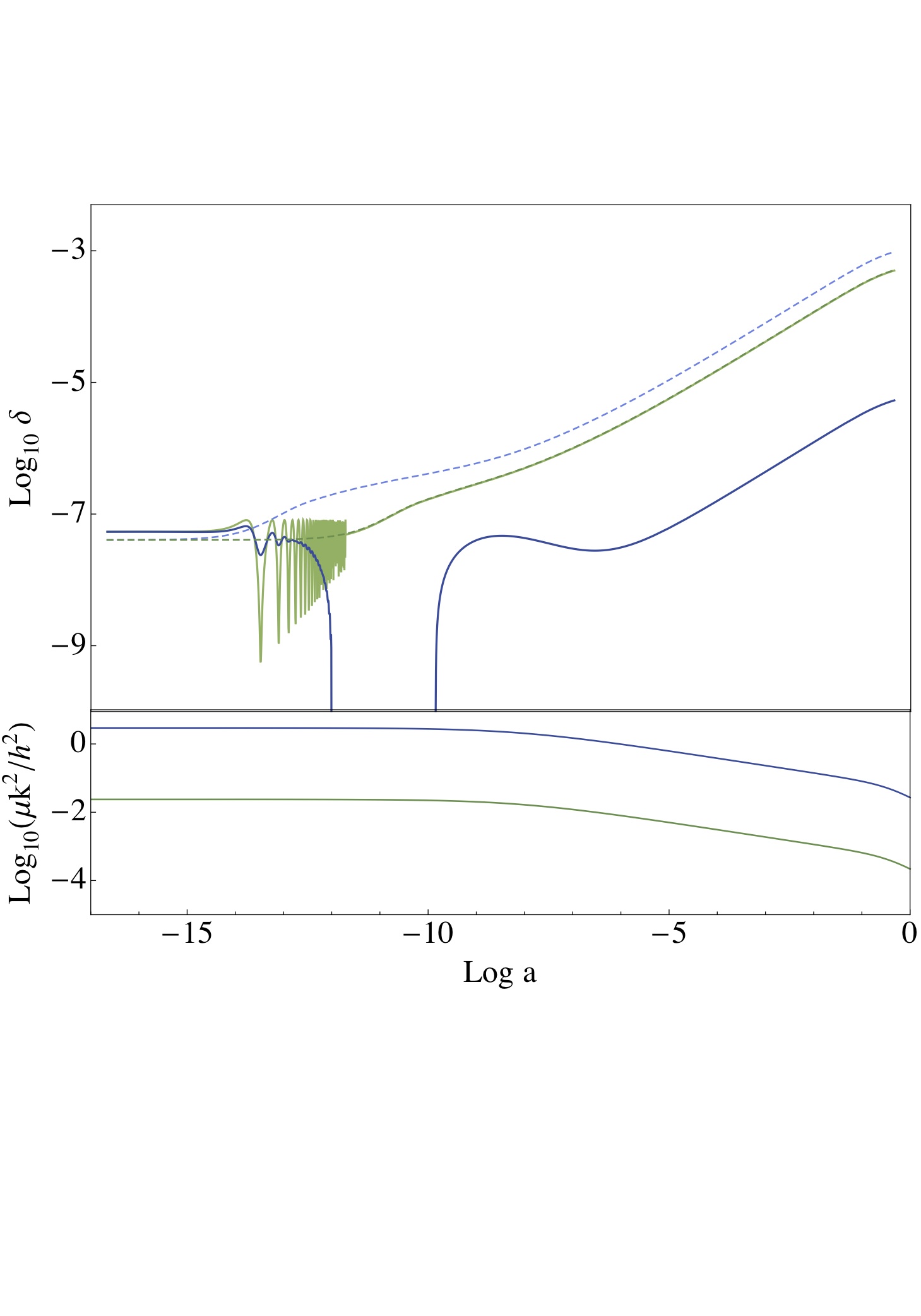}
	\caption{Bolon density contrast evolution. The upper panel shows the evolution of the linear density contrast for the bolon (solid lines) and a standard cold dark matter evolution (dashed lines) in the same background. The green lines show a $k=0.96$ h/Mpc mode, whereas the blue ones represent $k=10.7$ h/Mpc. In the lower panel we show the corresponding evolution of the quantity $\mu k^2/h^2$. The bolon exponential is $\lambda=30$ for this plot, the coupling is $\beta=0$.}
	\label{fig:kmodes1}
\end{figure}

The connection between the evolution of the bolon density contrast and the quantity $\mu h^2/k^2$ is shown in Figure \ref{fig:kmodes1}. We display numerical for two different wavenumbers, $k=0.96$ h/Mpc (green) and $k=10.7$ h/Mpc (blue). The solid lines represent the bolon evolution, the dashed lines standard cold dark matter modes with the same wavenumbers evolved in the same background, given here for comparison. The normalization for both modes is arbitrary. The first difference to note is the slightly elevated value for the initial density contrast in the adiabatic mode for the bolon compared to standard CDM. We also clearly see the onset of oscillations in both modes, but while the $k=0.96$ h/Mpc mode, when averaged, follows the evolution of the cold dark matter mode exactly, the growth of the $k=10.7$ h/Mpc mode is suppressed during a prolonged stage of its evolution. This was to be expected from the evolution of $\mu h^2/k^2$, which can be seen in the lower panel of the picture. For the $k=0.96$ h/Mpc mode we have $\mu h^2/k^2 \ll 1$ throughout the entire evolution, whereas for $k=10.7$ h/Mpc $\mu h^2/k^2>1$ initially, but the value decreases for later times and growth sets in.

\begin{figure}[t]
	\centering
  \includegraphics[width=1.0\linewidth]{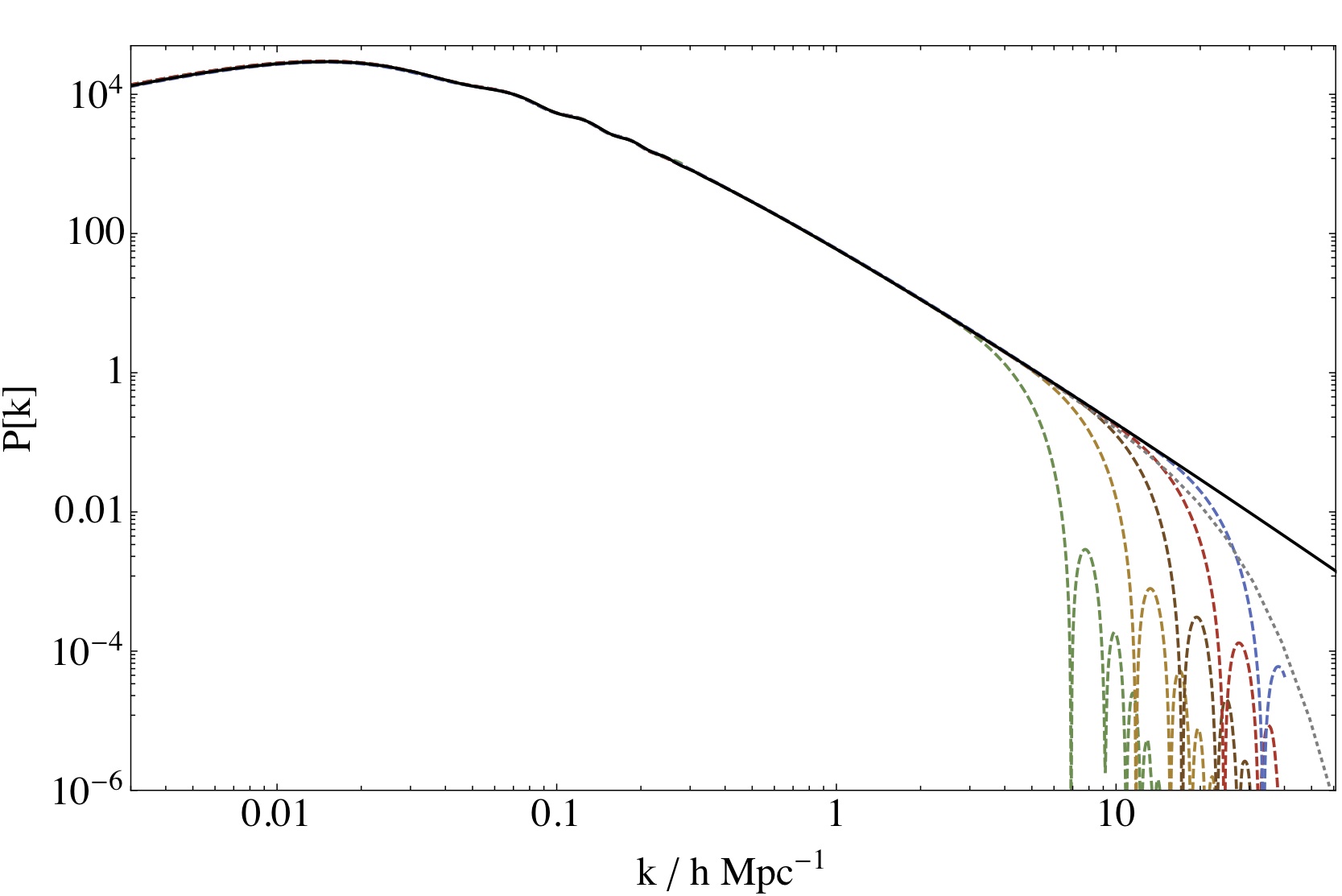}
	\caption{Power spectra for standard cold dark matter (black), warm dark matter with $m_{\rm wdm} = 2.284$ keV (grey, dotted) and the bolon (dashed) for different masses. The corresponding values are from left to right $\lambda=30$ (green line,  $N=0.3793$), $\lambda=40$ (orange line, $N=0.3804$), $\lambda=50$ (brown line, $N=0.3792$), $\lambda = 60$ (red line, $N=0.3791$) and $\lambda=70$ (blue line,  $N=0.3784$). The coupling is $\beta=0$, the scalar field transition values are $\varphi_0=13.806$ and $\varphi_1=15.493$ for all curves.}
	\label{fig:psUncoupled}
\end{figure}

The synchronous gauge power spectrum, which we reconstructed from our gauge-invariant quantities, resulting from this evolution is shown in Figure \ref{fig:psUncoupled} (dashed lines). For comparison we also show a number of additional power spectra. The solid black line represents the power spectrum obtained for a pure cold dark matter component, evolved in the same background cosmology. The dashed lines represent bolon power spectra for different choices of the bolon exponent, from $\lambda = 30$ on the left to $\lambda = 70$ on the right, but with $c^2$ adjusted to give the same late background cosmology. Furthermore we show a warm dark matter power spectrum, obtained by modifying the transfer function for the cold dark matter component (gray line) in a manner suggested in refs. \cite{Barkana:2001gr,Bode:2000gq}, i.e. for thermally produced warm dark matter. The mass of the wdm-particle used here is 
\beq
m_{\rm wdm} = 2.284 \; {\rm keV} \, .
\eeq 
All power spectra have been normalized to the value of $\sigma_8=0.801$ given above.

The cutoff in the bolon power spectrum is initially much steeper than in warm dark matter models, which is typical for scalar field dark matter. Furthermore the shift of the cutoff to smaller wavenumbers for smaller $\lambda$ can be easily understood by noting that decreasing the bolon-exponential $\lambda$ and adjusting the mass-parameter $c^2$ using equation (\ref{c2Guess}) leads to effective decrease of the bolon mass. To see this, simply evaluate equation (\ref{c2Guess}) for $\beta=0$ to see that $c^2 \propto \lambda^6$, i.e. $m_0 \propto \lambda^4$ in this case. The resulting effect is a larger sound speed and as a consequence a suppression of growth extending to smaller wavenumbers. The opposite effect for larger values of $\lambda$ is also clear.

We now move on to study the influence the coupling $\beta$ has on the evolution of the bolon density contrast. It can be summarized in four effects:
\begin{enumerate}
\item  The coupling causes the energy density of the bolon to scale slightly differently from a standard cold dark matter one. For a positive coupling it scales away faster then $a^{-3}$ and therefore the bolon energy density exceeds the one obtained for the $\beta=0$ case for $z>0$ after adjusting $\rho_{\chi,0}$, leading to shift of matter-radiation equality to earlier times. This results in a shift of the maximum in the matter-power-spectrum to larger wavenumbers, since the horizon size at matter-radiation-equality is suppressed, an effect only boosted by the effect the coupled evolution has on the Hubble parameter. Similarly, a negative coupling has the opposite effect, shifting the maximum to smaller wavenumbers.
\item The growth of bolon perturbations during bolon-domination also gets changed by the coupling. The evolution of the growth of linear perturbations in a coupled quintessence scenario was analyzed in ref. \cite{Amendola:1999dr} and the results can be applied to to our model in the $\mu k^2/h^2 \ll 1$ regime. For not too large coupling the effect can be summarized as follows: A positive coupling increases the growth of linear perturbations, whereas a negative coupling decreases the growth compared to the uncoupled case, but the growth is always suppressed when compared to the standard cold dark matter evolution in an Einstein-de-Sitter universe, where $\delta \propto a$. This effect leads to an increase of power for large $k$ modes in the power spectrum for positive coupling, and a suppression for a negative one.

For larger couplings this no longer holds true. The growth rate now starts to increase even for negative couplings and a growth faster than $\delta \propto a$ also becomes possible.
Note that this is not in conflict with the results found in ref. \cite{Amendola:1999er}, where a suppression of growth for all couplings was found. The difference arises from the background evolution, since in our model the quintessence field adjusts itself to a different fixed point than the one analyzed in ref. \cite{Amendola:1999er}.

\item Finally the coupling also changes the evolution of the bolon-mass and thus of the sound-speed present in the $\mu k^2/h^2 \gtrsim 1$ regime. 
Adjusting the background cosmology according to equation (\ref{c2Guess}) leads to an increase in the value of the bolon mass at $a=1$ (i.e. today) with a positive coupling, in contrast to what one might expect naively from the ${\rm e}^{-\beta \varphi/M}$-dependence. Similarly, it decreases for a negative coupling. This leads to a decrease of the soundspeed for a positive coupling and thus to the cutoff in the power spectrum shifted to larger wavenumbers. The opposite of course holds for a negative coupling.

\item As a last effect, the growth of perturbations during the radiation dominated in era and during transition from radiation domination to matter domination also gets affected by the coupling. This is difficult to access analytically, and the important modifications of the power spectrum can be understood in terms of the previously discussed points alone. We will therefore not go into this any further.
\end{enumerate}

\begin{figure}[t]
	\centering
  \includegraphics[width=1.0\linewidth]{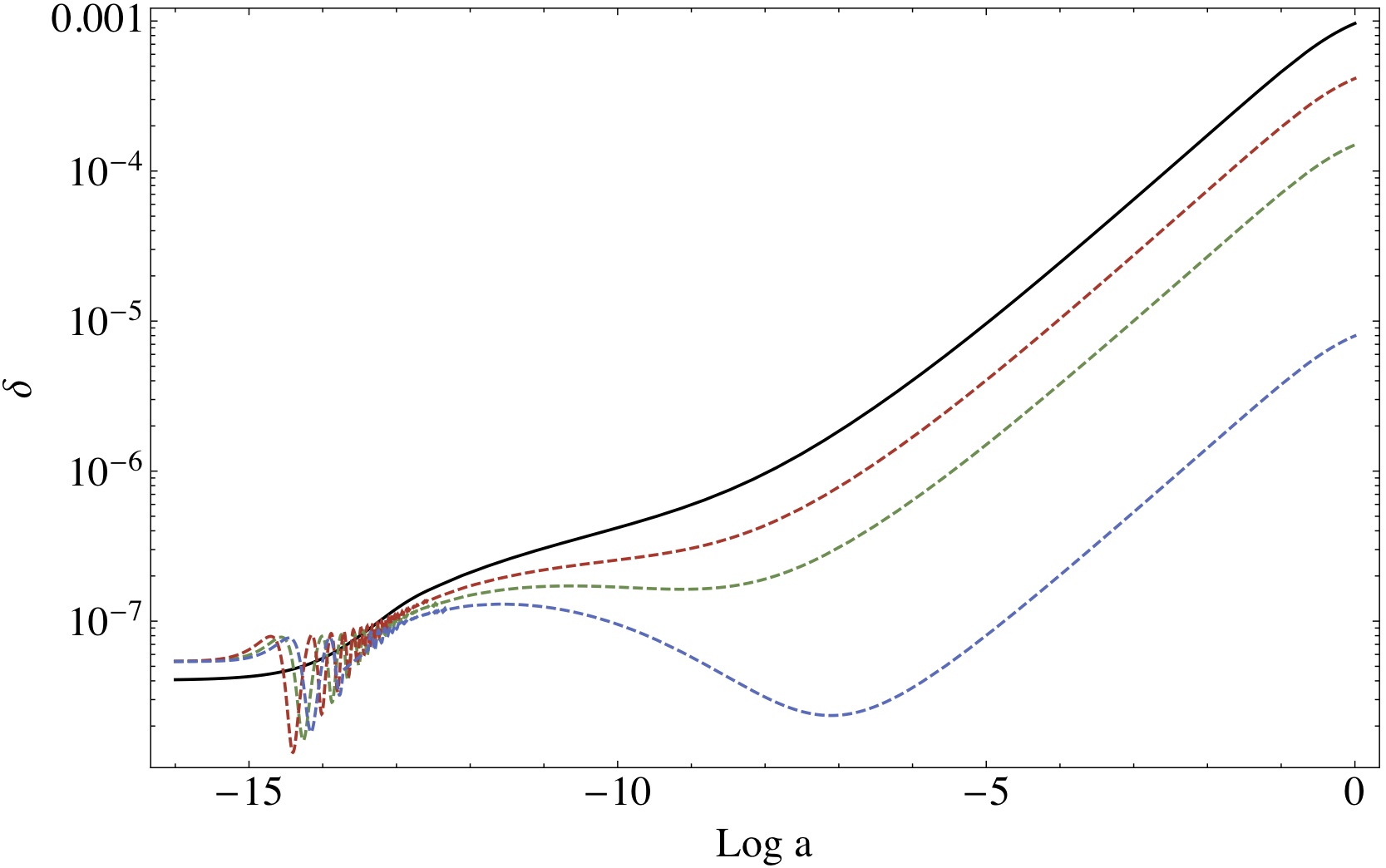}
	\caption{Evolution of the bolon density contrast with $k=10.7$ h/Mpc for varying couplings. The black line represents standard uncoupled cold dark matter, the dashed green line an uncoupled bolon model, the red line a model with $\beta=0.05$ ($N=0.3716$, $\varphi_0=13.786$ and $\varphi_1=16.248$) and the blue line $\beta=-0.1$ ($N=0.3777$, $\varphi_0=13.789$ and $\varphi_1=16.168$). The bolon exponential is $\lambda=40$.}
	\label{dcModes2}
\end{figure}

\begin{figure}[t]
	\centering
  \includegraphics[height=0.65 \linewidth]{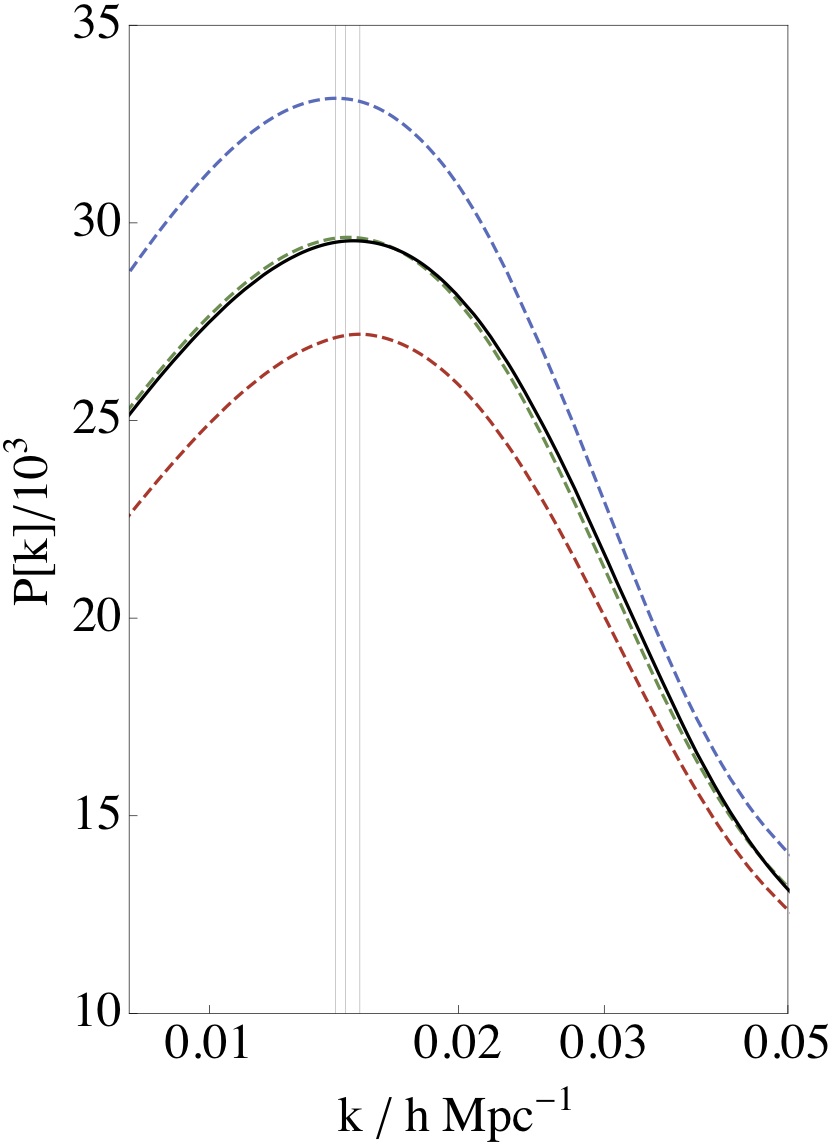}
  \includegraphics[height=0.65 \linewidth]{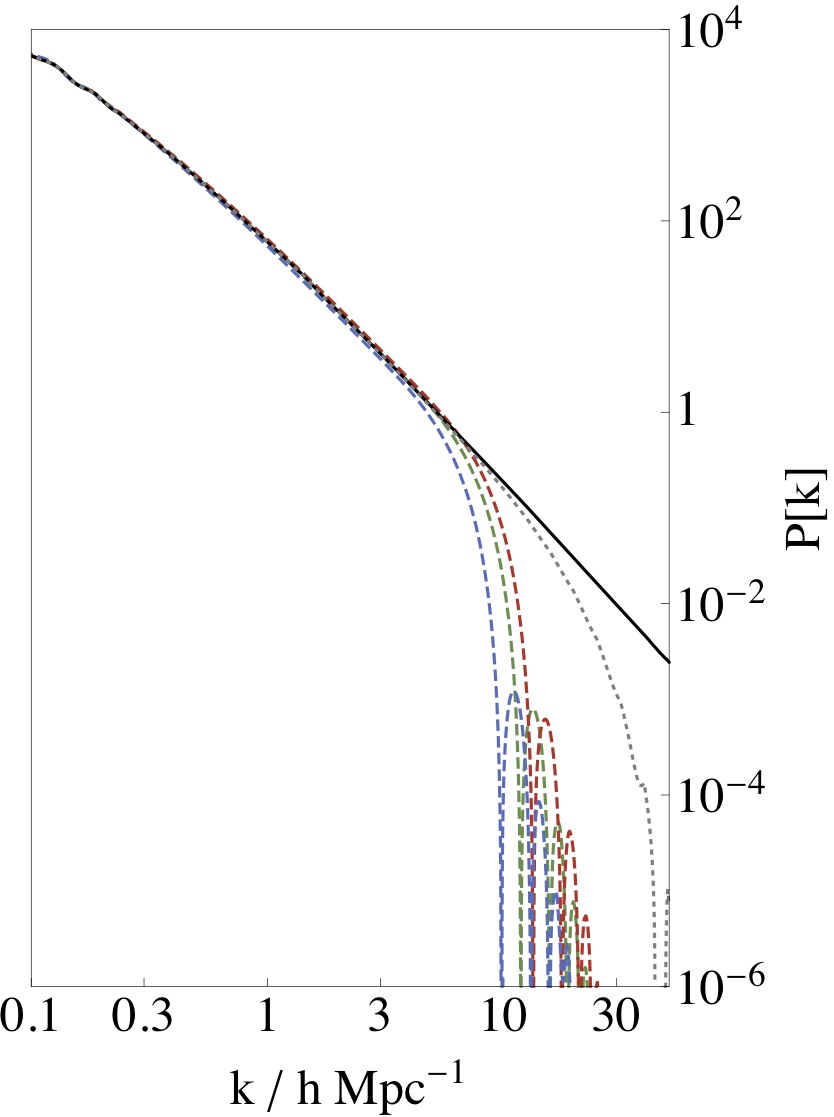}
	\caption{Power spectrum of the cosmon-bolon cosmology for different couplings. The black solid and gray dotted lines represent cold and warm dark matter spectra respectively, just as in Figure \ref{fig:psUncoupled}. The dotted lines represent cosmon-bolon models with $\lambda=40$ and $\beta=0, 0.05$ and $-0.1$ for the green, red and blue lines respectively. The three gridlines in the left hand figure show the positions of the maxima of the power spectra.}
	\label{fig:psCoupled}
\end{figure}

All these effects can be seen in Figures \ref{dcModes2} and \ref{fig:psCoupled}. In Figure \ref{dcModes2} we show the evolution of the bolon density contrast with wavenumber $k=10.7$ h/Mpc for a model with $\lambda=40$ and three different choices of the coupling plus a standard CDM evolution evolved in the same background (black solid line). 
The matter power spectra for the same models are shown in Figure \ref{fig:psCoupled}, where we have also added the WDM-modification (gray dotted line). Otherwise the coloring is the same as in Figure \ref{dcModes2}. One can clearly see the small shift of the maximum of the power spectrum depending on the coupling, as well as the growth modification, epitomized by the different slopes in the power spectra for large $k$. The difference in the spectra for smaller wavenumbers are an effect resulting from the normalization in conjunction with the different high-$k$ slopes and position of the maximum. The position of the cutoff (as long as it is at large enough $k$) has very little influence on the normalization.

Finally we have fitted the form of the cutoff by
\beq
\label{goodFit}
 P_{\chi} (k) = P_{\rm cdm} (k) \left\{ \frac{{\rm cos} \left[ (b \, x)^a \right]}{1+\sqrt{c} \, \left( x^{d_1} + x^{d_2} \right)} \right\}^2 \, ,
\eeq
where $x=k/k_J$ and the parameters $a,b,c,d_1$ and $d_2$ are functions of the model parameters $\lambda$ and $\beta$. We assumed a linear dependence for all functions here, optimized the parameters numerically and eliminated all terms the inclusion of which would not yield a significant improvement of the fitting. The resulting best estimate is given by
\begin{align}
a(\beta) & = 4.105 + 0.428 \beta \, , \\
b(\lambda,\beta) & = 0.827 + 0.098 \beta + 0.006 \lambda + 0.0025 \lambda \beta \, , \\
c(\lambda,\beta) & = -0.46 -1.9 \beta + 0.053 \lambda + 0.152 \lambda \beta \, , \\
d_1(\beta) & = 4.31-1.2\beta \, , \\
d_2(\beta) &= 7.66 + 1.49 \beta \, .
\end{align}
This gives a fitting better than 18\% up to suppression of $1/500$. A simpler fitting was proposed in similar models in refs. \cite{Hu:2000ke,Matos:2000ss}, but as we have checked such a simple fitting function does considerably worse. More details on the fitting procedure can be found in appendix \ref{app:CutoffFitting}. 

\section{Halo abundances}
\label{sec:PressSchechter}

In this section we go beyond the theory of linear perturbations and study the distribution of halos in our model. The approach we employ is known as the extended Press Schechter excursion set formalism, a theory originally developed in ref. \cite{Press:1973iz} and later refined and extended in several works \cite{Bond:1990iw,Bower:1991kf,Lacey:1993iv,Sheth:1999su,Sheth:2001dp}. It allows for the prediction of halo mass functions and merger histories, quantities which require knowledge about the highly non-linear regime of cosmic perturbations, from the linear power spectrum. 

The basic idea of Press and Schechter was to identify regions of space with an averaged density contrast above a certain threshold with collapsed objects. To put this in more precise terms: One averages the linearly evolved matter density contrast field over a radius $R$ and identifies the fraction of space which lies above a given threshold $\omega$ with the fraction of mass of the universe which is bound in objects with a mass greater than the mass associated with the size of the region, denoted by $M(R)$. 
The basic problem with this approach is the so called ''cloud in cloud problem'', a term which describes the following effect: Since large mass halos consist of a number of smaller mass halos, a region might switch back and forth from being considered collapsed (i.e. above the threshold) and non-collapsed (below the threshold) depending on the filtering radius $R$ and it becomes unclear which mass it should be assigned to. While this problem cannot be resolved in a unique fashion, the most commonly used approach is the prescription described in ref. \cite{Bond:1990iw} and now known as the excursion set formalism. \footnote{Probably the most well known alternative attempt to address this problem is the peak theory, originally developed by Bardeen et al. in ref. \cite{Bardeen:1985tr}. We will not investigate this approach further here.} In this approach one starts to filter the density contrast field with very large radii, leading to an effectively vanishing averaged density contrast everywhere, and then decreases the filter radius step by step. This creates a random walk in $R$-space for each spatial point. Now the fraction of mass in the universe which is bound in collapsed objects of mass bigger than $M(R)$, denoted by $\Omega(\omega;R)$, is simply given by the fraction of trajectories which have crossed the threshold at some radius greater than $R$. This amounts to solving a so called \textit{absorbing barrier problem}, as the random walk trajectories get absorbed by the threshold $\omega$ when they cross it for the first time.

As is common practice, we always work with the density contrast field at $z=0$ and put the entire time evolution into the barrier $\omega(z)=\omega(z=0)/D(z)$, where $D(z)$ is the linear growth function. 
Furthermore, instead of using the radius $R$, we rewrite everything in terms of the variance at a given filtering radius, which can be calculated from the power spectrum as follows:
\beq
\label{sigmaR}
S(R) =  \int \frac{\d k }{2\pi^2} \, k^2 \, P(k,t) \left| \widehat{W}_R(k) \right|^2 \, ,
\eeq
where $\widehat{W}_R(k)$ is the Fourier transform of the filtering function. This function is of course always invertible, with large radii corresponding to small variances. We will therefore from now on work with a mass-assignment $M(S) \equiv M(R(S))$ and a mass fraction $\Omega(\omega;S) \equiv \Omega(\omega;R(S))$. The number density of objects of mass $M$ is then given by
\beq
\label{totalNumberDensity}
n(\omega; M) \d \, {\rm ln} M = \bar{\rho}_m f(\omega; S)\frac{\d \, S}{\d M} \, \d \, {\rm ln} M
\eeq
where $f(\omega;S)$ is the first crossing rate at variance $S$, which is given by
\beq
\label{totalFCR}
f(\omega; S) \equiv \frac{\d \, \Omega (\omega; S)}{\d S} \, . 
\eeq
The details of this calculation depend on the choice of the filtering function, the mass assignment and the threshold $\omega$, which we will now discuss.

\subsection{Filter choices and mass assignments}

As a first step we calculate the variance of the matter density fluctuations 
for some choice of filtering function $W_R(r)$ with Fourier transform $\widehat{W}_R(k)$. The most common choices are a real space tophat window function with Fourier transform given by
\beq
\label{tophatForm}
\widehat{W}^{th}_R(k) = \frac{3 ({\rm sin}(k R) - k R \, {\rm cos}(k R))}{k^3 R^3}
\eeq
and a sharp filter in k-space given by
\beq
\widehat{W}^{sh}_R(k) =  \Theta (1-k R) \, .
\eeq
In a second step we need to assign a mass to each filtering radius $R$. For the tophat filter the obvious choice is to simply take the mass enclosed within the filter, but for the sharp-k filter there is no such canonical option. In fact, the integral over the spatial filter-function is not even well-defined in this case (see e.g. \cite{Maggiore:2009rv}). The only reasonable assumption one can make is that the mass should scale like $R^3$, i.e.
\beq
\label{massAssignment}
M(R) = \frac{4 \pi}{3} \rho_m (A R)^3 \, .
\eeq
The normalization $A=1$ corresponds to the tophat-filter choice, we will describe how we adjust $A$ for the sharp-k filter below.

Finally, the calculation of the first crossing rate $f(\omega; S)$ also depends critically on the choice of the filter. For a sharp k-filter one can easily see that the random walks in $S$-space consist of uncorrelated steps, which makes the problem considerably easier to tackle. In the case of very simple thresholds one can solve the first crossing rate analytically (see ref. \cite{Bond:1990iw}), and for the most straightforward case of a constant barrier one obtains
\beq
f(\omega; S) = \frac{\omega}{\sqrt{2 \pi} S^{3/2}} {\rm exp}\left( -\omega^2/2S \right) \; .
\eeq
When using a tophat filter the random walk becomes correlated, which complicates the problem considerably. A numerical procedure to calculate correlated random walks has been proposed in ref. \cite{Farahi:2013fca}, based on a set of earlier papers by Maggiore and Riotto \cite{Maggiore:2009rv,Maggiore:2009rw,Maggiore:2009rx}. In what follows below we will ignore this issue. While this is strictly speaking incorrect, there are two arguments in favor of this approach: First, the corrections are expected to be small for the spectra we employ here, and second, the elliptical barrier modification widely employed today has been matched to the results of CDM N-body simulations in precisely this fashion, and changing the calculation of the first crossing rate would require a re-adjustment of this barrier as well.

In the following sections we will portray results for both a sharp-k filter and a tophat filter. 
While the differences for CDM-spectra are minute once the mass-assignment and the barrier for the sharp-k filter has been adjusted correctly, the results for spectra exhibiting a sharp cutoff in the power spectrum (i.e. saturation in the variance function $S(R)$) are very different. At first glance, the sharp-k filter seems to have some advantages.

First, it allows for a relatively simple and yet rigorous calculation of the first crossing rate. Second, it is known that number densities for models exhibiting a cutoff in the linear power spectrum do not always show a cutoff in the number densities calculated with the Press-Scechter excursion set approach when a tophat-filter is used \cite{Benson:2012su} (see e.g. Figure \ref{fig:totalNumbers}). This is due to the oscillatory form of equation (\ref{tophatForm}), which leads to much milder decrease in $\d S/\d M$ compared to sharp-k filter results (see Figure \ref{fig:dsdm}). One should however note that this shortcoming can at least in some cases be cured by using a suitable barrier shape (e.g. in WDM-models, see ref. \cite{Benson:2012su}).

The downside (or upside, depending on your point of view) of using a sharp-k filter, in addition to the ambiguity when assigning masses, is that we cannot readily generalize the barriers deduced for a tophat filter to this approach. On the one hand this makes the formalism less deterministic, on the other hand it gives us some additional freedom. 

As we will see from the results below, when we extend the ePS formalism to the prediction of substructure abundances within a galaxy-sized dark halo, the advantages of the sharp-k filter cannot make up for its one huge problem, which is the mass assignment. 

\subsection{Spherical collapse}
Usually barriers used in the Press-Schechter formalism are derived from the spherical collapse model or generalizations thereof \cite{Abramo:2007iu,Pace:2010sn,Basse:2010qp}. This originally very simple formalism has been extended to include CDM models with a coupling to a quintessence field \cite{Wintergerst:2010ui}, but it runs into serious problems for a model such as ours, where the dynamics of the collapsing component are strongly scale-dependent already in the linear regime. Let us quickly outline why this is.

First we should note that spherical collapse itself is of course an approximation. One assumes an initial overdensity which is spherically symmetric and homogeneous (i.e. has a tophat-profile) and evolves this overdensity. Depending on the model investigated, at this point one either invokes Birkhoff's theorem or some incarnation of Newtonian cosmology in order to simplify the equations of motion. In the first case one effectively treats the overdensity as an independent FLRW-universe, the only difference to the background is a different curvature constant arising from the overdensity. In the Newtonian approach one has to take care of gradient terms appearing in the equations, which are generally ignored, usually with reference to either the lack of gradients in the initial density profile or to the size of the overall perturbation. Whichever approach one employs, it has to fulfill (at least) two basic criteria:
\begin{enumerate}
\item The perturbation should retain its (tophat-)profile, at least to very good approximation.
\item The evolution of the perturbation should agree with linear cosmological perturbation theory at early times.
\end{enumerate}
Both of these demands cannot be met by either formalism throughout the mass-range which interests us. To see why, let us investigate the first condition.

The evolution of linear perturbations in a model such as ours depends strongly on the wavenumber of the perturbation, even in a simplified cosmology which is bolon-dominated throughout its evolution. This is simply due to the scale-dependent sound-speed present in the the averaged equations and a fundamental difference to CDM-models, where we have a universal growth proportional to the scale factor during matter domination. Thus it is immediately clear that an initially tophat-shaped perturbation will not retain its shape even in the linear regime and not even approximately, if modes where the soundspeed is relevant make up a sizable contribution to the Fourier-decomposition of the perturbation. This is clearly the case if the size of the perturbation is close to the Jeans mass, which is the most interesting regime. For much larger masses however, the suppressed modes play (almost) no role for the evolution of the profile, as they are almost irrelevant in the Fourier-decomposition. Thus for large masses the spherical profile is stable to very good approximation and gives, in the uncoupled case, the same result as standard CDM spherical collapse.

One might think that a useful way to get around this problem is to employ Birkhoff's theorem, as it treats the overdensity as an independent universe and thus forces it to stay spherically symmetric. But in doing this, one quickly runs into conflicts with the second demand. As the soundspeed present in the averaged equations is non-adiabatic, it is a purely perturbative quantity (in contrast to the adiabatic sound-speed, which can be calculated from the background evolution only). Thus the background equations can not reflect the suppression of growth present in large-k modes in our model. Quite the opposite, we have verified numerically that one recovers a collapse model very similar to standard CDM collapse when evolving an overdensity as an independent universe in our model, as was to be expected from the fact that the averaged equations give rise to an $\omega=0$ evolution in the background. One can obtain a slightly delayed collapse time compared to a standard CDM collapse if one adjusts the initial conditions to fit the linear evolution at some early initial time, but not by much, and the delay depends on the initial time chosen. Simply put, the reason for this is that the spherical collapse model in its simplest form (i.e. with only one component collapsing) is determined completely by two parameters, the initial overdensity and its initial time derivative. When trying to adjust the time derivative to an initially suppressed evolution for small masses, one finds numerically that the evolution of the density contrast quickly adjusts itself to the standard CDM evolution, thus resulting in only a small delay in the collapse time. This is of course very different from the linear evolution recovered in cosmological perturbation theory and thus unrealistic.

The approach from Newtonian cosmology essentially suffers from the same shortcomings. One can easily find a version of Newtonian cosmology that recovers the cosmological linear perturbation equations in the subhorizon regime, but this just puts one back to the problem of stability of the tophat shape.

The problems just outlined indicate that finding the correct barrier for the bolon model could be a highly non-trivial issue. The one thing we can say for sure is that for large masses (corresponding to large radii), the modes which are suppressed in the linear regime play (almost) no role in the evolution of a tophat-perturbation, and we can thus employ the usual spherical collapse model and recover the standard barrier. For smaller masses close to the Jeans mass however, things seem to be much more complicated, and one might have to resort to a comparison with N-body or fluid simulations to tackle this issue. To our knowledge, no suitable data from such simulations of scalar field dark matter exists at this point.

\subsection{Barrier for a sharp-k filter}
The CDM spherical collapse model gives rise to a constant collapse barrier $\omega = \delta_{sc} \approx 1.686$, which was used in the first works employing the Press-Schechter formalism. While this very simple assumption already gives very useful results, comparisons with very accurate N-body simulations led to modifications motivated by the elliptical collapse model. As was shown in ref. \cite{Sheth:2001dp}, a remapping of the barrier according to
\beq
\label{ellipticalBarrier}
\omega(S,z)=\sqrt{A} \delta_{sc}(z) \left[ 1+b\left( \frac{S}{A \delta_{sc}^2(z)} \right)^c \right]
\eeq
with $A=0.707$, $b=0.5$ and $c=0.6$ gives a more accurate fit. The parameters $b$ and $c$ are a result of the elliptical collapse model, whereas the parameter $A$ has to be put in by hand. It can however be argued that it is related to the way in which structures in N-body simulations are identified, a method which allows for some variability with influences on the halo abundances. The barrier was derived using a spatial tophat filter, but the corresponding first crossing rate was calculated using uncorrelated random walks. This is strictly speaking inconsistent, but the barrier has been adjusted to fit N-body simulations in ref. \cite{Sheth:2001dp} in precisely this fashion and therefore gives correct results. 

The number densities obtained when using a sharp-k filter however have a different shape when the same barrier is used, due to a different mass-dependence of the variance $S(M)$. This discrepancy cannot be fixed by adjusting the mass-assignment alone. In order to get correct results, we also have to modify the barrier associated with a sharp-k filter. This is not inconsistent, as the barrier motivated by spherical (or elliptical) collapse can only be expected to give good results when a tophat-filter is used. The possibilities to do this are of course plentiful, but one can get very good agreement already for a very simple modification. Following \cite{Benson:2012su}, we simply shift the barrier upwards by multiplying it with a constant factor
\beq 
\label{barrierScaling}
\omega \rightarrow B \omega \, .
\eeq
This rescaling is applied after other barrier modifications such as the elliptical modification in equation (\ref{ellipticalBarrier}). As it turns out, by adjusting the parameters $A$ in eq. (\ref{massAssignment}) and $B$ in the above rescaling, we can get a good fit to the total number densities obtained for a CDM spectrum.\footnote{As mentioned above, for spectra exhibiting a cutoff in the power spectrum the tophat and sharp-k filter results differ considerably for masses close to and below the cutoff scale. This discrepancy cannot be erased by a simple scaling of the barrier, but this might not even be desirable. In the WDM case the turnover seen in the halo mass function in N-body simulations cannot be reproduced with a tophat filter and the elliptical barrier \cite{Benson:2012su}.} For the choices $A=2.27$ and $B=1.1$, the relative discrepancy is below $10\%$ everywhere in the mass range from $10^6 M_\odot $ to $10^{16} M_\odot $. \footnote{Other mass assignments suggested for the sharp-k filter correspond to $A=2.5$ in \cite{Benson:2012su} or $A=2.42$ in \cite{Lacey:1993iv}, which are not too different from our fit. These assignments are however motivated by different considerations.}

\subsubsection{Modifications near the Jeans mass}

In a further extension of the formalism, several authors have studied how to apply the Press Schechter excursion formalism to WDM models \cite{Barkana:2001gr,Benson:2012su}. The differences here are twofold.
First, the power spectrum exhibits a cutoff at some wavenumber, leading to suppression of number densities for the corresponding mass through the $dS/dM$ term in equation (\ref{totalNumberDensity}). Second, the velocity dispersion of WDM particles leads to modifications in the spherical and elliptical collapse model. As was shown in ref. \cite{Barkana:2001gr}, these effects lead to later virialization times and larger virialization radii, which need to be taken into account in the Press Schechter formalism by modifying the collapse barrier. A fit for this modification (which is accurate for masses not too far below the Jeans mass) was given in ref. \cite{Benson:2012su} and reads
\begin{align}
\label{wdmBarrier}
\omega_{WDM} (M,z) = & \omega_{sc}(z) \left[ h(x)\,  \frac{0.04}{{\rm exp(2.3x)}} \right. \nonumber \\
& \left. + (1-h(x)) \, {\rm exp} \left( \frac{0.31687}{{\rm exp}(0.809 x)} \right) \right]
\end{align}
where $x={\rm ln} (M/M_J)$, with $M$ the halo mass in question and $M_J$ the Jeans mass, defined in ref. \cite{Benson:2012su} as the halo mass for which pressure and gravity balance initially (in the linear regime). The function $h(x)$ is given by
\beq
h(x) = \left[ 1+{\rm exp}[(x+2.4)/0.1] \right]^{-1} \, .
\eeq
This modification effectively increases the barrier dramatically for masses below the Jeans mass and thus suppresses the first crossing rate $f(\omega;S(M))$ in that regime, which has a non-negligible effect on the predicted number densities, as was shown in ref. \cite{Benson:2012su}.

\begin{figure}[t]
	\centering
  \includegraphics[width=1.0\linewidth]{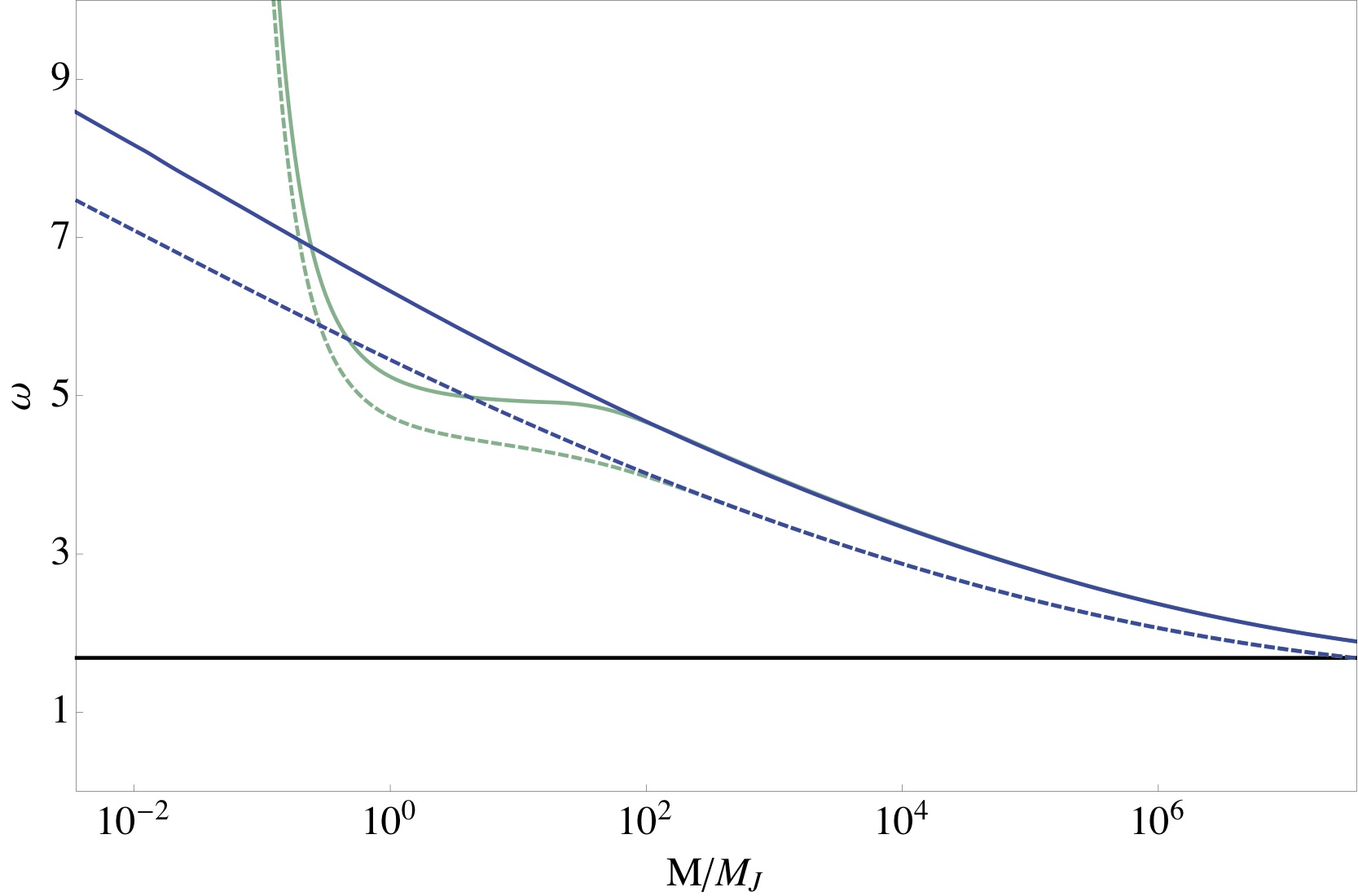}
	\caption{Different barriers used in this work. We show the spherical collapse barrier (black), the elliptical collapse barrier for the tophat-filter given in equations (\ref{ellipticalBarrier}) (blue dashed) and the elliptical collapse barrier shifted by a factor of 1.1 (blue solid) for the sharp-k filter. Finally we show the barriers modified through equation (\ref{wdmBarrier}) to get a sharp upturn near the Jeans mass for both the tophat elliptical barrier (green, dashed) and its raised version for the sharp-k filter (green, solid). All modification have been calculated from a bolon power spectrum with $\lambda=65$, $\alpha=20$ and $\beta=0$.}
	\label{fig:barriers}
\end{figure}

\begin{figure}[t]
	\centering
  \includegraphics[width=1.0\linewidth]{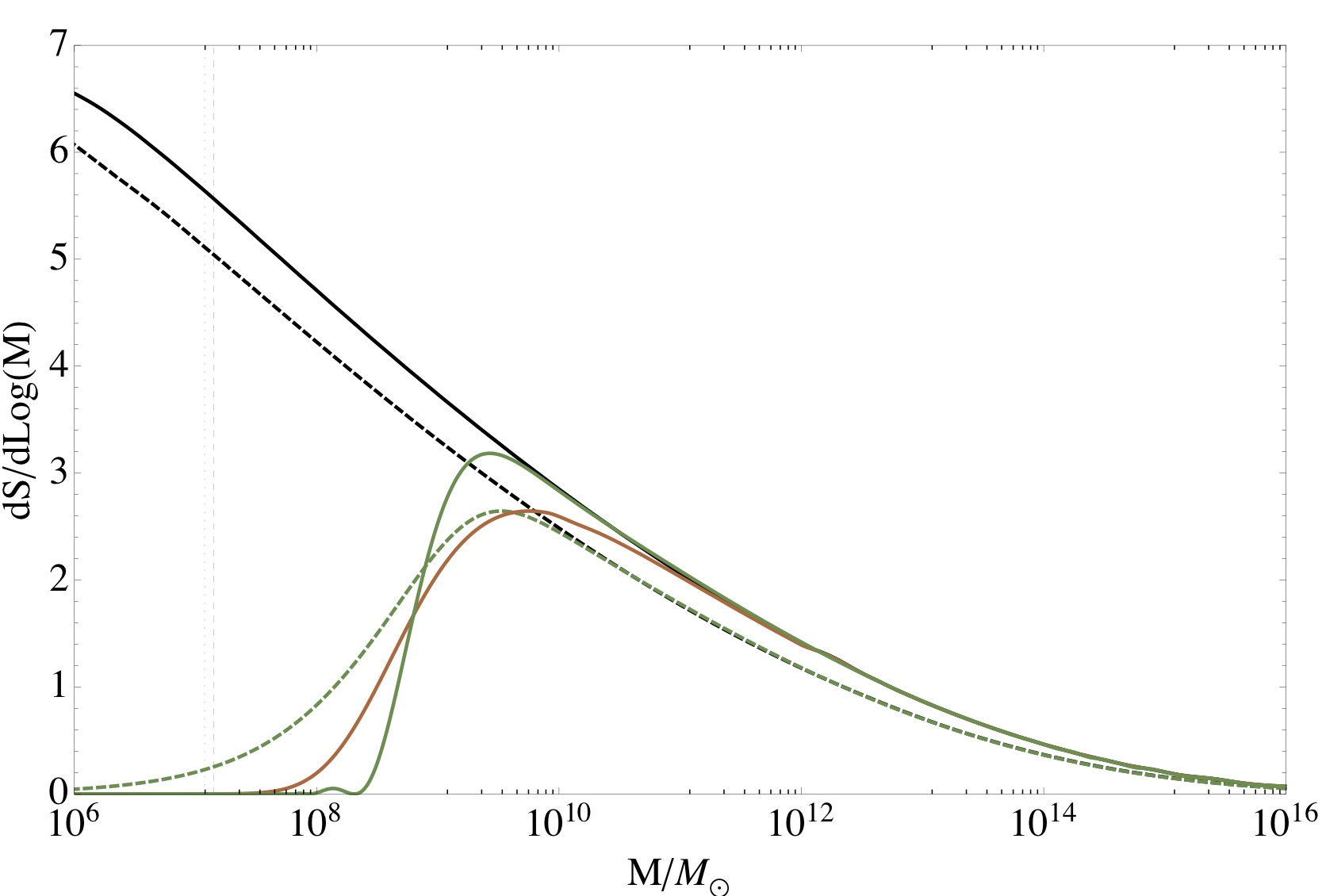}
	\caption{dS/dLog(M) for a tophat filter (dashed) and a sharp-k filter (solid) with the mass assignment given in eq. (\ref{massAssignment}) for a standard CDM power spectrum (black) and the bolon model (green) with $\lambda=65$, $\alpha=20$ and $\beta=0$. The red line represents a thermal WDM model with a $2.284$ keV mass for a sharp-k filter.}
	\label{fig:dsdm}
\end{figure}

One should note that the modification given in equation (\ref{wdmBarrier}) was obtained from a 1-dimensional baryonic simulation, where the temperature was adjusted to fit the WDM velocity dispersion, and not from some spherical collapse model of WDM in the stricter sense, which would suffer from the same fundamental difficulties we mentioned above.
It is unclear to us what the correct shape of the barrier should be in our model, in particular whether or not it exhibits an increase around the Jeans mass or not. In order to show the full range of possibilities and the effect of the barrier choice, we will present all calculations for both cases, once for the modified elliptical collapse barrier described above and once for a barrier where the WDM-modification given in equation (\ref{wdmBarrier}) is included. In the latter case we do of course insert the Jeans mass appropriate for our model, which is given by equations (\ref{jeansWavenumber}) and (\ref{jeansMass}). Different barriers are shown in Figure \ref{fig:barriers}.

\subsection{Total number counts}

\begin{figure}[t]
	\centering
  \includegraphics[width=1.0\linewidth]{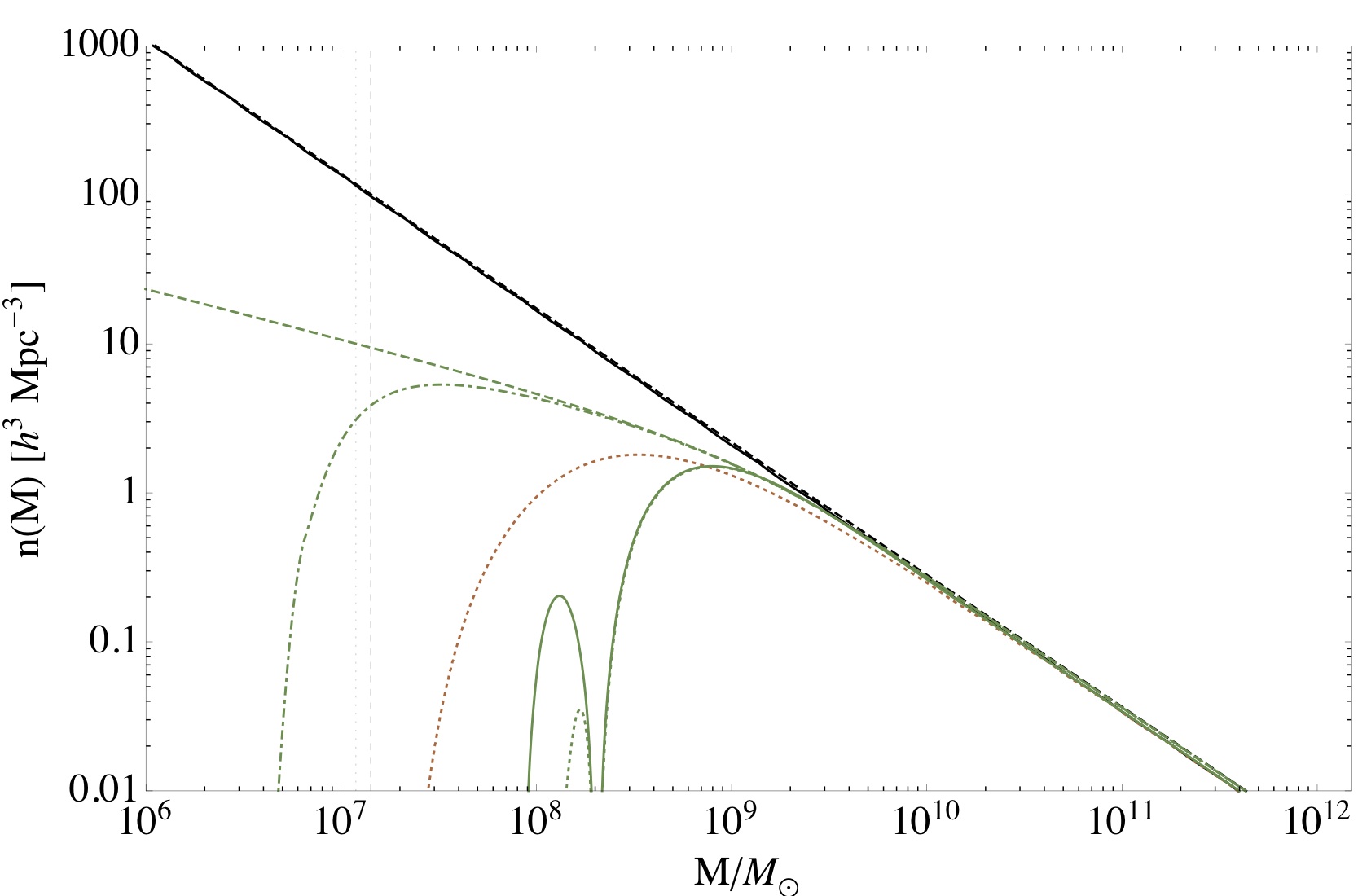}
	\caption{Total number densities predicted by the Press-Schechter excursion set formalism for different models and different filters. The black lines show the results for a CDM power spectrum with a tophat (dashed) and a sharp-k filter (solid) respectively. The elliptical barrier and mass assignment for the sharp-k filter has been successfully adjusted to fit the tophat results, both lines are (almost) indistinguishable. The green lines show the results for a bolon model with $\lambda=65$, $\alpha=20$ and $\beta=0$. A tophat filter was used for the dashed line, and an additional upturn of the barrier near the Jeans mass inserted for the dashed-dotted one. The solid line shows the sharp-k filter results with an elliptical barrier, whereas the dotted line shows what happens if an upturn in the barrier is included. Finally the red line shows the WDM model with a sharp-k filter and Jeans mass modification already displayed in Figure \ref{fig:dsdm} for comparison.}
	\label{fig:totalNumbers}
\end{figure}

The total number densities resulting from the Press-Schechter excursion set formalism are shown in Figure \ref{fig:totalNumbers}. The $\Lambda$CDM model is shown in black, with the dashed curve representing the tophat filter with the elliptical collapse barrier adjusted to fit N-body results. The solid curve shows the sharp-k filter results, with the mass assignment and the barrier shift adjusted to fit the dashed curve. Both lines are almost indistinguishable. The green lines are results obtained for the bolon-model with $\lambda=65$, $\alpha=20$ and $\beta=0$, where a tophat filter was used for the dashed line and a sharp-k filter for the solid one. For the dotted and dashed-dotted lines the additional modification of the barrier due to a possible upturn at the Jeans-mass was put in, for a sharp-k filter and a tophat filter respectively. Finally the red dotted line represents the WDM model from the previous section, here we used the fully modified barrier, including the modification from equation (\ref{wdmBarrier}) and a sharp-k filter. All first-crossing rates were calculated numerically using the method presented in the appendix of ref. \cite{Benson:2012su}.

One can clearly see the effects the choices of filter, mass assignment and barrier have on the predicted number densities in the bolon model. First, when using a tophat filter, a strong suppression is only present if an upturn of the barrier near the Jeans mass included, for the standard elliptical collapse barrier the suppression is much weaker. Second, for a sharp-k filter, this is not the case, both curves are almost identical. The only difference is a slightly quicker suppression of the oscillatory part, resulting from the oscillatory shape of the power-spectrum cutoff, if the barrier is raised near the Jeans mass. This can be easily understood and nicely visualizes the main issue for the sharp-k filter: For this filter choice, the cutoff in the number densities is determined by the very sharp cutoff in the function $\d S/ \d M$ alone. For our mass-assignment, this happens at masses above the Jeans mass, which makes a possible barrier modification irrelevant. A much lower choice of the parameter $A$ would of course again lead to a bigger difference between the two barrier choices and at very low $A$ the first crossing rate would once again determine the cutoff position, similar to the tophat filter. From these simple observations it is clear that the mass-assignment for the sharp-k filter introduces huge uncertainties for models such as ours, which will translate from the total number densities to substructure abundances (which we treat below) as well. We know currently of no way to resolve this issue and stick to the tophat filter from now on.

\subsection{Progenitor mass functions}

Finally we move on to show what is probably the most interesting aspect of small scale power suppression, and that is the abundance of substructure within a typical galaxy such as ours. The extended Press-Schechter-formalism allows for the calculation of the so called conditional mass function, denoted by 
\beq
g (M_1,z_1 | M_2,z_2) = - f(\omega_1;S_1 | \omega_2;S_2) \frac{\d S_1}{\d {\rm log} M} \Big|_{M=M_1} \, .
\eeq 
This describes the fraction of mass from halos of mass $M_2=M(S_2)$ at redshift $z_2$ corresponding to the barrier $\omega_2$ which is contained in progenitor halos of mass $M_1=M(S_1)$ at redshift $z_1$ corresponding to the barrier $\omega_1$ per log M-interval. It is determined by the variance as a function of mass and $f(\omega_1,S_1| \omega_2,S_2)$, which corresponds to the first crossing rate of Langevin-trajectories through the barrier $\omega_1$ at $S_2$, for trajectories which originated at $\omega_2(S_2)$ at variance $S_2$.

This function can again be obtained analytically in the case of a constant barrier \cite{Bond:1990iw}, but for the complicated barrier shapes needed here we have to resort to numerics again. Luckily one can employ the same strategy as in the case of total number densities, because this conditional first crossing rate is the same as the unconditional first crossing rate with an adapted barrier shape \cite{Benson:2012su}:
\beq
f(\omega_1;S_1 | \omega_2;S_2) = f(\tilde{\omega};S_1-S_2) \, ,
\eeq
where
\beq
\tilde{\omega}(S) = \omega_1(S+S_2) - \omega_2(S_2) \, .
\eeq
The change of the barrier with redshift is of course determined by the linear growth function $D(z)$, which we obtain numerically from our Boltzmann-code. It can be somewhat approximated by $D(z)=(1+z)$, but this fails for low $z$ due to dark energy and for medium $z$ due to the modified structure growth if $\beta$ is non-zero. Furthermore, as mentioned before, the growth in our model is non-universal for all k-modes, which gives a $k$-dependent linear growth function as well. As this is difficult to incorporate into the extended Press-Schechter formalism, we simply indicate indicate its possible effect, as we did when discussing total number densities, by a barrier-modification near the Jeans mass as given in equation (\ref{wdmBarrier}). 

The elliptical barrier modification given in equation (\ref{ellipticalBarrier}) has been shown to also give better results than the constant spherical collapse barrier for progenitor mass functions \cite{Sheth:2001dp}. One should note that the time-evolution of the elliptical barrier is not obtained by applying 
\beq
\omega(z) = \omega(z=0) / D(z) \, ,
\eeq
as one might expect from the Press-Schechter logic, but by replacing the critical overdensity $\delta_{sc} \rightarrow \delta_{sc}(z)$ and applying the elliptical barrier modification afterwards, as already denoted in eq. (\ref{ellipticalBarrier}).\footnote{Exchanging these two modifications would lead to a strong underprediction of subhalo abundances for higher redshifts, i.e. a strongly accelerated version of hierarchichal structure growth, which is not realistic.} 

\begin{figure*}[t]
	\centering
  \includegraphics[width=.44\linewidth]{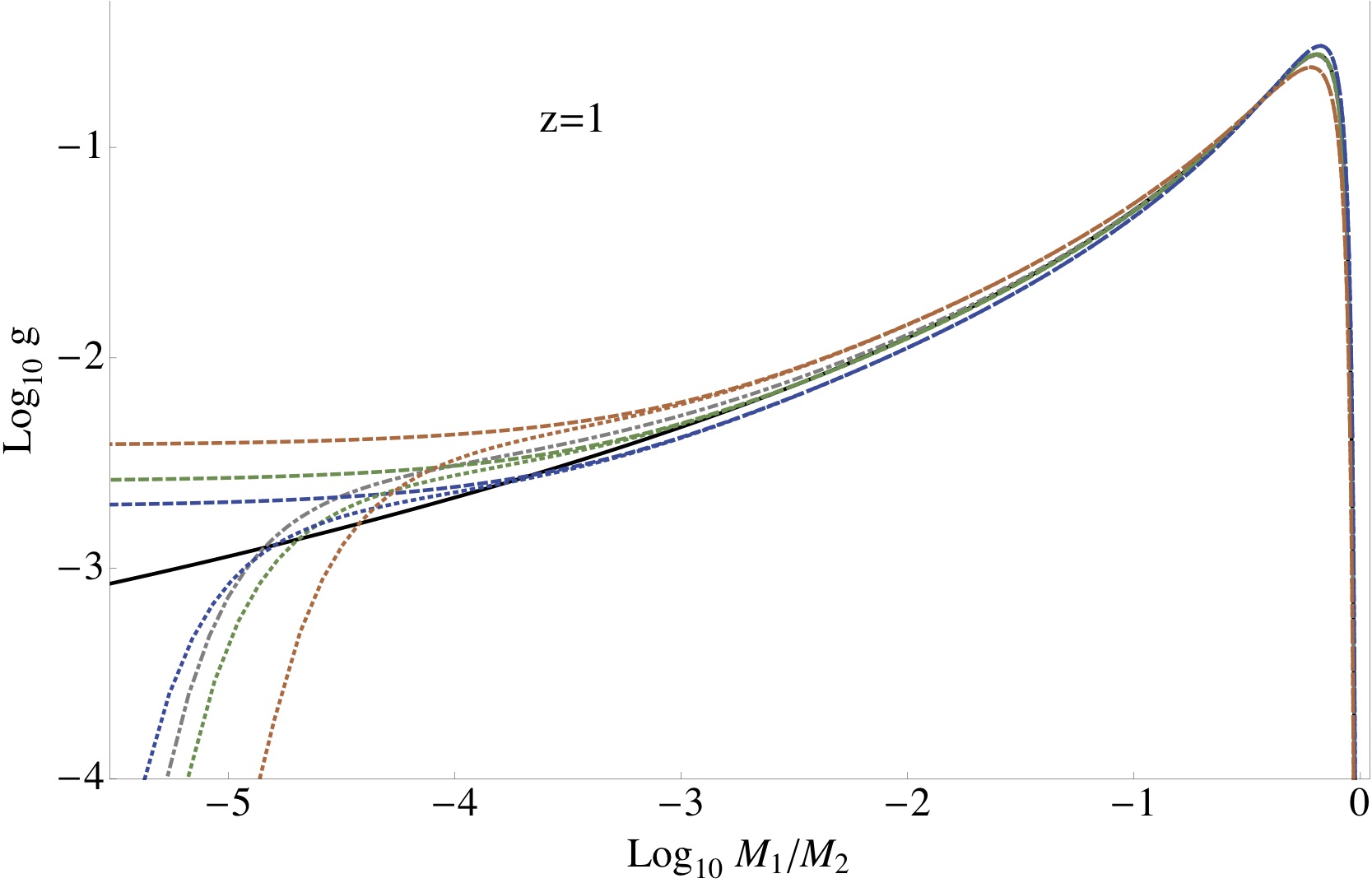}
  \includegraphics[width=.44\linewidth]{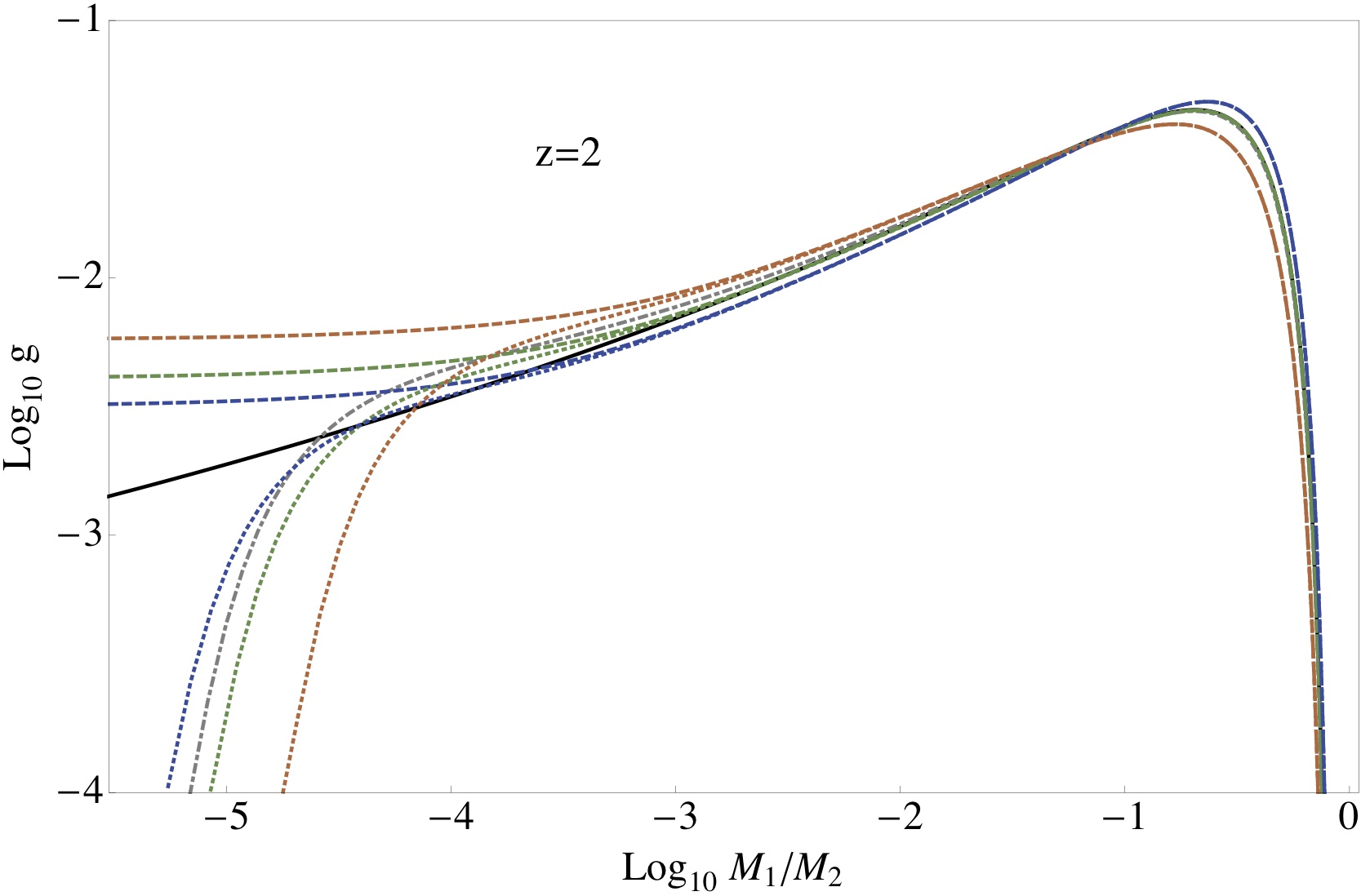}
  \includegraphics[width=.44\linewidth]{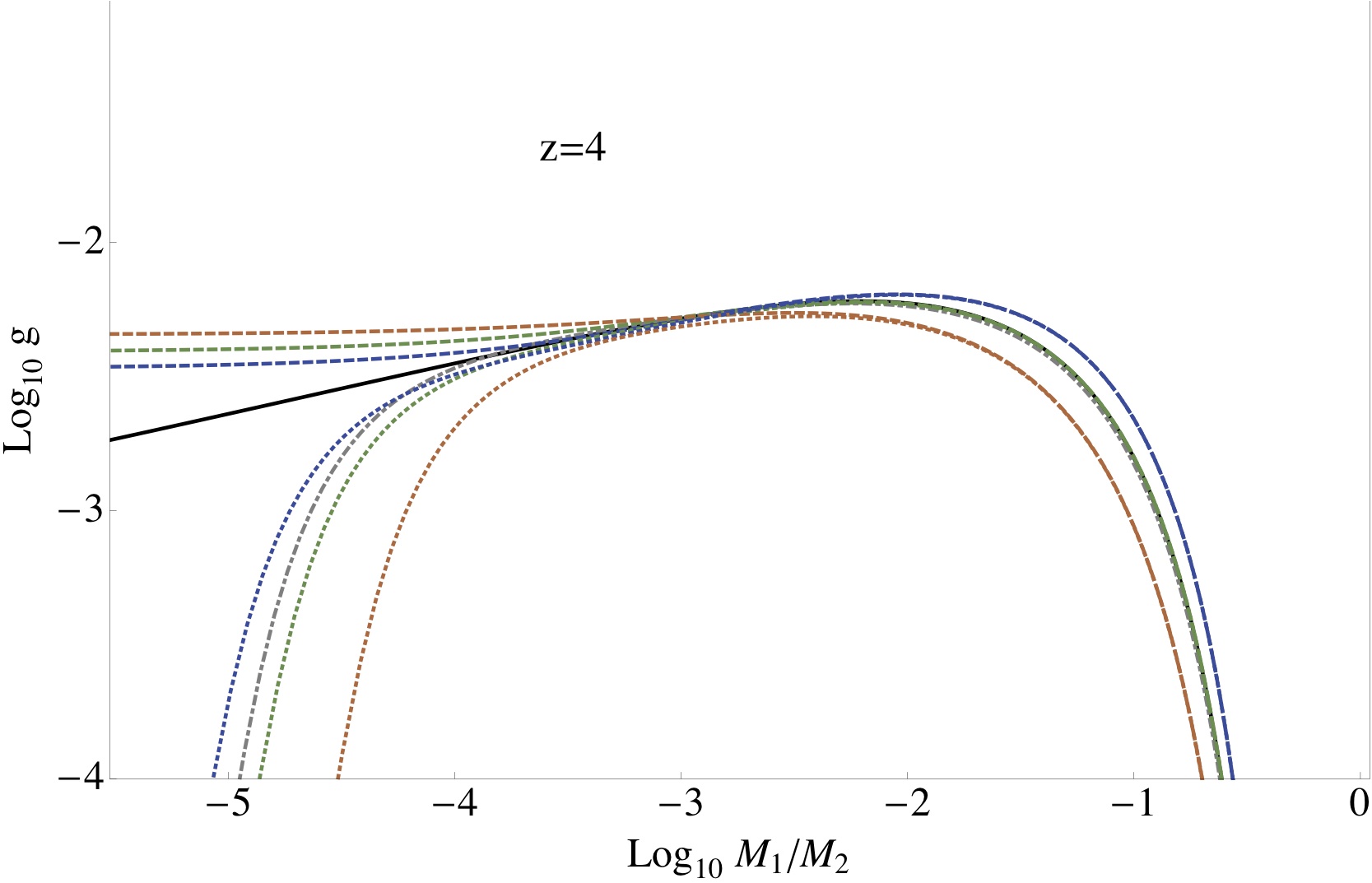}
  \includegraphics[width=.44\linewidth]{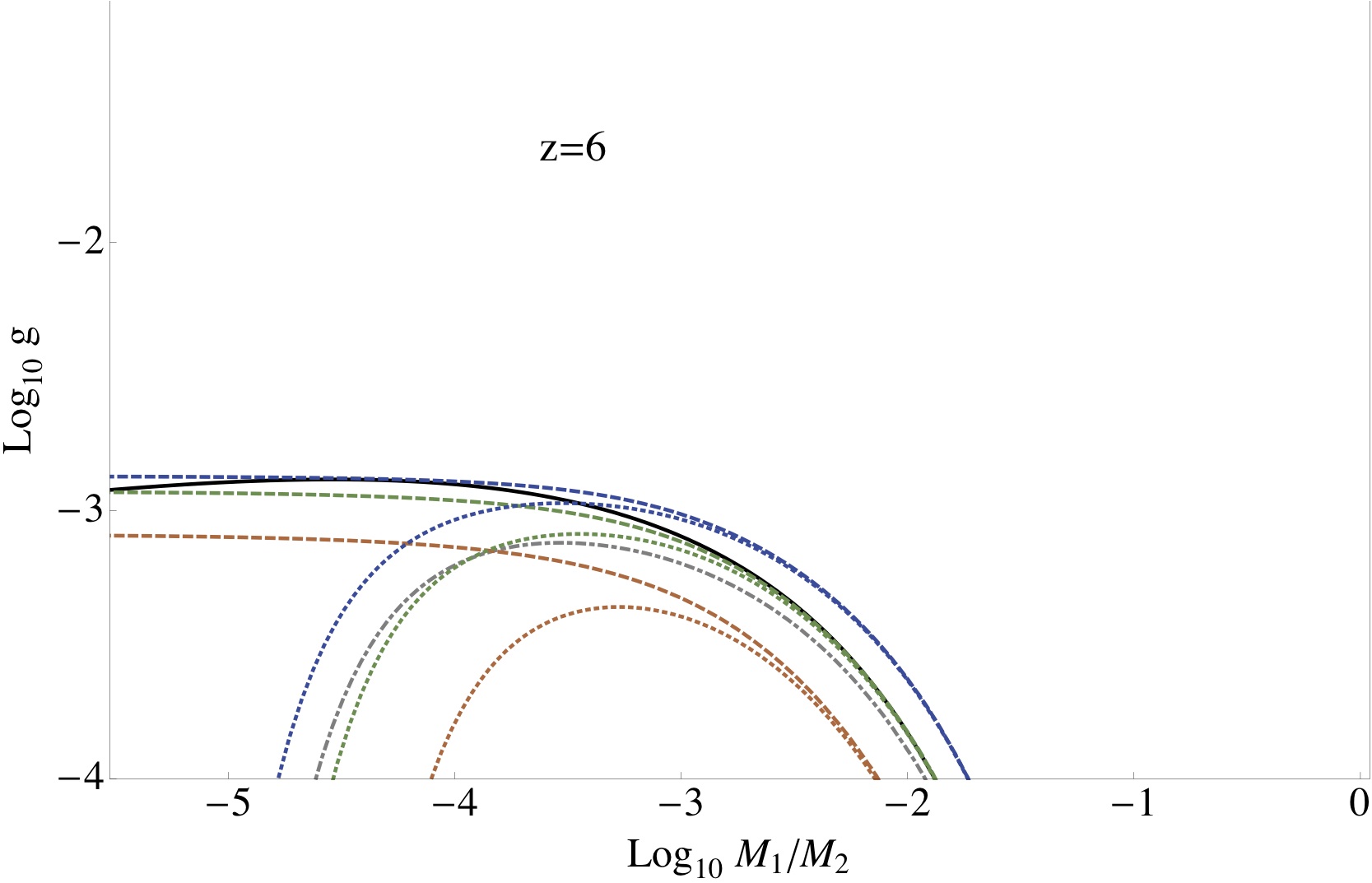}
	\caption{Conditional mass functions for different power spectra. The ePS-result for a tophat filter with the shifted barrier are shown in solid black for CDM and in dash-dotted gray for a wdm model with particle mass of 2.284 keV. The dashed lines represent the bolon-model with $\lambda=65$, $\alpha=20$ and $\beta=-0.1,0.,0.05$ for the red, green and blue lines respectively, where the elliptical barrier was used. The dotted lines stand for the same models, but this time with an additional upturn of the barrier near the Jeans mass included.}
	\label{fig:fullCondFCR}
\end{figure*}

The conditional mass function can be seen for a number of redshifts in Fig. \ref{fig:fullCondFCR}. 
The reference mass in this figure is of the order of a typical galaxy mass, we chose
\beq
M_2=1.8 \times 10^{12} M_\odot \, ,
\eeq
and the reference redshift is $z_2=0$. We use these parameters as reference for all plots in this section.

We show the standard CDM prediction (black solid line), together with the WDM-model already used above (gray dash-dotted line) and three bolon models with modified barrier (dotted) and without (dashed). The three colors represent different couplings of $\beta=-0.1$ (red), $\beta=0$ (green) and $\beta=0.05$ (blue). The bolon exponent $\lambda$ is 65 for all curves, while $\alpha=20$. We used a tophat filter for all the calculations.

\subsection{Current number of subhalos}

As a last step, we now calculate the number of subhalos we expect in a typical galaxy. This is not a quantity directly accessible through the ePS-approach and we have to do some additional work. Following ref. \cite{Giocoli:2007gf}, we calculate the current number of subhalos by a simple integration in barrier space, i.e.
\beq
\frac{\d n}{\d M_1} = \int_{\delta_0}^{\infty} \frac{M_2}{M_1} f(\omega(\delta_{sc,1});S_1|\omega(\delta_{sc,2});S_2) \d \delta_{sc,1} \, ,
\eeq
where $\delta_0$ denotes the spherical collapse barrier at redshift 0.
Afterwards we calculate the cumulative number of subhalos through a simple integration in $M_1$:
\beq
n(>M_1) = \int_{M_1}^{M_2} \frac{\d n}{\d M_1} \d M_1 \, .
\eeq
A few comments are in order at this point. First, the integration over barrier-space necessarily overcounts halos, as one halo might retain its mass over a prolonged period of time. Therefore we lose the overall normalization of the cumulative number density. We solve this issue by adjusting the CDM result to N-body simulations and use the same normalization for WDM and the bolon model. Second, one should note that an additional assumption going into this ansatz is that the current distribution of subhalos corresponds directly to the distribution of progenitor halos when averaged over redshift. This is far from obvious, but seems to hold, as we recover the CDM N-body results to good accuracy (see below). 

\begin{figure}[t]
	\centering
  \includegraphics[width=1.0\linewidth]{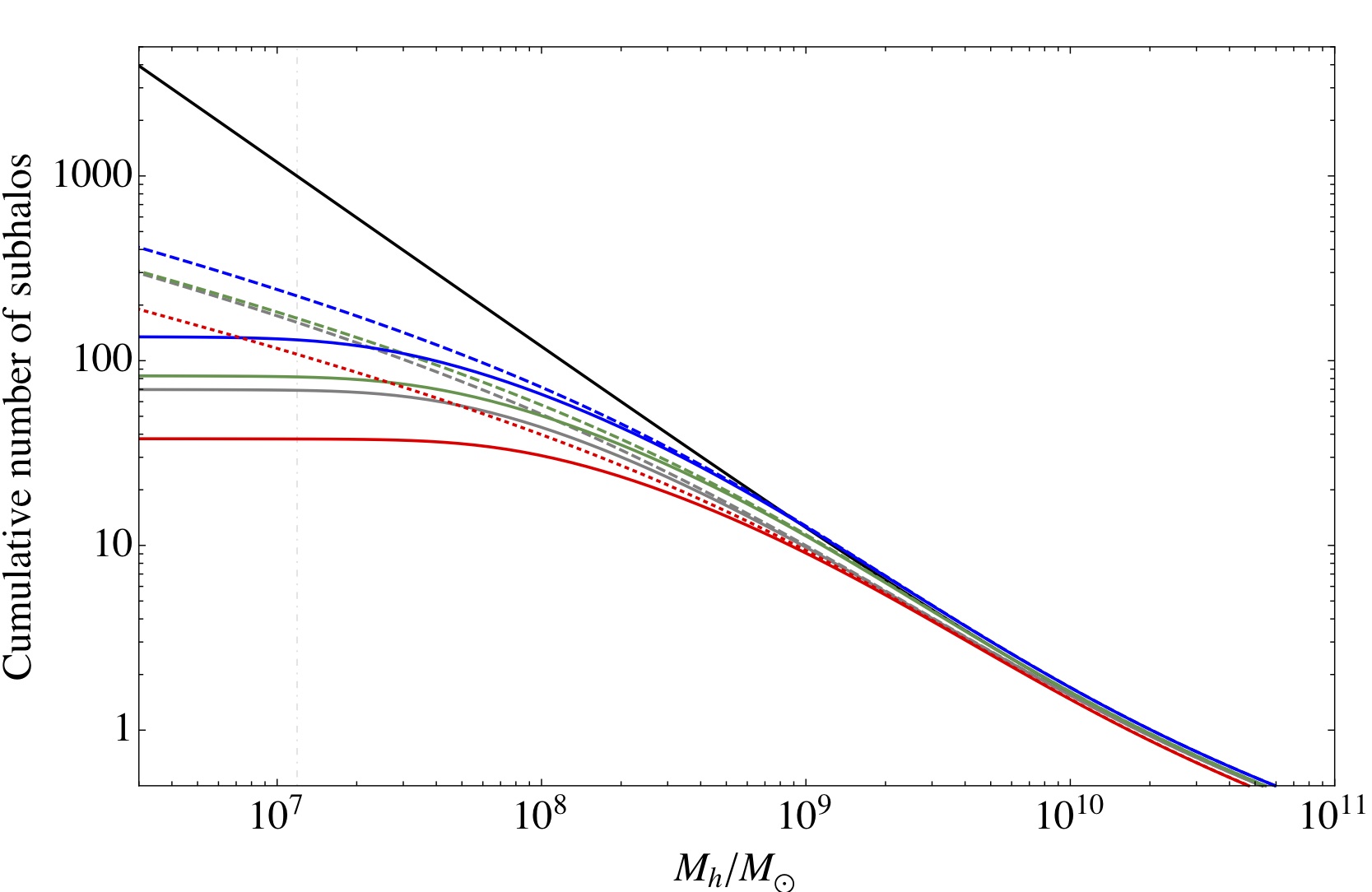}
	\caption{Cumulative number of Milky Way subhaloes as a function of halo mass $M_h$. The CDM power spectrum is represented by a solid black line, the gray lines show a WDM modification for a thermally produced WDM particle of mass $m_{\rm wdm} = 2.284$ keV. The green lines stands for a bolon power spectrum with $\beta=0$ and $\lambda=69$, the blue and red lines correspond to the same $\lambda$ but with $\beta=0.05$ and $\beta=-0.1$ respectively. For all WDM and bolon models, the dashed lines are results calculated without an upturn of the barrier near the Jeans mass, whereas such a modification is included for the solid lines. As an additional orientation we added the dashed-dotted gridline at the WDM Jeans mass.}
	\label{fig:cumNums}
\end{figure}

Results for cumulative number counts of subhalos can be seen in Figure \ref{fig:cumNums}. A comparison with the N-body results in Figure 11 in ref. \cite{Lovell:2013ola} shows that the CDM result agrees to excellent accuracy (as expected, as we have adjusted our normalization accordingly), whereas the WDM curve looks somewhat different. It seems to agree rather well with the N-body results down to masses slightly above the Jeans mass, but then the ePS results stagnate more quickly, whereas the N-body results continue to rise a while longer. We underestimate the asymptotic value by a factor of about 2. Two reasons for this come to mind. First, structure formation is not strictly hierarchical, there are violent mergers and disruptions which can form halos even below the Jeans mass. Such effects can not be properly included in the purely hierarchical ePS approach. Second, this is also the regime where spurious halos start to play a role, and issues with the identification of such halos could introduce additional uncertainties.

In the light of recent observations of ultra-faint dwarf-galaxies \cite{2009ApJS..182..543A}, these results can be used to put constraints on the current bolon mass $m_\chi (t_0)$. We simply demand that the number of subhalos should not fall below the number of dwarf galaxies estimated from observations. These estimates are still the subject ob ongoing debate, current numbers range from 66 \cite{Lovell:2013ola} to several hundred \cite{Tollerud:2008ze}. Here we choose the lower value of 66 in order to remain cautious. One should point out that additional uncertainties get introduced by baryonic physics. Galaxy formation on such small scales appears to be a highly stochastic process \cite{Strigari:2008ib}, potentially leaving a large number of halos void of stars. This effect may contribute to raising the number of dark halos required to explain current observations and the bounds we set here are therefore very conservative. 

\begin{figure}[t]
	\centering
  \includegraphics[width=0.92\linewidth]{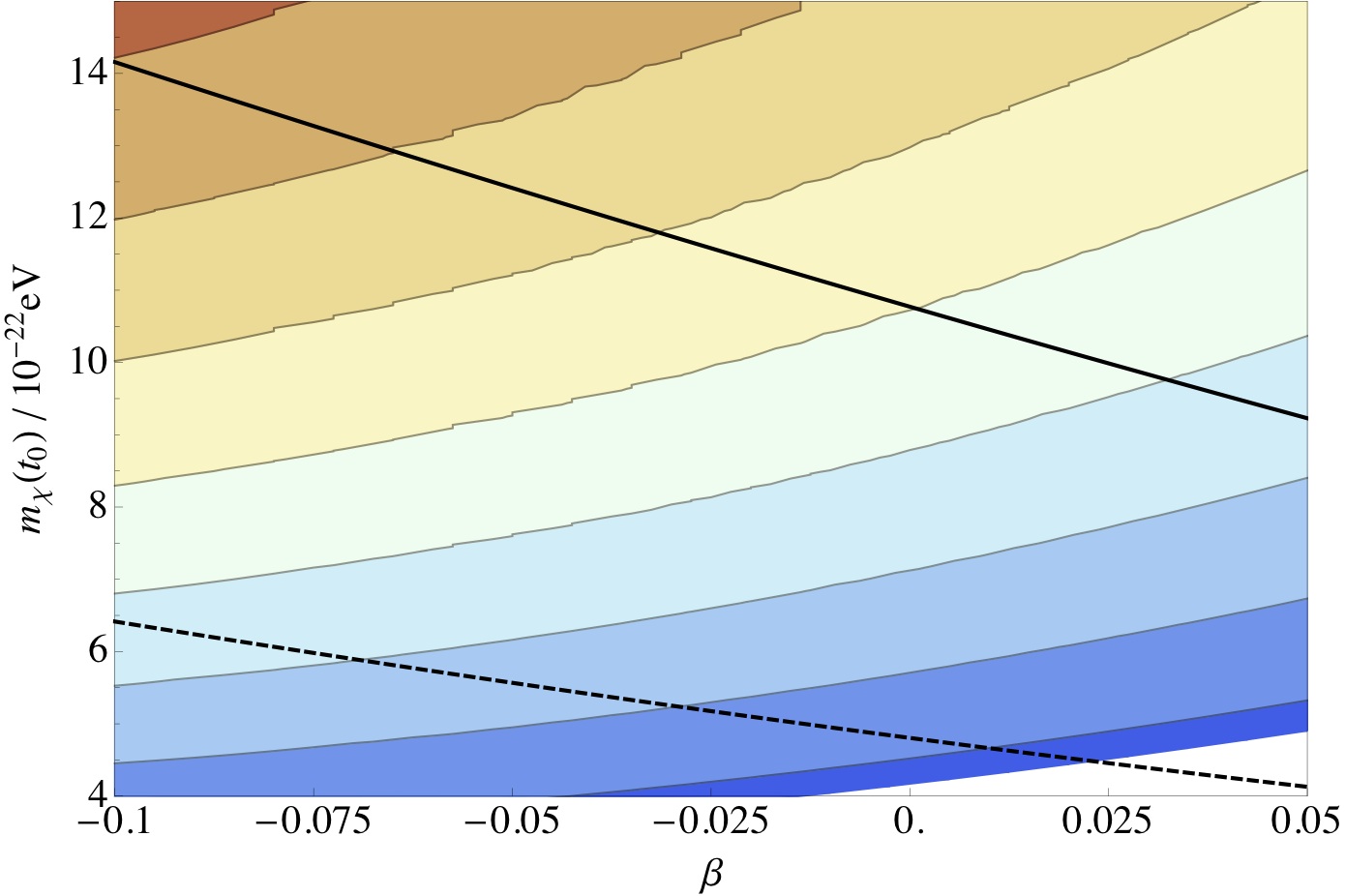}
  \raisebox{.5\height}{\includegraphics[width=0.06\linewidth]{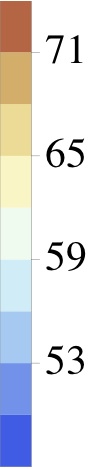}}
	\caption{Allowed parameter range for the cosmon-bolon model. The colored contours show different bolon exponents $\lambda$. The solid line displays the lower boundary of parameters which yield more than 66 subhalos in the Milky way if the modified barrier is used, the dashed line shows the same exclusion curve for the standard elliptical barrier.}
	\label{fig:lbplot}
\end{figure}

The masses of the dark matter halos containing ultra-faint dwarf-galaxies appear to have a common lower mass bound at around $ 10^7 M_\odot$, which is the benchmark mass we employ. As we already underestimate the WDM N-body results already by a factor of roughly $1.5$ at this mass, we artificially remedy this fact by raising our obtained number counts by this factor when we use the modified spherical collapse barrier in order to remain extra cautious. The results can be seen in Figure \ref{fig:lbplot}.

Clearly the smallest current bolon masses $m_\chi (t_0)$ are possible for the largest couplings $\beta$. However, we expect current coupling constraints from CMB observations to apply to our model as well, as the wavenumbers for which the CMB has the most constraining power are much lower than the ones where linear structure formation is modified compared to CDM. In refs. \cite{Pettorino:2012ts,Pettorino:2013oxa} the couplings bounds are roughly $|\beta| <  0.1$. From this constraint we derive an upper bound for current bolon mass, which we estimate by evaluating the boundaries presented in Figure \ref{fig:lbplot} for large couplings, we choose $\beta=0.05$. The resulting bound is
\beq
m_{\chi}(t_0) \gtrsim 9.2 \, (4.1) \times 10^{-22} {\rm eV}
\eeq
for the modified (ellitpical) barrier. 
This bound lies at the larger end of typical ultra-light scalar field dark matter masses.

\section{Conclusion and Outlook}
\label{sec:Conclusion}
In this work we have studied the evolution of linear-perturbations in the coupled cosmon-bolon model in some detail. We built our analysis on the study of linear perturbations in the very early universe in coupled two scalar field models, which we provide in an accompanying paper \cite{EarlyScalings}. Our work has given a detailed procedure to average out the quick oscillations at both the background and the perturbative level to arrive at an effective theory of the evolution of linear perturbations, which we treated numerically. As a result we provide a reasonably accurate fitting formula for the power-spectrum modification when compared to standard CDM coupled to quintessence. 

We then move on to investigate phenomenologically interesting predictions of our model by employing the Press-Schechter excursion set approach. We have discussed in some detail the different approaches one can take here, in particular with respect to the choice of the filter and the barrier employed, both of which can have influential consequences on the results. We have shown how the coupling influences the abundance of progenitor halos in a typical galaxy such as ours and how to translate this into a prediction for the number of virialized dark matter subhalos expected to be present today. 

Our analysis does however have one shortcoming: As we have discussed, the spherical collapse model, which serves as a basis for the entire Press-Schechter-excursion approach, can not be easily generalized to our model. This problem can only be addressed by studying non-linear perturbations in our model, a work which is currently in progress. We hope to generalize the mechanism presented here to average out the linear perturbations to the non-linear regime and arrive at an effective theory which could be used not only to study spherical collapse in out model, but also serve as a basis for large numerical simulations of cosmological structure formation in coupled scalar field dark matter models. 

Pending the results of this work, we still want to make one last comment concerning the small scale problem of the cosmological standard model. Traditionally this problem has been divided into two parts, the \textit{missing satellite problem} and the \textit{cusp-core problem}. While the missing satellite problem may well have a solution in the baryonic sector alone (see e.g. ref. \cite{Weinberg:2013aya} and references therein), the cusp-core problem (and the - possibly strongly related - \textit{too big to fail problem}) might still require some modification of the dark matter sector. As has been recently pointed out, warm dark matter, at least in the simplest version of being a single thermally produced component, can not resolve these issues consistently \cite{Schneider:2013wwa}. The problem here is essentially that existing constraints on WDM particle masses from Lyman-$\alpha$ constraints \cite{Viel:2013fqw} and ultra-faint dwarf galaxies \cite{Polisensky:2010rw,Polisensky:2013ppa} lead to an allowed range where the core structure of both host galaxies (the cusp-core problem) and of massive subgalaxies (the too big to fail problem) can not be explained anymore. As the WDM model appears to somewhat similar to our model in the linear regime one might be tempted to draw similar conclusions here. This would however be premature. All the constraints derived for WDM models rely on simulations of non-linear structure growth, and, as mentioned above, the non-linear dynamics of our model are still under investigation and might turn out to be somewhat different from WDM. Furthermore the presence of a coupling enlarges the parameter space and therefore could point towards a way out of this possible dilemma.

Furthermore, despite the similarities between our model and WDM models, there are observable differences. In particular, the oscillations present in the scalar field sector are expected to translate to the gravitational potential in virialized structures, similar to what happens in oscillatons or boson stars \cite{Lee:1995af,UrenaLopez:2001tw}. These oscillations are in principle observable. In a recent paper Khmelnitsky and Rubakov investigated the effects scalar field dark matter wold have on variations in pulsar timing signals and whether or not the resulting signatures are detectable in the near future \cite{Khmelnitsky:2013lxt}. The bounds we set on the current bolon mass exclude a possible near future detection by more than one order of magnitude, as can be seen by a comparison with Figure 1 in ref. \cite{Khmelnitsky:2013lxt}.

Finally, even if the small scale problems of the cosmological standard model should find a full solution in the baryonic sector, it certainly does not invalidate the scalar field dark matter model. It has a strong theoretical motivation in its connection to a possible solution of the cosmological constant problem and therefore remains an interesting alternative.

\acknowledgments{The author would like to thank Valery Rubakov for useful discussions. This work is supported by the grant ERC-AdG-290623.}

\appendix

\section{Averaging the bolon perturbations}

\label{app:bgExpansion}

In this section we present the full averaging procedure for the bolon oscillations at both the background level and at the level of linear perturbations. We start with the background.

\subsection{Background}

As already mentioned in the main text, we expand all dynamical quantities in a Taylor series according to equation (\ref{MuExpansion}), and each coefficient in a Fourier-type expansion as presented in equation (\ref{FourierTypeExpansion}). Comparing coefficients at each order in $\tilde{\mu}$ yields the following decomposition for the scalar fields:
\begin{align}
\label{BolonDecomposition}
\chi &= \chi_0 {\rm cos}(x) + \chi_1 {\rm sin}(x) + \chi_2 {\rm cos} (3x) + \chi_{\rm ig} \, , \\
\label{CosmonDecomposition}
\varphi &= \bar{\varphi} + \varphi_1 {\rm cos}(2x) + \varphi_2 {\rm sin}(2x) + \varphi_{\rm ig} \, ,
\end{align}
and consistency requires that $\chi_1$ is of the order $\mathcal{O}(\tilde{\mu} \chi_0)$, $\chi_2$ is of the order $\mathcal{O}(\tilde{\mu}^2 \chi_0)$, $\varphi_1$ is of the order $\mathcal{O}(\tilde{\mu}^2 \bar{\varphi})$ and $\varphi_2$ is of the order $\mathcal{O}(\tilde{\mu}^3 \bar{\varphi})$. The leading order contributions are formally related by setting $\chi_0$ to be of the order $\mathcal{O}(\tilde{\mu} \bar{\varphi})$, so that the energy densities for both scalar fields are of the same order, as is appropriate for a scaling scenario. The terms $\chi_{\rm ig}$ and $\varphi_{\rm ig}$ represent higher order contributions, they will not play a role in our analysis. 

Here we used a notation already introduced in the main text: For all quantities which are slowly evolving to leading order we will mark the leading order quantity with a bar (like $\bar{\varphi}$ above). This will turn out to be helpful in the final equations, where we will also encounter quantities averaged over one oscillation period, which we will indicate using triangular brackets: $< . >$. This may seem unnecessary, since of course $\bar{\varphi} = <\varphi>$ holds, but we find it useful to make this distinction in order to indicate quantities which evolve adiabatically at leading order.

The oscillatory terms present in (\ref{BolonDecomposition}) and (\ref{CosmonDecomposition}) do of course leave an imprint on the scale factor and we can deduce from Einsteins equations that the expansion for $a(\eta)$ reads
\beq
a = \bar{a} + a_{\rm osc} + a_{\rm ig} \, ,
\eeq
where $a_{\rm osc}$ is of the order $\mathcal{O}(\tilde{\mu}^2 \bar{a})$ and $a_{\rm ig}$ represents higher order contributions which will not concern us. For the conformal Hubble rate this directly implies
\beq
h = \bar{h} + h_{\rm osc} + h_{\rm ig} \, , \quad {\rm where}  \quad h_{\rm osc} = \frac{a_{\rm osc}'}{\bar{a}} \, .
\eeq
Note that taking the derivative of an oscillatory quantity decreases the order in $\tilde{\mu}$ by one and therefore $h_{\rm osc}$ is of order $\mathcal{O}(\tilde{\mu} \bar{h})$ ($h_{\rm ig}$ again stands for higher order terms).
In terms of the dimensionless coupling parameter $q_{\alpha}$ the expansion directly gives
\begin{align}
\label{BackgroundCoupling}
(1&+\omega_\chi) q_\chi = -\beta  \frac{\bar{\varphi}'}{3 \bar{h} M} \left( 1 + {\rm cos}(2x) \right) \, 
\end{align}
to leading order. Averaging this expression over one oscillation period gives:
\beq
\label{avBGCoupling}
<(1+\omega_\chi) q_\chi> = -\beta  \frac{\bar{\varphi}'}{3 \bar{h} M} \, .
\eeq
Plugging the ansatz into Friedmann's equation gives to leading order:
\beq
\bar{h}^2 = \frac{\bar{a}^2}{3M^2} \left( \bar{\rho}_\chi + \bar{\rho}_\varphi + \bar{\rho}_{\rm ext} \right) \, ,
\eeq
where $\bar{\rho}_\chi$ and $\bar{\rho}_\varphi$ denote the (non-oscillatory) leading order contributions to the bolon- and cosmon energy-densities, respectively. These are given by
\begin{align}
\bar{\rho}_\chi &= \frac{1}{2} \bar{m}_\chi^2(\bar{\varphi}) \chi_0^2 \, , \\
\bar{\rho}_\varphi &= V_1(\bar{\varphi}) + \frac{1}{2 \bar{a}^2} \bar{\varphi}'^2  \, ,
\end{align}
where $\bar{m}_\chi(\bar{\varphi})$ simply denotes the leading order term in the $\tilde{\mu}$-expansion for the mass, which is of course only $\bar{\varphi}$-dependent:
\beq
\bar{m}_\chi(\bar{\varphi}) = m_0 {\rm e}^{-\beta \bar{\varphi}/M} \, .
\eeq
We will drop the explicit $\bar{\varphi}$-dependence of $\bar{m}$ in all the formulas below. The additional quantity $\bar{\rho}_{\rm ext}$ labels the sum of all energy densities which are present in addition to the scalar fields, in particular photons, neutrinos and baryons. We assume all these quantities to evolve adiabatically to subleading order on a timescale $1/\bar{m}_\chi$. This is consistent with the equations of energy conservation for all standard cosmological components to the order which we consider here.

Evaluating the scalar field equations to leading order yields:
\begin{align}
\label{dchi0Equation}
\chi_0' &= -\frac{3}{2} \bar{h} \chi_0 + \frac{\beta}{2} \frac{\chi_0}{M} \bar{\varphi}' \, , \\
\varphi_1 &= - \frac{\beta}{8} \frac{\chi_0^2}{M} \, , \\
\bar{\varphi}'' &= -2 \bar{h} \bar{\varphi}' - \bar{a}^2 V_{1,\bar{\varphi}} + \frac{\beta}{M} \bar{a}^2 \bar{\rho}_\chi \, .
\end{align} 
From eq. (\ref{dchi0Equation}) we can directly see that ${\rm ln}(\chi_0) \propto {\rm ln}(\bar{a}^{-3/2}) + \frac{\beta}{2} \frac{\bar{\varphi}}{M}$, i.e. $\chi_0 \propto \bar{a}^{-3/2} {\rm exp}(\beta \bar{\varphi} / 2M) $ and therefore
\beq
\bar{\rho}_\chi \propto \bar{a}^{-3} {\rm exp} (-\beta \bar{\varphi}/M) \, .
\eeq
Note that this is prefectly consistent with equation (\ref{avBGCoupling}).

In order to average the perturbations we need the next to leading order equations. Using the above results from Einsteins equations
\begin{align}
h_{\rm osc} &= \frac{1}{8} \bar{a} \bar{m}\frac{\chi_0^2}{M^2} {\rm sin}(2x) \, ,\\
a_{\rm osc} &= -\frac{1}{16} \bar{a}  \frac{\chi_0^2}{M^2} {\rm cos}(2x) \, ,
\end{align}
and the field equation for the bolon gives at this order
\begin{align}
\chi_2 =& \frac{3}{128} \frac{\chi_0^3}{M^2} \left( 1 - \frac{2}{3} \beta^2 \right) \, , \\
\chi_1' =& - \left( \frac{3}{2} \bar{h} - \frac{\beta}{2} \frac{\bar{\varphi}'}{M}\right)\chi_1- \frac{1}{2\bar{m} \bar{a}} \left( \chi_0'' + 2 \bar{h} \chi_0' \right) \nonumber \\
&+ \frac{3-2\beta^2}{32 M^2} \bar{m} \bar{a} \chi_0^3 \, ,
\end{align}
while the cosmon field equation gives
\beq
\varphi_2 = - \frac{\chi_0}{16 \bar{m} \bar{a} M } \left( 4 \beta \bar{m} \bar{a} \chi_1- 2 \beta \bar{h} \chi_0 - \chi_0 \bar{\varphi}' /M \right) \, .
\eeq

\subsection{Linear Perturbations}

We will now extend the procedure given above to the linear perturbations.
As already mentioned in the main text, the problem that arises when treating perturbations in this model lies in the fact that quick oscillations are present in both the background and the perturbative fields, and it is a priori unclear how they interfere with each other. Resolving these oscillations numerically is computationally very demanding and it is therefore desirable to find a way to analyze the long term behavior of the amplitudes, averaged over one oscillation period, in particular if one has future parameter constraints from MonteCarlo runs or Likelyhood-analyses in mind. Some of the approaches present in the literature deal with such an averaging (e.g. ref. \cite{Matos:2000ss}), but they treat only a single scalar field in a harmonic potential and the procedures they apply, while yielding correct results, do not lend themselves well to a generalization to two coupled scalar fields. We will therefore use the same procedure we just applied to the background evolution, i.e. we expand all dynamical perturbative quantities first in a Taylor-series in $\tilde{\mu}$ and then each coefficient in a Fourier-type sum of harmonic functions with time-dependent frequencies, where all occurring frequencies are multiples of a base-frequency (see equations (\ref{MuExpansion}) and (\ref{FourierTypeExpansion})). By plugging the results into the linearized field equations (\ref{CosmonPEquation}) - (\ref{PsiEquation}) and comparing coefficients we see which frequencies are present at which order and can then use the results to calculate the evolution equations for the energy density and momentum perturbations for the bolon averaged over one oscillation period. 

As mentioned in the main text, our starting point are the exact gauge-invariant linearly perturbed scalar field equations (our conventions concerning perturbation theory can be found in ref. \cite{EarlyScalings})
\begin{align}
\label{CosmonPEquationA}
X'' + 2hX' + k^2 X + a^2 V_{,\varphi \varphi} X + a^2 V_{,\varphi \chi} Y \nonumber \\
+ 2 a^2 V_{,\varphi} \Phi - \varphi' \Phi' - 3 \varphi' \Psi' & = 0 \, , \\
\label{BolonPEquationA}
Y'' + 2hY' + k^2 Y + a^2 V_{,\varphi \chi} X + a^2 V_{,\chi \chi} Y \nonumber \\
+ 2 a^2 V_{,\chi} \Phi - \chi' \Phi' - 3\chi' \Psi' & = 0 \, ,
\end{align}
and the Poisson-equation
\begin{align}
\label{PsiEquationA}
k^2 \Psi &= -\frac{a^2}{2M^2} \sum_\alpha \left( \delta \rho_\alpha - 3 h \right[(\rho + p)v\left]_\alpha \right) \, .
\end{align}
The two gravitational potentials can be related via 
\beq
\label{phiPsiRelation}
\Phi = \Psi - a^2 \Pi_{\rm tot}/M^2 \, .
\eeq
Even though scalar fields do not produce anisotropic stress, we differentiate between $\Psi$ and $\Phi$, since anisotropic stress can be introduced through neutrino- and photon-contributions. The Boltzmann equations for those components as well as the fluid equations for baryons do of course have to be added to the equations derived here in order to complete the set of equations.

The remaining Einstein equations are not necessary to analyze the evolution of linear perturbations, but can be used to provide trivial checks of the calculations. We will use equation for the derivative of the gravitational potential several times, as it simplifies some steps. Using equation (\ref{phiPsiRelation}) it reads:
\beq
\label{dPsiEquation2}
\Psi'+h\Psi = -\frac{a^2}{2 M^2} \sum_\alpha \left[ (\rho + p) v \right]_\alpha + a^2 \Pi_{\rm tot} / M^2 \, .
\eeq

As we will see below, the combined Taylor-Fourier expansion looks different depending on the size of the wavenumber $k$ and we have to split up our analysis into two regimes: Large wavelength perturbations for which $\mu k^2/h^2 \ll 1$ and small wavelength perturbations which have $\mu k^2/h^2 \gtrsim 1$. 

\subsubsection{$\mu k^2/h^2 \ll 1$}
For large wavelengths the only consistent expansion in terms of trigonometric functions is given by
\begin{align}
X &= P + Q \, {\rm cos}(2x) + R \, {\rm sin}(2x)  + X_{\rm ig} \, , \\
Y &= A\, {\rm cos}(x) + B\, {\rm sin}(x) + C\, {\rm sin}(3x) + Y_{\rm ig} \, , \\
\Psi &= \bar{\Psi} + \Psi_{\rm osc} + \Psi_{\rm ig} \, ,
\end{align}
where $x$ is defined as in section \ref{sec:LateUniverse} 
\beq
x(\eta) \equiv \int_{t_0(\eta_0)}^{t(\eta)} m(\varphi(t')) dt'
\eeq
and the linearized field equations require $Q$ to be of the order $\mathcal{O}(\tilde{\mu}^2 P)$, $R$ to be of order $\mathcal{O}(\tilde{\mu} P)$, $A$ to be of order $\mathcal{O}(\tilde{\mu} B)$, $C$ to be of order $\mathcal{O}(\tilde{\mu}^2 B)$, $\Psi_{\rm osc}$ to be of order $\mathcal{O}(\tilde{\mu} \bar{\Psi})$ and finally $P$ to be of order $\mathcal{O}(B)$ and $\bar{\Psi}$ to be of order $\mathcal{O}(B/M)$. The terms $X_{\rm ig}$, $Y_{\rm ig}$ and $\Psi_{\rm ig}$ again represent higher order contributions which we ignore.
We always assume that all additional energy densities, i.e. the ones for photons, neutrinos and baryons evolve slowly to leading order in $\tilde{\mu}$ on a timescale $1/\bar{m}$, the corresponding velocity potentials as well as the higher momenta in the multipole-expansion of the Boltzmann-equations for photons and neutrinos even to subleading order. This assumption directly implies that $\Phi$ can be expanded just as $\Psi$
\begin{align}
\Phi &= \bar{\Phi} + \Phi_{\rm osc} + \Phi_{\rm ig} \, ,
\end{align}
and the terms are related by
\begin{align}
\label{PhiPsiRelation}
\bar{\Phi} &= \bar{\Psi} - \bar{a}^2 \bar{\Pi}_{\rm tot} / M^2 \quad {\rm and} \quad \Phi_{\rm osc} = \Psi_{\rm osc} \, .
\end{align}
We emphasize that this is consistent with all the evolution equations for non-scalar field quantities, since the quick oscillations only enter these indirectly through the background and the gravitational potentials. Furthermore it is also justified for $k \ll \bar{m}_\chi$ because the only additional energy-scales present in the equations are the Hubble parameter and the wavenumber, both of which are much smaller than the bolon mass. The fact that this reasoning breaks down eventually does not concern us, since we do not consider such large wavenumbers.

Evaluating the scalar field equations (\ref{CosmonPEquationA}) and (\ref{BolonPEquationA}) to leading order gives
\begin{align}
B' =& \bar{a} \bar{m} \chi_0 (\beta P/M - \bar{\Phi})- \frac{1}{2} B (3 \bar{h} - \beta \bar{\varphi}'/M) \, , \\
R =& -\frac{\beta}{4} B \chi_0 / M \, ,
\end{align}
while the Poisson equation (\ref{PsiEquationA}) yields
\begin{align}
\bar{\Psi} =& \frac{-\bar{a}^2}{2 k^2 M^2 - \bar{\varphi}'^2} \left( \overline{\delta \rho}_{\rm ext} - 3 \bar{h} \overline{ \left[ (\rho + p)v \right] }_{\rm ext} \right. \nonumber \\
&+ \bar{m}^2 A \chi_0 - \frac{3}{2} \bar{m} \frac{\bar{h}}{\bar{a}}  B \chi_0 + \bar{m}^2 \chi_1 B + V_{1,\bar{\varphi}} P \nonumber \\
&\left.- \beta \bar{m}^2 \chi_0^2 \frac{P}{M}
+ 3 \frac{\bar{h} \bar{\varphi}'}{\bar{a}^2} P + \frac{\bar{\varphi}' P'}{\bar{a}^2} + \bar{\varphi}'^2 \frac{\bar{\Pi}_{\rm tot}}{M^2} \right) \, ,
\end{align}
where $\overline{\delta \rho}_{\rm ext}$ and $\overline{\left[ (\rho + p)v \right]}_{\rm ext}$ denote the summed up energy- and momentum density perturbations of all additional components of the cosmic fluid. 

Unfortunately the leading order results are not sufficient (as we will see below), and we have to go to subleading order. Doing this we obtain from the Poisson equation 
\begin{align}
\label{PsiOsc1}
\Psi_{\rm osc} &= \frac{1}{8} \frac{\chi_0 B}{M^2} {\rm sin} (2x) \, .
\end{align}
Note that this expression is most easily obtained by evaluating equation (\ref{dPsiEquation2}) to leading order, instead of going to next to leading order in the Poisson equation. Both ways are of course equivalent, and we did both calculations in order to verify our results.

The bolon field equation yields at this order
\begin{align}
A' =& -\frac{3}{2} \bar{h} A - \frac{3}{32} \bar{a} \bar{m} \frac{\chi_0^2}{M^2} B + \bar{a} \bar{m} \chi_1 \bar{\Phi} \nonumber \\
&+ \frac{\bar{h}}{\bar{a} \bar{m}} B' + \frac{1}{2 \bar{m} \bar{a}} B'' + \frac{k^2}{2 \bar{a} \bar{m}} B \nonumber \\
&+ \frac{\beta^2}{16} \bar{a} \bar{m} \frac{\chi_0^2}{M^2} B - \beta \bar{a} \bar{m} \frac{\chi_1}{M} P + \frac{\beta}{2} \frac{\bar{\varphi}' A}{M}  \, , \\
C =& \frac{13 + 2 \beta^2}{128} \frac{\chi_0^2}{M^2}B \, ,
\end{align}
whereas the cosmon field equation gives
\begin{align}
Q =& -\frac{\chi_0 \bar{\varphi}' }{4 \bar{a} \bar{m} M^2} \left( 1+\frac{\beta^2}{2} \right) B - \frac{\beta}{4} \left( \frac{\chi_0}{M} A - \frac{\chi_1}{M} B \right) \nonumber \\
&  + \frac{3 \beta}{8} \frac{\bar{h}}{\bar{a} \bar{m}} \frac{\chi_0}{M} B \, , \\
P'' =& - 2 \bar{h} P'  -k^2 P - \bar{a}^2 V_{1,\bar{\varphi} \bar{\varphi}} P - 2 \bar{a}^2 V_{1,\bar{\varphi}} \bar{\Phi} \nonumber \\
& + \bar{\varphi}' \bar{\Phi}' + 3 \bar{\varphi}' \bar{\Psi}'  + \beta \bar{a}^2 \bar{m}^2 \left( \frac{\chi_0}{M} A + \frac{\chi_1}{M} B \right) \nonumber \\
&- \beta^2 \bar{a}^2 \bar{m}^2 \frac{\chi_0^2}{M^2} P + \beta \bar{a}^2 \bar{m}^2 \frac{\chi_0^2}{M} \bar{\Phi} \, .
\end{align}

\vspace{10 pt}

Now we can use the equations given above to average the equations of energy- and momentum-conservation for the bolon. We could obviously do the same for the cosmon, but it is more common in coupled scenarios to treat the cosmon field perturbations directly and we will adopt the same approach here. The field perturbations for the bolon can be mapped directly onto a fluid description with an energy density perturbation, a pressure perturbation and a momentum perturbation. The issue here is, of course, whether or not we can simply average these quantities and recover the standard evolution equations (with averaged background quantities $\bar{h}$ and $\bar{a}$), or if some unexpected interference between oscillatory terms at the background and the perturbative level modifies the results. As it will turn out they don't to subleading order, but this is not obvious since the oscillations present in background quantities, even if they are subleading contributions, can make a difference in conjunction with perturbative quantities which are oscillatory at leading order. We will perform the procedure once in detail. 

We start by calculating the averaged perturbative fluid quantities for the bolon to subleading order:
\begin{align}
&<\delta \rho_\chi> = \bar{m}^2 \left( \chi _0 A + \chi_1 B \right) - 2 \beta \bar{\rho}_\chi \frac{P}{M} \, , \\ 
&<\delta p_\chi > = 0 \, , \\
&<\left[ (\rho + p)v \right]_\chi> = \frac{\bar{m} \chi _0}{2 \bar{a}} B \, .
\end{align}
Note that all these quantities have oscillatory contributions at leading order. Now we average over the equation of energy conservation for the bolon term by term to obtain:
\begin{align}
&<3 h \delta \rho_\chi> = 3 \bar{h} <\delta \rho_\chi> \, , \\
&<k^2 \left[ (\rho + p)v \right]_\chi> = k^2 <\left[ (\rho + p)v \right]_\chi> \, , \\
&<3h\delta p_\chi> = -\frac{3}{16} \bar{a} \bar{m}^3 \frac{\chi_0^3}{M^2} B \, , \\
&<3(1+\omega_\chi) \rho_\chi \Psi'> = 3 \bar{\rho}_\chi \bar{\Psi}' -\frac{3}{16} \bar{a} \bar{m}^3 \frac{\chi_0^3}{M^2} B \, , \\
&<3 h(1+\omega_\chi) \rho_\chi q_\chi \Phi > =  -\beta \bar{\rho}_\chi \frac{\bar{\varphi}'}{M} \bar{\Phi} \, , \\
&<3 h(1+\omega_\chi) \rho_\chi q_\chi \tau_\chi > = - \beta \frac{\bar{\varphi}'}{M} <\delta \rho_\chi> \nonumber \\
& \hspace{110 pt} - \beta \frac{\bar{\rho}_\chi}{M}\left(P' - \bar{\varphi}' \bar{\Psi} \right) \, .
\end{align}
Note that the "interference" terms arising during the averaging procedure in lines 3 and 4 of the above set of equations precisely cancel each other. 

Since
\beq
<\delta \rho_\chi'> = <\delta \rho_\chi>' 
\eeq
is obvious to all orders, we can directly write down the equations of energy conservation in terms of the density contrast $\delta_\chi$ and velocity potential $\bar{\Theta}_\chi$ as defined in the main text:
\begin{align}
<\delta_\chi>' =& -\bar{\Theta}_\chi + 3 \bar{\Psi}' - \beta P'/M + \beta \frac{\bar{\varphi}'}{M} \left( \bar{\Psi} - \bar{\Phi} \right) \, .
\end{align}

Moving on to the equation of momentum conservation for the bolon we get to subleading order:
\begin{align}
&<4 h \left[ (\rho + p)v \right]_\chi> = 4 \bar{h} <\left[ (\rho + p)v \right]_\chi> \, , \\
&<\delta p_\chi> = 0 \, , \\
&<(1+\omega_\chi) \rho_\chi \Phi> = \bar{\rho}_\chi \bar{\Phi} \, , \\
&<3 h q_\chi \frac{(\rho_\chi + p_\chi)}{(\rho_{\rm tot}+p_{\rm tot})} \left[ (\rho + p)v \right]_{\rm tot} + a f_\chi> = \beta \bar{\rho}_\chi P/M \, . 
\end{align}
Since again
\beq
<\left[(\rho+p)v\right]_\chi'> = < \left[(\rho+p)v\right]_\chi >'
\eeq
is obvious to all orders we can easily obtain the following equation for $\bar{\Theta}_\chi$:
\beq
\bar{\Theta}_\chi' = - \bar{h} \bar{\Theta}_\chi + k^2 \bar{\Phi} + \beta \frac{\bar{\varphi}'}{M} \bar{\Theta}_\chi - \beta k^2 P/M \, .
\eeq
Rewriting the cosmon field equation and the Poission equation in terms of the averaged bolon density contrast and velocity potential yields:
\begin{align}
P'' =& - 2 \bar{h} P'  - \left( k^2 + \bar{a}^2 V_{1,\bar{\varphi} \bar{\varphi}} \right) P \nonumber  - 2 \bar{a}^2 V_{1,\bar{\varphi}} \bar{\Phi} \nonumber \\
& + \bar{\varphi}' \left( \bar{\Phi}' + 3 \bar{\Psi}' \right)
+ \frac{\beta \bar{a}^2}{M} \bar{\rho}_\chi \left( <\delta_\chi> +2 \bar{\Phi} \right) \, ,
\end{align}
and
\begin{align}
\label{poissonEquationA}
\bar{\Psi} =\frac{-\bar{a}^2}{2 M^2 k^2 - \bar{\varphi}'^2} & \left[\bar{\rho}_{\rm ext}  \left( \bar{\delta }_{\rm ext} + \frac{3 \bar{h}}{k^2} (1+\bar{\omega}_{\rm ext}) \bar{\Theta}_{\rm ext} \right) \right. \nonumber \\
&+ \bar{\rho}_\chi \left( <\delta_\chi> + \frac{3 \bar{h}}{k^2} \bar{\Theta}_\chi \right) + \bar{\varphi}'^2 \frac{\bar{\Pi}_{\rm tot}}{M^2}\nonumber \\
&\left. + V_{1,\bar{\varphi}} P + 3 \bar{h} \frac{\bar{\varphi}' P}{\bar{a}^2
} + \frac{\bar{\varphi}' P'}{\bar{a}^2}  \right] \, .
\end{align}

\subsubsection{$\mu k^2/h^2 \gtrsim 1$}

In the regime of smaller wavelengths the general decomposition of the scalar fields yields a different result. The expansion in terms of trigonometric functions remains the same, i.e.
\begin{align}
X &= P + Q \, {\rm cos}(2x) + R \, {\rm sin}(2x)  + X_{\rm ig} \, , \\
Y &= A\, {\rm cos}(x) + B\, {\rm sin}(x) +C \, {\rm cos}(3x)+ D \, {\rm sin}(3x) + Y_{\rm ig} \, , \\
\Psi &= \bar{\Psi} + \Psi_{\rm osc} + \Psi_{\rm ig} \, , \\
\Phi &= \bar{\Phi} + \Phi_{\rm osc} + \Phi_{\rm ig} \, ,
\end{align}
but the orders of magnitude change. This time we have $Q$ and $R$ both of order $\mathcal{O}(\tilde{\mu} P)$, while $Y_{\rm ig}$ is still of order $\mathcal{O}(\tilde{\mu}^3 P)$, $A$ and $B$ are of the order $\mathcal{O}(P)$ while $C$ and $D$ are of the order $\mathcal{O}(\tilde{\mu}^2 P)$ and $X_{\rm ig}$ is of the order $\mathcal{O}(\tilde{\mu}^3 P)$. For the gravitational potential we again have $\bar{\Psi}$ of the order $\mathcal{O}(P/M)$, but this time with subleading order contributions as well, $\Psi_{\rm osc}$ of the order $\mathcal{O}(\tilde{\mu} \bar{\Psi})$ and $\Psi_{\rm ig}$ of the order $\mathcal{O}(\tilde{\mu}^2 \bar{\Psi})$. Furthermore our assumptions about the additional (non scalar field) components of the cosmic fluid remains the same, which means the the relations (\ref{PhiPsiRelation}) still hold. Note that $A$, $B$, $Q$ and $R$ all have corrections at next to leading order which will be relevant below, but we do not write them here explicitly. Instead we we will split up $A$ according to $A=A_0 + A_1$ with $A_1$ of the order $\mathcal{O}(\tilde{\mu} A_0)$ and accordingly for all other variables.

When evaluating the field equations we have to deal with the fact that there are oscillations with frequency k present in the cosmon perturbations (and possibly in the other components of the cosmic fluid as well). Since we are setting $k^2$ to be of the order $\mathcal{O}(\bar{h}^2/\tilde{\mu})$, we restrict ourselves to the case $k^2 \ll \bar{m}^2$, so that we can still consider all the the quantities $P$, $Q$, $R$, $A$, $B$, $C$ and $\bar{\Psi}$ to be almost constant on the timescale $1/\bar{m}$. As we will discuss further below, our approximation becomes inconsistent for very small wavelengths, but we are not concerned about this. Due to these additional (lower frequency) oscillations one has to take into account that $P'$ is of the order $\mathcal{O}(\tilde{\mu}^{-1/2} \bar{h} P)$ and $P''$ is of the order $\mathcal{O}(\tilde{\mu}^{-1} \bar{h}^2 P)$. When evaluating the field equations we will always include the next ''half order'' in the lower order in $\tilde{\mu}$, e.g. if we evaluate an equation at the order $\mathcal{O}(\tilde{\mu}^0 M \bar{h})$ we will also include terms of the order $\mathcal{O}(\sqrt{\tilde{\mu}} M \bar{h})$.

 \begin{comment}
{\bf Somehow the whole approximation goes awry if we assume the same for $\bar{\Psi}$. We have to assume that the gravitational potential does not follow these oscillations in order for our approximation to work out.}
\end{comment}

Applying this scheme to the cosmon field equation (\ref{CosmonPEquationA}) yields to leading order
\begin{align}
Q_0 &= -\frac{1}{4} \beta \frac{\chi_0}{M} A_0 \, , \\
R_0 &= -\frac{1}{4} \beta \frac{\chi_0}{M} B_0 \, , \\
P'' &= - 2 \bar{h} P' - k^2 P + \beta \bar{a}^2 \bar{m}^2 \frac{\chi_0}{M} A_0 \, .
\end{align}
A comparison of coefficients in the bolon field equation (\ref{BolonPEquationA}) results in the following two equations:
\begin{align}
A_0' =& -\frac{3}{2} \bar{h} A_0 +\frac{k^2}{2 \bar{a} \bar{m}} B_0 + \beta \frac{\varphi _0'}{2 M} A_0 \, ,\\
B_0' =& -\frac{3}{2} \bar{h} B_0 - \frac{k^2}{2 \bar{a} \bar{m}} A_0 -\bar{a} \bar{m} \chi _0 \bar{\Phi } \nonumber \\
&+ \beta \bar{a} \bar{m} \frac{\chi _0}{M} P + \beta  \frac{\bar{\varphi}'}{2 M} B_0 \, .
\end{align}
Evaluating the Poisson-equation (\ref{PsiEquationA}) to leading order gives
\begin{align}
\bar{\Psi} =& -\frac{1}{2} \frac{\bar{a}^2}{k^2 M^2} \left( \overline{\delta \rho}_{\rm ext} - 3 \overline{\left[ (\rho + p)v \right]} _{\rm ext}  \right) \nonumber \\
& - \frac{\bar{a}^2 \bar{m}^2 \chi _0}{2 M^2 k^2} A_0 -\frac{\bar{\varphi}'}{2 M^2 k^2} P' - \frac{1}{2} \frac{\bar{a}^2 \bar{\varphi}'^2}{k^2 M^4} \bar{\Pi}_{\rm tot}\, .
\end{align}
We can now again evaluate equation (\ref{dPsiEquation2}) to leading order and identify the oscillatory part of the gravitational potential to obtain
\beq
\Psi_{\rm osc}' = \frac{\bar{a} \bar{m} \chi _0}{4 M^2} A_0 {\rm sin}(2x) - \frac{\bar{a} \bar{m} \chi _0}{4 M^2}B_0 {\rm cos}(2x) \, ,
\eeq
i.e.
\beq
\Psi_{\rm osc} = \frac{\chi _0}{8 M^2} A_0 {\rm cos}(2x) + \frac{\chi _0}{8 M^2}B_0 {\rm sin}(2x) \, .
\eeq
As a trivial check one can easily verify that the slowly evolving part of equation (\ref{dPsiEquation2}) is indeed satisfied with the $\bar{\Psi}$-solution given above. When doing that, on should however take into account next-to-leading order terms in $P$, since they yield $P'$-terms which have to be included in the $\bar{\Psi}'$-expression in order to remain consistent (and make this calculation work).

Since we will need it for the averaging procedure below, we will also introduce the results from the next-to-leading order calculation. Here the bolon field equation yields
\begin{align}
A_1'  =&  -\frac{3}{2} \bar{h} A_1 + \frac{k ^2}{2 \bar{a} \bar{m}} B_1-\frac{3}{32 M^2} \bar{a} \bar{m} \chi _0^2  B_0 + \frac{\bar{h}}{\bar{a} \bar{m}} B_0' \nonumber \\
&  +\bar{a} \bar{m} \chi _1 \bar{\Phi } + \frac{1}{2} \chi _0 \left( 3 \bar{\Psi}' + \bar{\Phi}' \right) + \frac{1}{2 \bar{a} \bar{m}} B_0'' \nonumber \\
& + \frac{\beta \bar{\varphi}'}{2 M} A_1 + \frac{\beta ^2}{16 M^2} \bar{a} \bar{m} \chi _0^2 B_0 - \frac{\beta}{M}  \bar{a} \bar{m} \chi _1 P \, , \\
B_1' =& -\frac{3}{2} \bar{h} B_1 -\frac{k ^2}{2 \bar{a} \bar{m}} A_1 + \frac{9}{32 M^2} \bar{a} \bar{m} \chi _0^2 A_0 - \frac{\bar{h}}{\bar{a} \bar{m}}  A_0' \nonumber \\
&- \frac{3 \beta ^2}{16 M^2} \bar{a} \bar{m} \chi _0^2 A_0 + \frac{\beta  \bar{\varphi}'}{2 M} B_1 - \frac{1}{2 \bar{a} \bar{m}} A_0'' \, , \\
C =& \frac{\left(13+2 \beta ^2\right)}{128 M^2} \chi _0^2 A_0 \, , \\
D =& \frac{\left(13+2 \beta ^2\right)}{128 M^2} \chi _0^2 B_0 \, . \\
\end{align}
The cosmon field equation gives:
\begin{align}
R_1 =& -\frac{\beta}{4 M} \chi _0 B_1- \beta \frac{k ^2}{16 \bar{a}^2 \bar{m}^2} \frac{\chi _0}{M} B_0  + \frac{\beta}{4 M \bar{a} \bar{m}} \chi_0' A_0 \nonumber \\
&+ \frac{1}{4 \bar{a} \bar{m}} \frac{\chi _0 \bar{\varphi}'}{M^2} A_0 + \frac{\beta}{4 \bar{a} \bar{m} M} \chi _0 A_0' + \frac{1}{16 \bar{a} \bar{m}} \frac{\chi _0^2}{M^2} P' \nonumber \\
&- \frac{\beta}{4 M}  \chi _1 A_0 - \frac{\beta ^2}{8 \bar{a} \bar{m}} \frac{\chi _0 \bar{\varphi}'}{M^2} A_0 +\frac{3 \beta  \bar{h}}{8 \bar{a} \bar{m}} \frac{\chi _0}{M} A_0 \, , \\
Q_1 =& -\frac{\beta}{4 M} \chi _0 A_1 - \beta \frac{k ^2}{16 \bar{a}^2 \bar{m}^2} \frac{\chi_0}{M} A_0 
- \frac{3 \beta  \bar{h}}{8 \bar{a} \bar{m} M} \chi _0 B_0 \nonumber \\
&+\frac{\beta ^2}{4 M^2} \chi _0^2 P 
-\frac{\beta }{4 M} \chi _0^2 \bar{\Phi }
+\frac{\beta }{4 M} \chi _1 B_0 \nonumber \\
& -\frac{\beta}{4 \bar{a} \bar{m} M} \chi _0 B_0'
-\frac{1}{4 \bar{a} \bar{m} M^2} \chi _0 \bar{\varphi}' B_0 \nonumber \\
&+\frac{\beta ^2}{8 M^2 \bar{a} \bar{m}} \chi _0 \bar{\varphi}' B_0 
-\frac{\beta}{4 \bar{a} \bar{m} M} \chi_0' B_0 \, .
\end{align}

Now we can use these results to perform the same averaging we executed for large wavelengths. We will however only quote the results here, the details of the calculation are nor particularly enlightening. The averaged quantities now read (to subleading order):
\begin{align}
<\delta \rho_\chi> =& \bar{m}^2 \left( \chi _0 A_0 + \chi_1 B_0 \right) - 2 \beta \bar{\rho}_\chi \frac{P}{M}  \nonumber \\
&+ \frac{k^2}{4 \bar{a}^2} \chi_0 A_0 + \bar{m}^2 \chi_0 A_1 \, ,  \\
<\delta p_\chi > =& \frac{k^2}{4 \bar{a}^2} \chi_0 A_0 \, , \\
<\left[ (\rho + p)v \right]_\chi> =& \frac{\bar{m}}{2 \bar{a}} \left( \chi_0 B_0 - \chi_1 A_0 \right) +  \frac{3 \bar{h}}{4 \bar{a}^2} \chi _0 A_0 \nonumber \\
&- \frac{\beta  \chi _0 \left(\varphi _0\right)'}{4 \bar{a}^2 M} A_0 + \frac{\bar{m}}{2 \bar{a}} \chi_0 B_1 \, .
\end{align}
Again, all these quantities contain oscillatory parts at the order considered here, but in the case of $\delta \rho_\chi$ only at subleading order. Note that since $<\delta p_\chi>$ is surpressed by a factor of $\tilde{\mu}$ compared to $<\delta \rho_\chi>$, we have to the order considered here:
\beq
\label{ssEQ}
<\delta p_\chi> = \frac{k^2}{4 \bar{a}^2 \bar{m}^2} <\delta \rho_\chi> \, .
\eeq
So we have apparently reproduced a small, scale-dependent sound speed $c_{s,\chi}^2 = k^2/4\bar{a}^2\bar{m}^2$, which generalizes earlier findings in refs. \cite{Hu:2000ke,Matos:2000ss}. This should however be verified by averaging the entire equations of energy-and momentum conservation, taking care of all possible interferences between background and perturbative oscillations to subleading order in $\tilde{\mu}$, as we did above in the case of large wavelength. Doing this yields the following final equations for the density contrast and the velocity potential:
\begin{align}
\label{deltaChiEquationA}
<\delta_\chi>' =&- \bar{\Theta}_\chi + 3 \bar{\Psi}' - \beta \frac{P'}{M} + \beta \frac{\bar{\varphi}'}{M} \left( \bar{\Psi} - \bar{\Phi} \right) \nonumber \\
& - \left( 3 \bar{h} - \beta \frac{\bar{\varphi}'}{M} \right) c_{s,\chi}^2 <\delta_\chi>  \, , 
\end{align}
\begin{align}
\label{thetaChiEquationA}
\bar{\Theta}_\chi' =& - \bar{h} \bar{\Theta}_\chi + k^2 \bar{\Phi} + \beta \frac{\bar{\varphi}'}{M} \bar{\Theta}_\chi - \beta k^2 \frac{P}{M} \nonumber \\
& + k^2 c_{s,\chi}^2 <\delta_\chi> \, .
\end{align}
If we include subleading non-oscillatory terms in the Poisson-equations it again reads as in equation (\ref{poissonEquationA}). 

As was already mentioned in the main text, this set of equations is can not be derived from equations (25) to (30) in ref. \cite{Amendola:1999dr}. This is due to the fact that the equations in ref. \cite{Amendola:1999dr} were derived for a barotropic fluid, i.e. for a fluid without non-adiabatic pressure perturbations (see appendix of ref. \cite{EarlyScalings} for more details). The adiabatic sound speed is a background quantity which can be derived from the equation of state. Since we have $<\omega_\chi>=<\omega_\chi'>=0$ to all orders, we can directly deduce that the adiabatic sound speed is zero, and our sound speed is therefore non-adiabatic. As can be seen from comparing equations (\ref{deltaChiEquationA}) and (\ref{thetaChiEquationA}) to ref. \cite{Amendola:1999dr}, this makes no difference in the uncoupled scenario, as expected. The coupling however enters the equations differently.

\section{Comparison with CLASS}
\label{app:ClassComp}
\begin{figure}[h]
	\centering
  \includegraphics[width=1.0\linewidth]{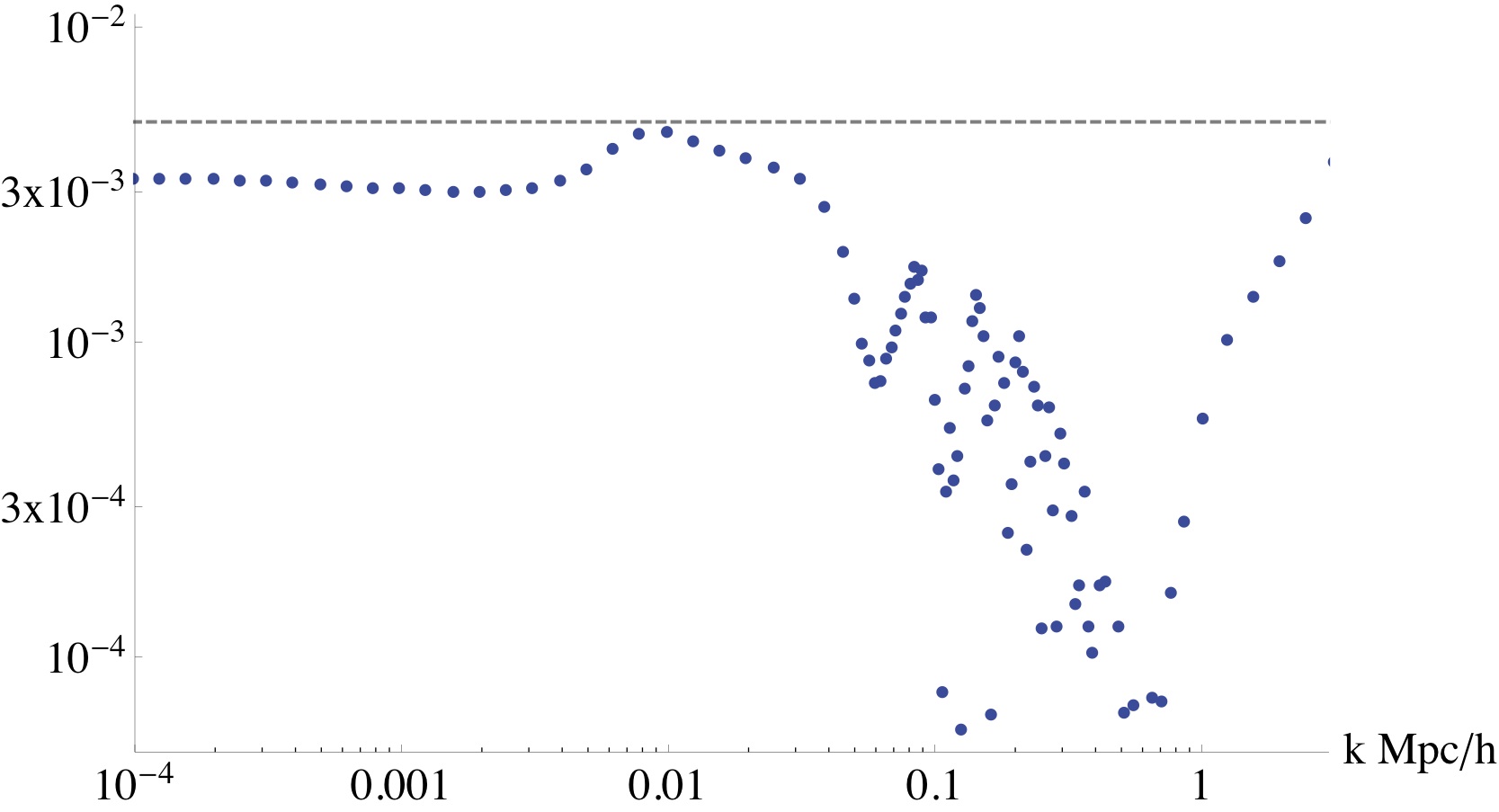}
	\caption{The absolute relative error for our matter power spectrum when compared to the results of the CLASS-code. The gray line is placed at $0.005$.}
	\label{fig:relClassComp}
\end{figure}

In this section we quickly compare the results of our Boltzmann-Code for the standard $\Lambda$CDM model with the results of the recently released CLASS-code \cite{Lesgourgues:2011re,Blas:2011rf}. To see that both codes agree to excellent accuracy, we simply plot the absolute relative error for the matter power-spectrum at $z=0$, which is the quantity we are most interested in in this work, i.e. we plot
\beq
\frac{\left| P_m^{\rm our} (k) - P_m^{\rm CLASS} (k) \right|}{P_m^{\rm CLASS}(k)} \, .
\eeq
The results for a run with the standard settings of CLASS and the WMAP-7 parameters (which we use throughout this work) are shown in Figure \ref{fig:relClassComp}, and we see that it is better than half a percent everywhere. This is the same order of magnitude as the error that gets introduced by switching between different approximation schemes within CLASS (see ref. \cite{Blas:2011rf}), and thus more than good enough for our purposes.

\section{Fitting the power-spectrum cutoff}
\label{app:CutoffFitting}
In this appendix we quickly provide a short overview over the procedure employed to find the analytical cutoff fitting and the corresponding accuracy. As basic data we used numerically evolved power spectra for a coupled CDM model and the coupled bolon model for a wide range of parameters. The evolution equations for the coupled CDM model are similar to those for presented for the averaged (late-time) coupled bolon model in section \ref{sec:LinearPerturbations}, except that the sound-speed is set to zero. The initial conditions for the adiabatic mode in the coupled CDM model are of course different, but easy to derive or even guess from the results of ref. \cite{EarlyScalings} and of no particular interest to us here. 
\begin{figure}[t]
	\centering
  \includegraphics[width=1.0\linewidth]{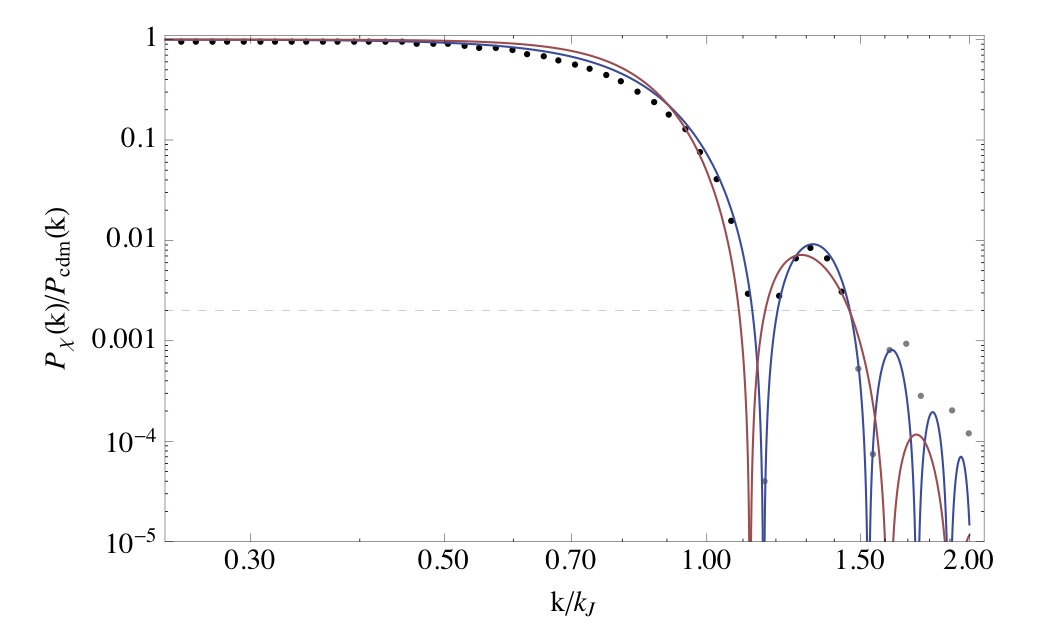}
	\caption{Fitting the power spectrum cutoff for $\lambda = 28$ and $\beta=-0.2$. The simple fitting (\ref{simpleFit}) is given in red, the optimized one (\ref{goodFit}) in blue.}
	\label{fig:relPS}
\end{figure}
\begin{figure}[t]
	\centering
  \includegraphics[width=1.0\linewidth]{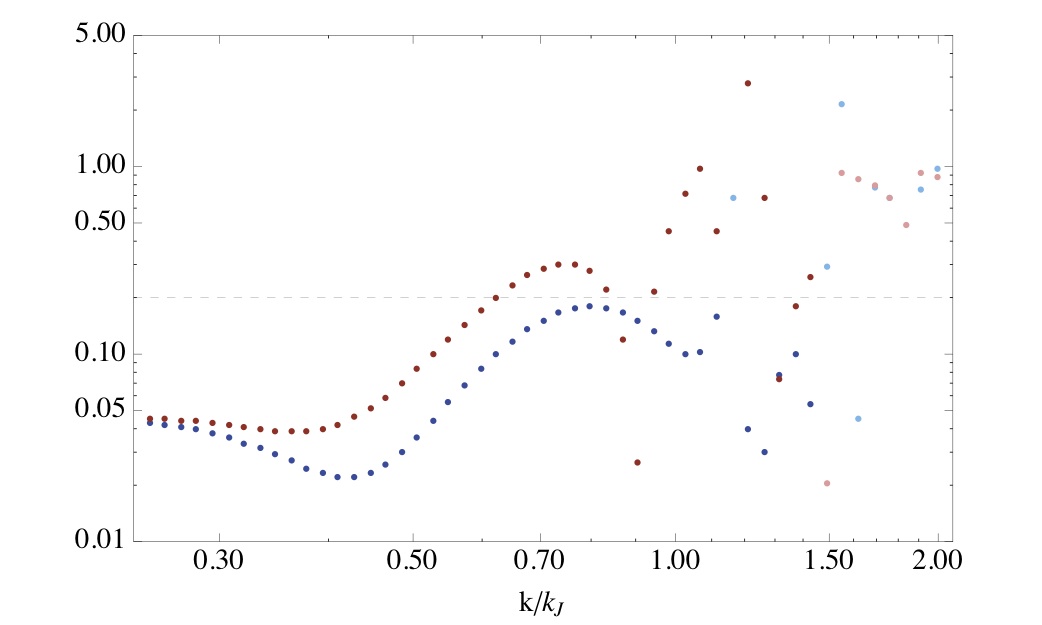}
	\caption{Relative errors for the fitting given in figure \ref{fig:relPS}. Mor information is given in the text.}
	\label{fig:relErr}
\end{figure}
The grid we investigated ran through the the parameters $\lambda$ and $\beta$, the boundaries we chose to investigate are $[20,30]$ for $\lambda$ and $[-0.2,0.1]$ for $\beta$. As was already mentioned earlier, larger couplings are observationally excluded and also in conflict with the early universe-evolution assumed in this paper.

Motivated by earlier results in refs. \cite{Hu:2000ke,Matos:2000ss}, we assumed that the cutoff can be well fit by the relation
\beq
\label{simpleFit}
 P_{\chi} (k) = P_{\rm cdm} (k) \left\{ \frac{{\rm cos} \left[ (b \, x)^a \right]}{1+c \, x^{d}} \right\}^2 \, ,
 \eeq
 where $x=k/k_J$. In \cite{Matos:2000ss}, a good fit of the cutoff in a similar model was reported to be given by $a=3$, $c=1$ and $d=8$. Due to the slightly different definitions of the Jeans scale $k_J$, we adjusted the parameter $b$ to best fit the data. By best fit we understand an minimalization of the maximal relative error when the fitting is compared to the numerical results for the entire grid. Furthermore we did not use the complete power spectra, but cut off wavenumbers for which the suppression compared to the CDM result was below a given threshold, which we denote by $t_c$. 
 
For $t_c = 1/500$ the best fit is given by $b=1.0725$ , this does however already result in a maximal relative error of order $3$. Raising the threshold to $t_c=1/100$ gives  the same $b=1..0725$, but the maximal relative error is still $0.99$. 

As the relative errors are still very big, we have improved the fitting function of the following form:
\beq
\label{goodFit}
 P_{\chi} (k) = P_{\rm cdm} (k) \left\{ \frac{{\rm cos} \left[ (b \, x)^a \right]}{1+\sqrt{c} \, \left( x^{d_1} + x^{d_2} \right)} \right\}^2 \, ,
\eeq
where this time the parameters $a,b,c,d_1$ and $d_2$ are functions of the model parameters $\lambda$ and $\beta$. We assumed a linear dependence for all functions here, optimized the parameters numerically with $t_c=1/500$ and eliminated all terms the inclusion of which would not yield a significant improvement of the fitting. The resulting best estimate is given by
\begin{align}
a(\beta) & = 4.105 + 0.428 \beta \, , \\
b(\lambda,\beta) & = 0.827 + 0.098 \beta + 0.006 \lambda + 0.0025 \lambda \beta \, , \\
c(\lambda,\beta) & = -0.46 -1.9 \beta + 0.053 \lambda + 0.152 \lambda \beta \, , \\
d_1(\beta) & = 4.31-1.2\beta \, , \\
d_2(\beta) &= 7.66 + 1.49 \beta \, .
\end{align}
This gives a fitting better than 18\%. For $t_c=1/100$ a fitting better than 8\% can be achieved with the same parameterization. A visualization of the cutoff fitting can be seen in Figures \ref{fig:relPS} and \ref{fig:relErr}. Figure \ref{fig:relPS} shows the numerically obtained ratio $P_\chi/P_{\rm cdm}$ for $\lambda = 28$ and $\beta=-0.2$, represented by points in black above the threshold $t_c=1/500$ and grey below it. In addition to that we also show the estimated ratios given by equation (\ref{simpleFit}) in red, and the more complicated one originating from equation (\ref{goodFit}) in blue. The parameters we chose from our grid were the ones giving the biggest errors for the optimized fitting, but not for the simple one. The corresponding relative errors can be seen in Figure \ref{fig:relErr}, again in blue for the optimized fitting, and in red for the simple one. The points we cut off with our threshold are shown in lighter colors, they were not included in the fitting.

\bibliographystyle{unsrtnat}
\bibliography{bibCBPers}

\end{CJK*}
\end{document}